\documentclass[fleqn,usenatbib]{mnras}

\usepackage[T1]{fontenc}
\usepackage{ae,aecompl}

\usepackage{times}

\usepackage{graphicx}	
\usepackage{amsmath}	
\usepackage{amssymb}	

\usepackage{url}

\newcommand{\vect}[1]{\boldsymbol{#1}}
\newcommand{\diff}{\text{d}}
\newcommand{\imag}{\text{i}}
\newcommand{\e}{\text{e}}

\newcommand{\Imag}{\text{Im}}

\newcommand{\mratio}{y}
\newcommand{\Eoff}{\ensuremath{\overline{E}_{a, {\rm lin}}}}
\newcommand{\Edot}{\ensuremath{\dot{E}_{\rm diss}}}
\newcommand{\Porb}{\ensuremath{P_{\rm orb}}}

\date{}
\pubyear{2020}
\title[Nonlinear tides in white dwarf binaries]{Nonlinear dynamical tides in white dwarf binaries}

\author[H. Yu, N. N. Weinberg and J. Fuller]{Hang Yu$^{1, 2}$\thanks{E-mail: hangyu@caltech.edu}, Nevin N. Weinberg$^{1}$ and Jim Fuller$^{2}$\\
$^{1}$Department of Physics, and MIT Kavli Institute, Massachusetts Institute of Technology, Cambridge, MA 02139, USA \\
$^{2}$TAPIR, Walter Burke Institute for Theoretical Physics, Mailcode 350-17 California Institute of Technology, Pasadena, CA 91125}

\begin{document}

\defcitealias{Weinberg:12}{WAQB12}
\defcitealias{Essick:16}{EW16}
\defcitealias{Fuller:12a}{FL12}
\defcitealias{Burkart:13}{BQAW13}

\label{firstpage}
\pagerange{\pageref{firstpage}--\pageref{lastpage}}
\maketitle

\begin{abstract}
Compact white dwarf (WD) binaries are important sources for space-based gravitational-wave (GW) observatories, and an increasing number of them are being identified by surveys like ZTF. We study the effects of nonlinear dynamical tides in such binaries. We focus on the global three-mode parametric instability and show that it has a much lower threshold energy than the local wave-breaking condition studied previously. By integrating networks of coupled modes, we calculate the tidal dissipation rate as a function of orbital period. We construct phenomenological models that match these numerical results and use them to evaluate the spin and luminosity evolution of a WD binary. While in linear theory the WD's spin frequency can lock to the orbital frequency, we find that such a lock cannot be maintained when nonlinear effects are taken into account. Instead, as the orbit decays, the spin and orbit go in and out of synchronization.  Each time they go out of synchronization, there is a brief but significant dip in the tidal heating rate. While most WDs in compact binaries should have luminosities that are similar to previous traveling-wave estimates, a few percent should be about ten times dimmer because they reside in heating rate dips. This offers a potential explanation for the low luminosity of the CO WD in J0651. Lastly, we consider the impact of tides on the GW signal and show that LISA and TianGO can constrain the WD's moment of inertia to better than $1\%$ for deci-Hz systems.
\end{abstract}

\begin{keywords}
instabilities - white dwarfs - stars: oscillations (including pulsations)- binaries (including multiple): close - gravitational waves
\end{keywords}

\section{INTRODUCTION}
\phantom{put this here because if I don't I get some weird compiler error}

As binary white dwarfs (WDs) with short orbital periods inspiral due to the emission of gravitational waves (GWs), they can evolve into a variety of interesting systems, including AM CVn stars \citep{Nelemans:01}, R Cor Bor stars \citep{Clayton:12}, and rapidly rotating magnetic WDs \citep{Ferrario:15}.
Merging WDs may also explode as type Ia supernovae \citep{Webbink:84,Iben:84,Toonen:12,Polin:19} or in other types of luminous thermonuclear events \citep{Shen:18,Polin:20}. Compact WD binaries emit GWs with frequencies of $\approx 1-100\textrm{ mHz}$, which makes them prominent sources for proposed space-based GW observatories such as the Laser Interferometer Space Antenna (LISA, \citealt{Amaro-Seoane:17}), TianQin~\citep{Luo:16}, and TianGO~\citep{Kuns:19}. 

The tidal interaction between the binary components spins them up and heats their interiors. As they inspiral, the tide becomes progressively stronger and eventually their spin frequency nearly equals the orbital frequency. However, they never become perfectly synchronous because of the continual GW-induced orbital decay. The degree of spin asynchronicity affects the tidal heating rate and luminosity of the WDs \citep{Iben:98,Fuller:12a,Fuller:13,Piro:19} and the outcome of their potential merger \citep{Raskin:12,Dan:14,Fenn:16}. 

The dominant mechanism of tidal dissipation is most likely the excitation of internal gravity waves, either in the form of standing waves (i.e., g-modes; \citealt{Fuller:11,Burkart:13}), or traveling waves  \citep{Fuller:12a,Fuller:12b,Fuller:13,Fuller:14}. 
As we will show, for orbital periods between approximately $10\textrm{ min}$ and $150\textrm{ min}$, which describes  many of the observed WD binaries, the resonant g-modes excited by the tide have such large amplitudes that they cannot be considered small, linear perturbations to the background star. On the other hand, the amplitudes are not so large that the modes break due to strong nonlinearities. The tidal dynamics and dissipation in this intermediate, weakly nonlinear regime are complicated and depend on details of the nonlinear coupling between g-modes driven directly by the tide and the sea of secondary modes they excite. 

In this Paper, we apply the weakly nonlinear tidal  formalism developed in \citet{Weinberg:12} to study tides in WD binaries. 
Our study fills the gap between those that assume the excited modes are linear standing waves (e.g., \citealt{Fuller:11,Burkart:13}) and those that assume they break and form strongly nonlinear traveling waves \citep{Fuller:12a,Fuller:12b,Fuller:13,Fuller:14}.
In Section~\ref{sec:model}, we present the background WD model we use throughout much of our analysis.  In Section~\ref{sec:formalism}, we describe the mode coupling and tidal driving equations that governs the mode dynamics and in Section~\ref{sec:num} we describe our numerical method for solving these  equations.  In Section~\ref{sec:results}, we present our solutions of the mode dynamics and show how tidal dissipation and synchronization varies with orbital period in the weakly nonlinear regime. We also compare our results with the previous studies that assumed the tide was either linear or strongly nonlinear.  In Section~\ref{sec:obs_signatures}, we describe the observable electromagnetic and GW signatures of the tidal interaction, including the tidal heating luminosities, GW phase shifts, and projected constraints on the WD moment of inertia. In Section~\ref{sec:conclusion}, we summarize our key results and conclude.

\section{BACKGROUND MODEL}
\label{sec:model}

We use \texttt{MESA} (version 10398; \citealt{Paxton:11, Paxton:13, Paxton:15, Paxton:18}) to construct a WD model, whose key parameters are summarized in Table~\ref{tab:bgConfig}.  To construct this model, we adopt parameters similar to those used by \citet{Timmes:18}.  Specifically, we start with a pre-main sequence star with an initial mass of $2.8\,M_\odot$ and metallicity $Z=0.02$ and let it evolve to a CO WD with mass $M=0.6\,M_\odot$ and effective temperature $T_{\rm eff}=9000\,K$. We include element diffusion, semiconvection, and thermohaline mixing throughout the evolution. We use \texttt{GYRE}~\citep{Townsend:13, Townsend:18} to compute the model's eigenmodes and construct our mode networks.

\begin{figure}
   \centering
   \includegraphics[width=0.45\textwidth]{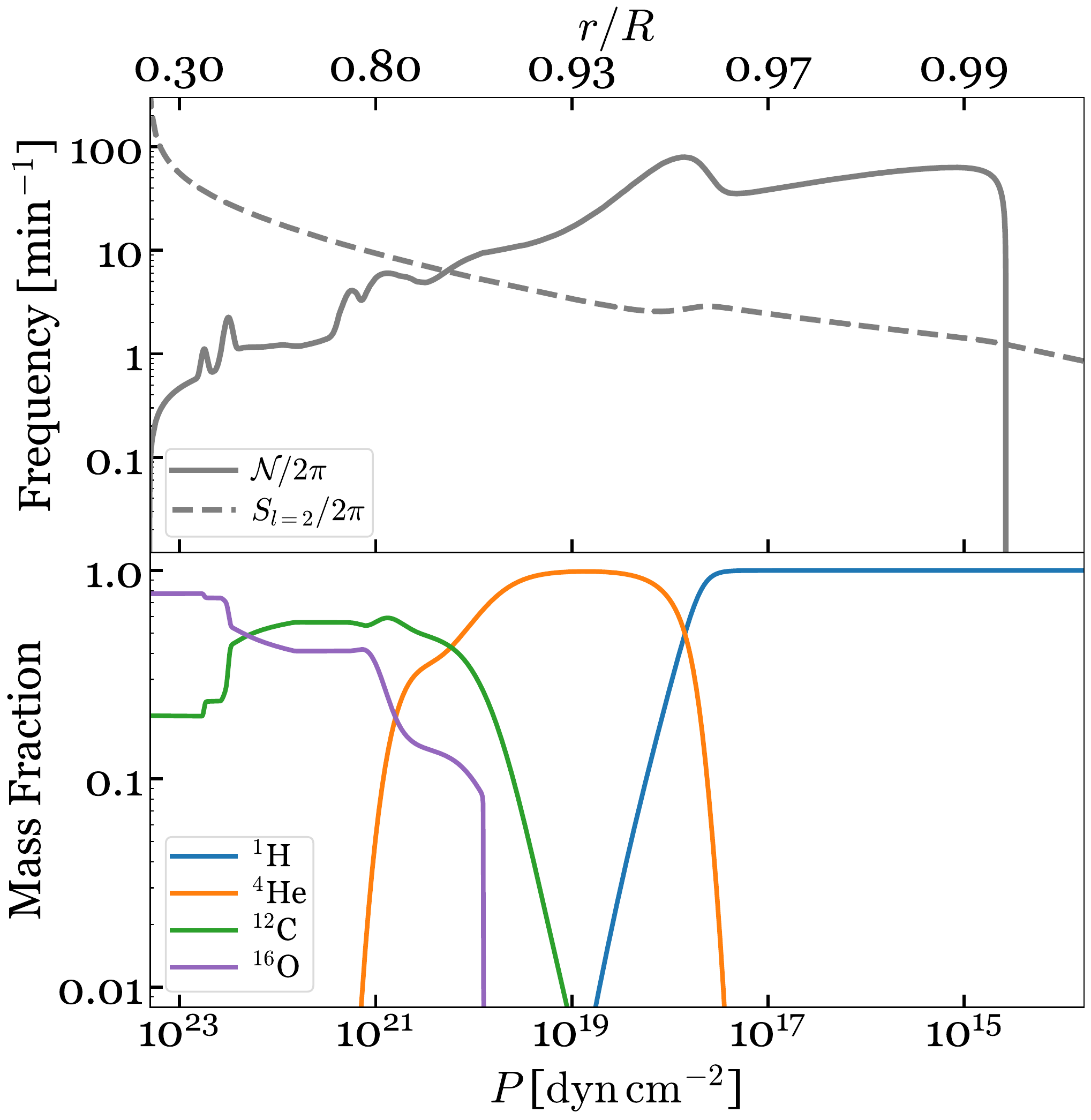} 
   \caption{Propagation diagram (top panel) and composition profile (bottom panel) of our WD model from the center to the surface. Note that we have oriented the bottom x-axis such that the radius increases to the right. }
   \label{fig:prop_diag}
\end{figure}

In the upper panel of Figure~\ref{fig:prop_diag}, we show the propagation diagram of our WD model. The solid line is the buoyancy frequency $\mathcal{N}$, where
\begin{equation}
\mathcal{N}^2 = g^2 \left(\frac{1}{c_{\rm e}^2} - \frac{1}{c_{\rm s}^2}\right),
\end{equation}
$c^2_{\rm e}=\diff P/\diff \rho$ is the equilibrium sound speed squared, $c_{\rm s}^2=\Gamma_1 P/\rho$ is the adiabatic sound speed squared, and $\Gamma_1$ is the adiabatic index. All other quantities have their usual meaning. The dashed line is the Lamb frequency $S_l$ for $l=2$, where 
\begin{equation}
S_l^2 = \frac{l (l+1) c^2_{\rm s}}{r^2}. 
\end{equation}
For the short-wavelength g-modes that comprise the dynamical tide, the square of the radial wavenumber  
\begin{equation}
k_r^2 = \frac{\omega^2}{c_{\rm s}^2}\left(\frac{S_l^2}{\omega^2} - 1\right)\left(\frac{\mathcal{N}^2}{\omega^2}-1\right),
\label{eq:kr_sq}
\end{equation}
where $\omega$ is the angular eigenfrequency of the mode. A g-mode  propagates where $k_r^2>0$, i.e., in regions where $\omega<\mathcal{N}$ and $\omega<S_l$, and is evanescent where $k^2_r<0$. 

The lower panel of Figure~\ref{fig:prop_diag} shows the composition profile of our model.  As is typical of stars supported by degeneracy pressure, the buoyancy is due largely to composition gradients, with peaks in $\mathcal{N}$ associated with sharp transitions in the internal composition.

\begin{table}
\begin{center}
\caption{\label{tab:bgConfig}The mass $M$, radius $R$, effective temperature $T_{\rm eff}$, and moment of inertia $I_{\rm WD}$, of our WD model. We will often express results in terms of the primary's natural units of energy  $E_0{\equiv} GM^2/R = 1.10\times10^{50}\textrm{ erg}$ and frequency $\omega_0{\equiv}\sqrt{GM/R^3}=0.346 \textrm{ rad s}^{-1}$.}
\begin{tabular}{cccc}
$M$              & $R$       
& $T_{\rm eff}$ &      $I_{\rm WD}$      \\
\hline
$0.6M_\odot$ & $8.75\times10^8\textrm{ cm}$ & 9000 K & $0.257 MR^2$ \\
\end{tabular}
\end{center}
\end{table}

\section{FORMALISM}
\label{sec:formalism}

\subsection{Equation of motion}
\label{sec:eom}
Consider a primary star of mass $M$ and a secondary star of mass $M'$ and choose a coordinate system whose origin is at the center of the primary and co-rotates with it. We assume that the orbit is circular and that the spin angular momentum of the primary is aligned with the orbital angular momentum.  For simplicity, we do not account for the effect of rotation on the mode dynamics except through the Doppler shift of the tidal driving frequency. The equation of motion governing the Lagrangian displacement field $\vect{\xi}(\vect{r}, t)$ of a perturbed fluid element at location $\vect{r}$ at time $t$ is then~(see, e.g., \citealt{Weinberg:12}, hereafter \citetalias{Weinberg:12})
\begin{equation}
\rho \ddot{\vect{\xi}} = \vect{f_1}[\vect{\xi}] + \vect{f_2}[\vect{\xi}, \vect{\xi}] + \rho \vect{a}_{\rm tide},
\label{eq:eom}
\end{equation}
where $\vect{f_1}$ and $\vect{f_2}$ represent the linear and leading-order nonlinear internal restoring forces,  and
\begin{equation}
\vect{a}_{\rm tide} = - \nabla U - \left(\vect{\xi} \cdot  \nabla \right) \nabla U
\end{equation}
is the tidal acceleration. The tidal potential can be expanded as
\begin{equation}
U(\vect{r}, t) = - \! \sum_{l\geq 2, m} \! \! W_{lm}\frac{G M'}{D(t)} \left[\frac{r}{D(t)}\right]^l \!\! Y_{lm}(\theta, \phi) {\rm e}^{-\imag m (\Omega_{\rm orb} - \Omega_{\rm s})t},
\label{eq:potential}
\end{equation}
where $Y_{lm}$ is the spherical harmonic function, and $D$, $\Omega_{\rm orb}$, and $\Omega_{\rm s}$ are the orbital separation, the orbital angular frequency, and the spin frequency of the primary, respectively. We focus on the leading order quadrupolar ($l=2$) tide, whose non-vanishing $W_{lm}$ coefficients  are $W_{2\pm2}=\sqrt{3\pi/10}$ and $W_{20}=-\sqrt{\pi/5}$.  It is useful to define
\begin{equation}
\epsilon = \left(\frac{M'}{M}\right) \left( \frac{R}{D}\right)^3 = \left(\frac{\mratio}{1+\mratio}\right) \left(\frac{\Omega_{\rm orb}}{\omega_0}\right)^2,
\label{eq:epsilon}
\end{equation}
where $\mratio=M'/M$ is the mass ratio. The quantity $\epsilon$ characterizes the overall tidal strength and will be useful when we want to distinguish the system's dependence on the tidal strength from its dependence on the driving frequency $2(\Omega_{\rm orb} - \Omega_{\rm s})$.

In order to solve Equation~(\ref{eq:eom}), we expand the six-dimensional phase space vector as
\begin{equation}
   \begin{bmatrix} 
      \vect{\xi} (\vect{r}, t) \\
      \dot{\vect{\xi}}(\vect{r}, t)  \\
   \end{bmatrix}
=\sum q_a(t) 
  \begin{bmatrix} 
      \vect{\xi}_a (\vect{r}) \\
      -\imag \omega_a \vect{\xi}_a(\vect{r})  \\
   \end{bmatrix},
   \label{eq:expansion}
\end{equation}
where $q_a(t)$, $\omega_a$, and $\vect{\xi}_a(\vect{r})$,  are the amplitude, frequency, and displacement of an eigenmode labeled by subscript $a$.    The frequency and displacement are found by solving the linear, homogeneous equation 
\begin{equation}
\vect{f_1}[\vect{\xi}_a] = -\rho \omega_a^2 \vect{\xi}_a,
\end{equation}
which we normalize as
\begin{equation}
2\omega_a^2 \int \diff^3 r \rho \vect{\xi}_{a}^\ast \cdot\vect{\xi}_{b}  = \frac{GM^2}{R} \delta_{ab} \equiv E_0 \delta_{ab}. 
\label{eq:xi_norm}
\end{equation}
Each eigenmode has a unique set of three quantum numbers: its angular degree $l_a$, azimuthal order $m_a$, and radial order $n_a$. The summation in Equation (\ref{eq:expansion}) runs over all mode quantum numbers and both signs of eigenfrequency in order to include each mode and its complex conjugate\footnote{If the amplitudes $q_{a+}$ and $q_{a-}$ correspond to eigenfrequencies $\omega_a$ and $-\omega_a$, respectively, then the reality of $\vect{\xi}$ requires $q_{a+} = q_{a-}^\ast$, where the asterisk denotes complex-conjugation. }. Using the orthogonality of the eigenmodes, Equation~(\ref{eq:eom}) can now be expressed as a set of evolution equations for the mode amplitudes 
\begin{equation}
\dot{q}_a + (\imag \omega_a + \gamma_a) q_a 
	= \imag \omega_a \left[U_a + \sum_b U_{ab}^\ast q_{b}^\ast + \sum_{bc}\kappa_{a b c} q_b^\ast q_c^\ast \right], \label{eq:ode_amp}
\end{equation}
where 
\begin{align}
\label{eq:Ua_def}
&U_a(t) = -\frac{1}{E_0} \int \diff^3 r \rho \, \vect{\xi}_a^\ast \cdot \nabla U, \\
&U_{ab}(t) = -\frac{1}{E_0} \int \diff^3 r \rho \,\vect{\xi}_a \cdot \left(\vect{\xi}_b \cdot \nabla\right) \nabla U, \\
&\kappa_{abc} = \frac{1}{E_0} \int \diff^3 r \, \vect{\xi}_a \cdot \vect{f_2}\left[\vect{\xi}_b, \vect{\xi}_c\right]. \label{eq:kappa_def}
\end{align}
The linear and nonlinear tidal coefficients $U_a$ and $U_{ab}$ characterize the strength of the coupling of modes to the tide,  and the three-mode coupling coefficient $\kappa_{abc}$ characterizes the strength of the coupling of modes to each other.

We further simplify Equation (\ref{eq:ode_amp}) by noting that the three-mode coupling involving the equilibrium tide cancels significantly with the nonlinear tide (i.e., $\sum_{c\in{\rm{eq}}} \kappa_{abc} q_c^\ast \simeq -U_{ab}$; \citetalias{Weinberg:12}). We therefore ignore $U_{ab}$ and the equilibrium tide and focus on the dynamical tide.\footnote{We note that the nonlinear driving by equilibrium tide might be unstable depending on the residual coupling. Roughly, the growth rate for the equilibrium-tide-driven instability is $\Gamma_{\rm nl}^{\rm (eq)}\sim \epsilon \kappa_{abc}^{\rm (eq)} \left(\Omega_{\rm orb}-\Omega_{\rm s}\right)$. If the residual coupling $\kappa_{abc}^{(\rm eq)}\sim 1$ after accounting for the cancellation with $U_{ab}$, then we have $\Gamma_{\rm nl}^{\rm (eq)} \sim 10^{-8}\,{\rm s}$ at $P_{\rm orb}=50\,{\rm min}$. This, while smaller than the nonlinear growth rate of the dynamical tide [Equation~(\ref{eq:tau_nl})], could be greater than the damping rate of the resonant $l=2$ modes at the same period. We defer the study of this effect to future work.} The latter is dominated by the linear driving of the most-resonant $l_a=|m_a|=2$ modes, for which $|\Delta_a / \omega_a| \ll 1$, where $\Delta_a=\omega-\omega_a$ is the linear detuning and $\omega=2(\Omega_{\rm orb}-\Omega_s)$ is the linear driving frequency.  We refer to such linearly resonant modes as parent modes.  By contrast, the other modes in our networks (the daughters, granddaughters, etc.)  are primarily excited through three-mode parametric resonances rather than direct driving by the tide since they have large $|\Delta_a|$ and smaller $U_a$ than the parents [since they have larger $n_a$ and $l_a$; see Equations (\ref{eq:WKB_Q_a}) and (\ref{eq:l_b_n_b_exp})].  In our mode network calculations, we  therefore solve a reduced set of amplitude equations in which the parent modes $\{a\}$ satisfy
\begin{equation}
\dot{q}_a + (\imag \omega_a + \gamma_a) q_a 
= \imag \omega_a U_a +  \imag \omega_a \sum_{bc}\kappa_{a b c} q_b^\ast q_c^\ast ,
\label{eq:parent_amp_eqn}
\end{equation}
and the daughter modes $\{b,c\}$ satisfy
\begin{equation}
\dot{q}_b + (\imag \omega_b + \gamma_b) q_b 
= \imag \omega_b \sum_{ac}\kappa_{a b c} q_a^\ast q_c^\ast, \end{equation}
and similarly for the granddaughters, great-granddaughters, etc.

The energy of a mode $E_a (t)$ is related to its amplitude by\begin{equation}
E_a(t) = q_a^\ast (t) q_a(t) E_0.
\label{eq:E_a_def}
\end{equation}
This neglects the energy in the three-mode coupling, 
\begin{equation}
\frac{1}{3} \sum_{b, c} \kappa_{abc} q_a q_b q_c + {\rm c.c.}, 
\label{eq:E_a_nl_def}
\end{equation}
where c.c. stands for complex conjugate. As we show in Section~\ref{sec:energy_and_AM_transfer}, 
this energy is much less than $\sum_a E_a$ and therefore, we will use Equation~(\ref{eq:E_a_def}) to represent the mode energy.

\subsection{Power-law relations for the coefficients}
\label{sec:powerlaw_relations}

\begin{figure}
   \centering
   \includegraphics[width=0.45\textwidth]{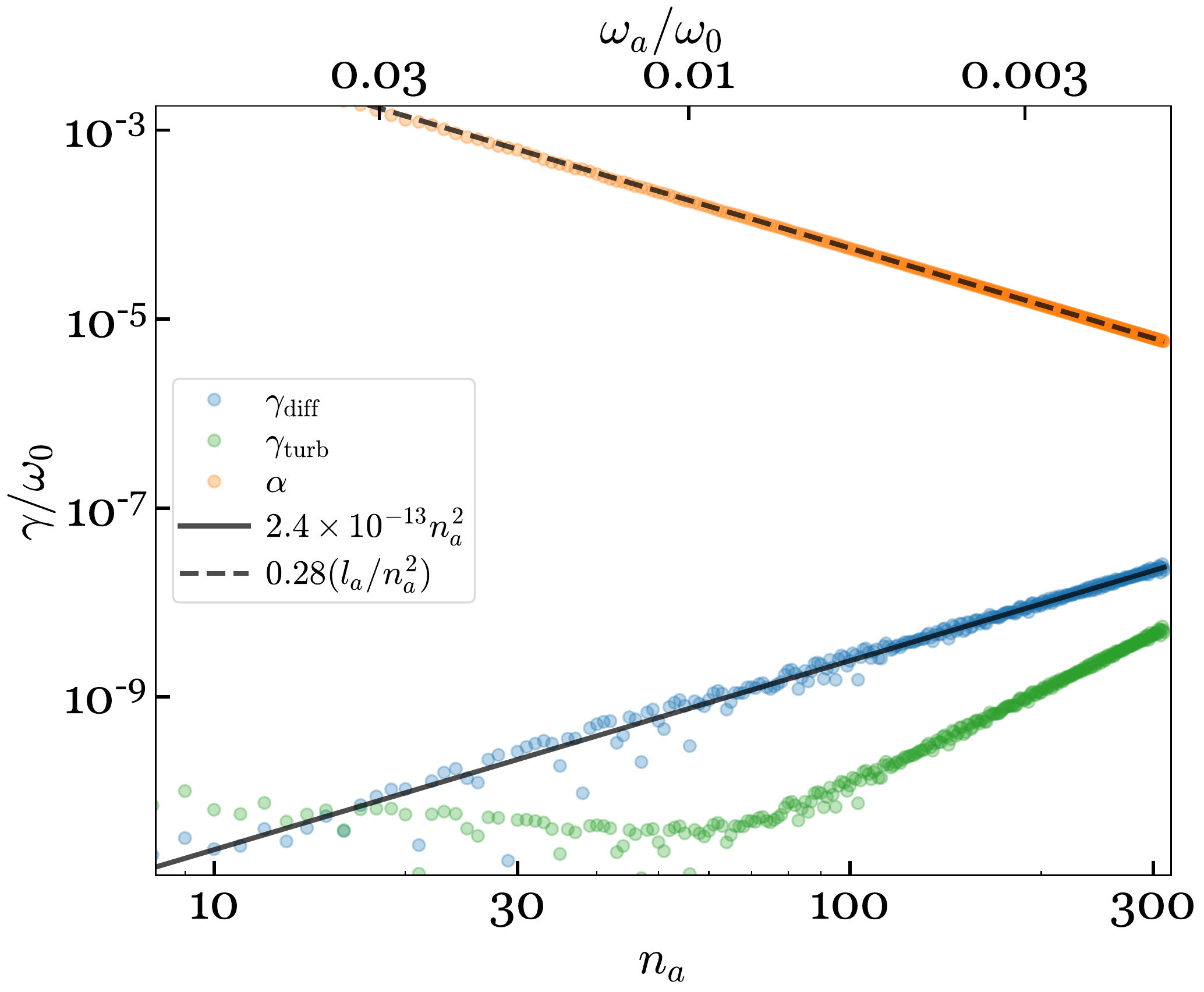} 
   \caption{Linear damping rate of $l_a=2$ modes. The blue and green circles represent the respective contributions of radiative diffusion $\gamma_{\rm diff}$ and convective turbulence $\gamma_{\rm turb}$. The orange circles are the inverse group traveling time $\alpha$ of each mode (see Section~\ref{sec:sw_vs_tw}). The solid and dashed black lines show the WKB scaling relations [Equations~(\ref{eq:WKB_gamma}) and (\ref{eq:WKB_alpha_a})].}
   \label{fig:gam}
\end{figure}

In Appendix~\ref{sec:WKB_relations} we describe our calculations of $\gamma_a$, $U_a$, and $\kappa_{abc}$ in detail. For the tidal synchronization problem, we are mostly interested in binaries with orbital periods in the range $P_{\rm orb}=[10, 100]\textrm{ min}$, which corresponds to $l_a=2$ parent modes with radial orders in the range $n_a\simeq [10,100]$ [see Equation (\ref{eq:WKB_Pa})]. For such high-order modes, we find that the coefficients follow simple power-law relations in $n_a$ and $l_a$.  

We find that the eigenfrequencies of our WD model are approximately given by
\begin{equation}
\omega_a \simeq 
0.28\frac{l_a}{n_a} \omega_0 = 
0.098 \frac{l_a}{n_a}\ {\rm rad\,s^{-1}},  \label{eq:WKB_omega}
\end{equation}
i.e., the mode periods are given by
\begin{equation}
P_a =\frac{2\pi}{\omega_a} \simeq 
1.1 \frac{n_a}{l_a}\ {\rm min},
\label{eq:WKB_Pa}
\end{equation}
where $\omega_0=\sqrt{GM/R^3}$ is the dynamical frequency of the WD.

In Figure~\ref{fig:gam}, we show the linear dissipation rates $\gamma_a$. The dissipation is dominated by electron conduction and radiative diffusion and for $n_a\gg l_a$ (as is true of all modes in our networks) is approximately given by
\begin{equation}
\gamma_a \simeq 2.4\times 10^{-13} n_a^2\omega_0 = 8.4\times10^{-14}n_a^2\ {\rm s^{-1}}. \label{eq:WKB_gamma}
\end{equation}
By comparison, the dissipation due to turbulent convective damping (green dots) is much smaller for the modes we are interested in (see Appendix~\ref{sec:WKB_relations} for details).

From Equations (\ref{eq:potential}) and (\ref{eq:Ua_def}), we can write the linear tide coefficient as
\begin{equation}
U_a =W_{lm} Q_{a} \left( \frac{M'}{M} \right)  \left(\frac{R}{D}\right)^{l+1} {\rm e}^{-\imag m (\Omega_{\rm orb} - \Omega_{\rm s})t},
\end{equation}
where the overlap integral
\begin{align}
Q_{a} &= \frac{1}{MR^l} \int \diff^3 r \rho \vect{\xi}^\ast \cdot \nabla \left(r^l Y_{lm}\right), \nonumber \\
&\simeq 2.1 n_a^{-3.7} \delta_{l_al} \delta_{m_am}.
\label{eq:WKB_Q_a}
\end{align}
In Figure~\ref{fig:Qn}, we show $Q_{a}$, calculated using the method described in Appendix~\ref{sec:tidal_overlap}, and the numerical fit above.  Note that the overlap is non-zero only if $l_a=l$ and $m_a=m$.  

 In Figure~\ref{fig:kappa_abc} we show the three-mode coupling coefficients as a function of the parent mode's radial order $n_a$. For high-order modes, we find 
 \begin{equation}
\kappa_{abc} \simeq  41 \left(\frac{T}{0.18}\right) \left(\frac{n_a}{l_a}\right)^2,
\label{eq:WKB_kap_abc}
\end{equation}
where $a$ is the parent mode. Here $T$ is an angular integral that depends only on each mode's angular quantum numbers and vanishes if the modes do not satisfy the angular selection rules: (i) $|l_b - l_c| \leq l_a \leq l_b+l_c$, (ii) $l_a+l_b+l_c$ is even, and (iii) $m_a+m_b+m_c=0$. Otherwise, it is of order unity for the typical triplets that we consider, e.g., $T\simeq 0.18$ for $l_a=l_b=l_c=2$ and $(m_a,m_b,m_c)=(2,-2,0)$.  In addition to these angular selection rules, the modes couple significantly only if their radial orders satisfy $|n_b-n_c| \lesssim n_a$ (\citealt{Wu:01},\citetalias{Weinberg:12}). 

\begin{figure}
   \centering
   \includegraphics[width=0.45\textwidth]{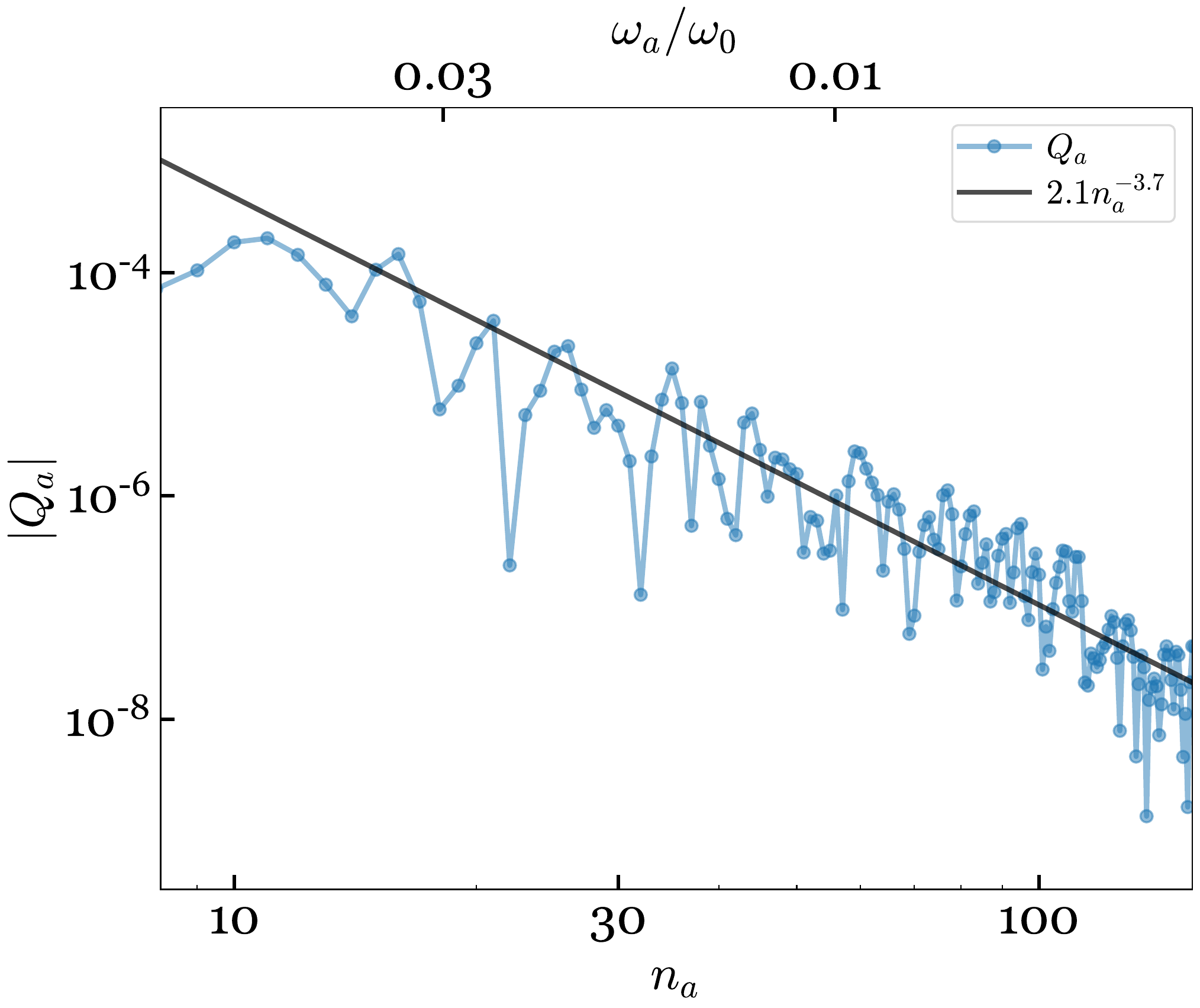} 
   \caption{Linear tidal overlap $Q_a$ (blue circles). The black line is the fit given by Equation~(\ref{eq:WKB_Q_a}). 
   }
   \label{fig:Qn}
\end{figure}

\subsection{Nonlinear instability threshold}
\label{sec:nonlinear_threshold}
In the absence of nonlinear interactions, a mode driven by the linear tide has an energy
\begin{equation}
\frac{E_{a, {\rm lin}}}{E_0} = \frac{\omega_a^2}{\Delta_a^2 + \gamma_a^2} U_a^2, 
\label{eq:E_lin}
\end{equation}
where $\Delta_a = \omega - \omega_a$ and $\omega = m (\Omega_{\rm orb} - \Omega_{\rm s})$. 
In linear theory, the parent's energy and dissipation rate are smallest when the parent is half-way between resonances, i.e., when the detuning is at a maximum $|\Delta_a|=|\partial \omega_a/\partial n_a|/2 \simeq \omega_a/2n_a$ ($\gg \gamma_a$ for the periods of interest). The linear energy of a parent half-way between resonances is 
\begin{equation}
\frac{\Eoff}{E_0} = 7.9\times10^{-18} \left(\frac{2\mratio}{1+\mratio}\right)^2\left(\frac{P_{\rm orb}}{50\, {\rm min}}\right)^{-9.3}
\label{eq:E_lin_off}
\end{equation}
assuming a non-rotating WD such that $P_a \simeq P_{\rm orb}/2$.

\begin{figure}
   \centering
   \includegraphics[width=0.45\textwidth]{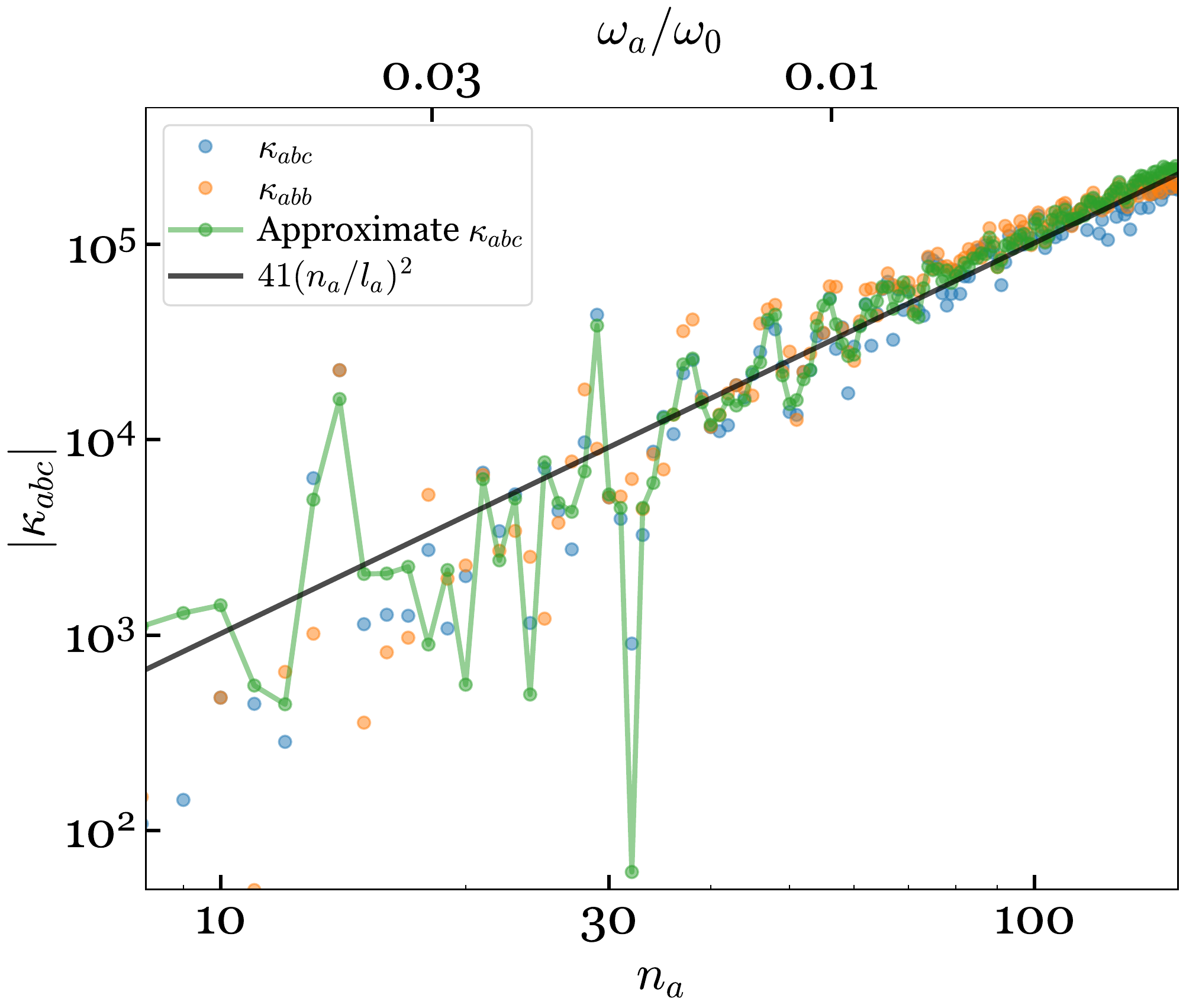} 
   \caption{Three-mode-coupling coefficient $\kappa_{abc}$ as a function of the parent mode's radial order. The blue circles are the coupling computed with daughter pairs  that have the smallest frequency detuning with respect to the parent mode, and the orange circles further restrict it to self-coupled daughters ($b=c$). The green circles connected with solid lines use the approximate expression for the coupling coefficient integrand given by Equation~(\ref{eq:kap_abc_approx}). Note that for a given parent mode $a$, its $\kappa_{abc}$ for different daughter pairs satisfying the selection rules are all  approximately equal as long as $|n_b-n_c| \lesssim n_a$.}
   \label{fig:kappa_abc}
\end{figure}

Now consider a simple three-mode system consisting of a parent mode driven by the tide coupled to a resonant daughter pair.  If $\Eoff > E_{\rm th}$ the parent is unstable even at maximum $\Delta_a$, where the energy threshold (see, e.g.,  \citetalias{Weinberg:12} and \citealt{Essick:16}, hereafter \citetalias{Essick:16})
\begin{equation}
\frac{E_{\rm th}}{E_0} =\frac{1}{4\kappa_{abc}^2}\left(\frac{\gamma_b\gamma_c}{\omega_b \omega_c} \right)\left[1 + \left(\frac{\Delta_{bc}}{\gamma_b + \gamma_c}\right)^2\right],
\label{eq:E_th_def}
\end{equation}
with $\Delta_{bc} = \omega + \omega_b + \omega_c$ the nonlinear detuning. Note that if $\omega>0$, we have $\omega_b, \omega_c<0$ according to our sign convention.    

\begin{figure}
   \centering
   \includegraphics[width=0.45\textwidth]{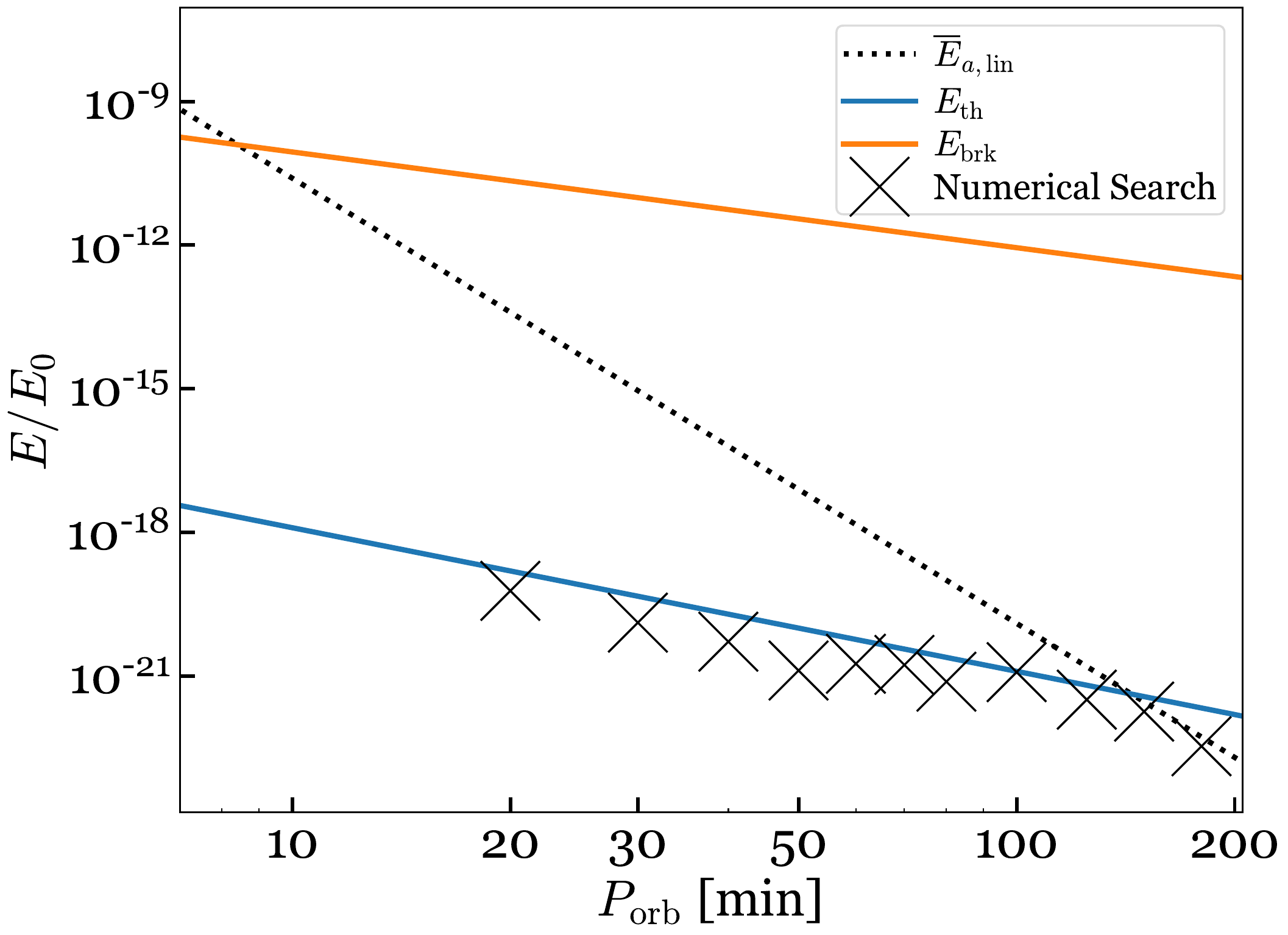} 
   \caption{Linear energy $\Eoff$ for a parent half-way between adjacent resonance peaks (Equation~\ref{eq:E_lin_off}; dotted black line),  the nonlinear threshold energy $E_{\rm th}$ based on the analytic scaling relations  (Equation~\ref{eq:E_th}; blue line), and the wave breaking energy $E_{\rm brk}$ (Equation~\ref{eq:Ebrk}; orange line) as a function of orbital period assuming a non-rotating WD. The black crosses show the minimum $E_{\rm th}$ obtained from a numerical search for daughter pairs. }
   \label{fig:E_th_vs_P}
\end{figure}

In Figure~\ref{fig:E_th_vs_P}, we show $\Eoff$ (dotted line) and the minimum $E_{\rm th}$ from a numerical search of daughter pairs (black crosses) assuming a non-rotating WD.  We also show an analytic estimate of the minimum $E_{\rm th}$ (blue line), whose calculation we describe below.  We see that for $P_{\rm orb}\lesssim 150\,{\rm min}$, even maximally detuned parent modes are parametrically unstable.  In fact, since $\Eoff \gg E_{\rm th}$ over much of this range, we will see that a single parent excites many unstable daughter pairs.

The daughter pairs that minimize $E_{\rm th}$ are those that satisfy the $\kappa_{abc}$ selection rules, have $|n_b-n_c| < n_a$, and nonlinear detunings $\Delta_{bc}\approx \gamma_b + \gamma_c$ since they minimize the sum in quadrature in the  brackets of Equation~(\ref{eq:E_th_def}); see also \citetalias{Essick:16}.  We can obtain an analytic estimate of the minimum $E_{\rm th}$ by using the scaling relations given above and an estimate for the minimum $\Delta_{bc}$.  Following the argument given by \cite{Wu:01}, we obtain an estimate for the minimum $\Delta_{bc}$ by noting that for a fixed parent mode $a$, there are $\sim n_a^2$ daughter pairs satisfying $|n_b-n_c| \lesssim n_a$ at fixed $l_b$ and $l_c$. As we allow the angular degree $l_b$ and $l_c$ to vary, we obtain an extra factor of $l_b l_a$ of mode pairs that satisfy the condition $|l_b - l_c| < l_a < l_b+l_c$. The eigenfrequencies of these potential daughter modes span a range of order $n_a|\partial \omega_b /\partial n_b|  \simeq n_a\omega_b/n_b$. Therefore, the typical minimum three-mode detuning assuming a non-rotating WD is of order\footnote{
For a rotating WD, the detuning is smaller by yet another factor of $l_b$ because the degeneracy between different combinations of  $m_b$ and $m_c$ is lifted (i.e., rotational splitting).}  
\begin{equation}
\Delta_{bc} \approx \frac{1}{l_b l_a n_a^2} \frac{n_a \omega_b}{n_b} \simeq \frac{0.07 l_a \omega_0}{l_b^2 n_a^3},
\label{eq:Deltabc_typical}
\end{equation}
where in the second approximation we first eliminated $n_b$ in terms of $(\omega_b, l_b)$ using Equation~(\ref{eq:WKB_omega}) and then assumed that $|\omega_b| \simeq \omega_a/2$.
The factor of 0.07 is from a fit to our numerical search for daughter pairs that minimize $E_{\rm th}$.  By using the scaling relations in Section~\ref{sec:powerlaw_relations} and Equation  (\ref{eq:Deltabc_typical}) and setting $\Delta_{bc}\simeq  \gamma_b+\gamma_c$, it follows that the minimum threshold energy  
\begin{equation} 
\frac{E_{\rm th}}{E_0} \simeq 1.0\times10^{-20} \left(\frac{0.18}{T}\right)^2 \left(\frac{\Porb}{50\,{\rm min}}\right)^{-3},
\label{eq:E_th}
\end{equation}
where this assumes a non-rotating WD and $P_a = \Porb/2$. As Figure~\ref{fig:E_th_vs_P} shows, this $E_{\rm th}$ estimate is in good agreement with that from the numerical search for daughter pairs.  
The daughters that minimize $E_{\rm th}$ typically have
\begin{equation}
l_b   \simeq 6.0 \left(\frac{\Porb}{50\,{\rm min}}\right)^{-5/4} 
 \textrm{and } 
\,\, n_b \simeq 281 \left( \frac{\Porb}{50\,{\rm min}} \right)^{-1/4}.
 \label{eq:l_b_n_b_exp}
\end{equation}

\subsection{Energy and angular momentum transfer}
\label{sec:energy_and_AM_transfer}

In this Section we derive the tidal power $\dot{E}_{\rm tide}$ and the tidal torque $\tau_{\rm tide}$  in the inertial frame. We define the torque to be from the orbit to the WD and when $\tau_{\rm tide}>0$ the tide spins up the WD.
Given the interaction Hamiltonian 
\begin{equation}
H_{\rm int} = -2E_0 \sum_{\omega_a > 0} \left(  q_a^\ast U_a + q_a U_a^\ast\right), \label{eq:Hamiltonian}
\end{equation}
the tidal torque acting on the WD is
\begin{align}
\tau_{\rm tide} &= \frac{\partial H_{\rm int}}{\partial \Phi} \simeq -4 \sum_{\omega_a > 0} {\rm Re}\left[q_a^\ast \frac{\partial U_a}{\partial \Phi}\right] E_0 \nonumber \\
& = -4 \sum_{\omega_a > 0 } m_a {\rm Im } \left[q_a^\ast U_a\right] E_0,
\label{eq:tau_tide_def}
\end{align}
where a factor of two arises from the sum over modes and their complex conjugate and another because we restricted the sum to positive frequencies, and the last equality follows because $\partial U_a / \partial \Phi = - {\imag} m U_a$. Note that we dropped the term $\propto U_{ab}$ in the interaction Hamiltonian because only the linearly resonant parents have a significant direct coupling to the tide (Section~\ref{sec:eom}).  

The associated tidal power, assuming a circular orbit, is given by 
\begin{equation}
    \dot{E}_{\rm tide} = \Omega_{\rm orb}\tau_{\rm tide}. 
\end{equation}
In general, this will power a combination of mode energy, tidal heating, and WD spin energy.  However, as we illustrate in Section~\ref{sec:saturation} (see, e.g., Figures~\ref{fig:E_Edot_60} and \ref{fig:E_Edot_30}), in steady state the time-averaged total mode energy is approximately constant and $\sum_a \dot{E}_a \simeq \sum_a \left(\dot{q}_a^\ast q_a+q_a^\ast\dot{q}_a \right)E_0 \simeq 0$.  
Using Equation~(\ref{eq:parent_amp_eqn}) we thus have, in a time-averaged sense,
\begin{equation}
\sum_a \omega_a {\rm Im} \left[ q_a^\ast U_a\right]  \simeq - \sum_b \gamma_b q_b^\ast q_b. 
\label{eq:sum_dEa_dt}
\end{equation}
The summation on the left-hand side is only over parent modes since only they feel a strong, direct driving by the tide (Section~\ref{sec:formalism}), whereas on the right-hand
side it is over all modes from all generations. We also dropped
the three-mode dissipation terms as they contribute little to the total dissipation.\footnote{
There are two dissipation terms that arise directly from three mode coupling: the first originates from $\sum_a \left( \dot{q}_a^\ast q_a + q_a^\ast\dot{q}_a\right)$ and contributes $\sum_{abc}2(\omega_a + \omega_b + \omega_c)\kappa_{abc}\Imag[q_aq_bq_c]$ and the second comes from the nonlinear piece in the total mode energy [Equation~(\ref{eq:E_a_nl_def})] and contributes $\sum_{abc} 2 \gamma_a \kappa_{abc} {\rm Re}[q_a q_b q_c]$. Since the detuning $\left(\omega_a+\omega_b+\omega_c\right) \sim \gamma_a$ for the most unstable daughters, the two terms are comparable.  Since $|\sum_{abc} \kappa_{abc} q_a q_b q_c|\ll \sum_a E_a$ (see last paragraph of this Section), the nonlinear dissipation is much smaller than the lower-order contribution $\sum_a \gamma_a E_a$.}

For the most-resonant parent modes with $\omega_a \geq 0$ and $l_a=2$, the azimuthal order $m_a=m=2$ and $\omega_a \simeq \omega = m \left(\Omega_{\rm orb}-\Omega_{\rm s}\right)$. We can therefore relate the tidal torque and power to the total dissipation rate inside the star, $\Edot$, as 
\begin{align}
&\tau_{\rm tide} = \frac{m }{\omega} \dot{E}_{\rm diss}, \label{eq:tau_tide}\\
&\dot{E}_{\rm tide} = \frac{\Omega_{\rm orb}}{\left(\Omega_{\rm orb}-\Omega_{\rm s}\right)} \dot{E}_{\rm diss},
\end{align}
where
\begin{equation}
\dot{E}_{\rm diss} =  4\sum_{\omega_b > 0} \gamma_b q_b^\ast q_b  E_0\simeq 4\sum_{\omega_b > 0} \gamma_b E_b.
\label{eq:dE_diss}
\end{equation}
If we assume that the WD rotates with uniform angular velocity $\Omega_{\rm s}$, then the tidal torque spins it up at a rate
\begin{equation}
\dot{\Omega}_{\rm s} = \frac{\tau_{\rm tide}}{I_{\rm WD}},
\label{eq:dOmega_s}
\end{equation}
and the orbital frequency changes at a rate 
\begin{equation}
\dot{\Omega}_{\rm orb} = \dot{\Omega}_{\rm orb,gw} + \frac{3 \tau_{\rm tide}}{\mu D^2}, \label{eq:dOmega_orb}
\end{equation}
where the GW induced orbital decay rate
\begin{equation}
\dot{\Omega}_{\rm orb, gw}= \frac{96}{5} \left(\frac{G\mathcal{M}_c}{c^3}\right)^{5/3} \Omega_{\rm orb}^{11/3}.
\label{eq:dOmega_gw}
\end{equation}
Here $\mu$ is the reduced mass and  $\mathcal{M}_c=\mratio^{3/5}(1+\mratio)^{-1/5} M$ is the chirp mass.   In our study, we obtain $\dot{E}_{\rm diss}$ from our mode network simulations according to Equation~(\ref{eq:dE_diss})    and thereby determine $\tau_{\rm tide}$, $\dot{\Omega}_{\rm s}$, and $\dot{\Omega}_{\rm orb}$.\footnote{Alternatively, we can compute $\tau_{\rm tide}$ by taking the time average of Equation~(\ref{eq:tau_tide_def}) in steady state. Although we verified that the two methods yield consistent results, in practice we find that ${\rm Im}[q_a^\ast U_a]$ of individual parents is much more oscillatory than $\sum_b \gamma_b E_b$ and thus Equation~(\ref{eq:dE_diss}) provides a more numerically accurate estimate of the torque.}  

Our large nonlinear networks display complicated  dynamics. Nonetheless, some insights can be gained by considering the nonlinear equilibrium of simple  three-mode systems (\citetalias{Weinberg:12}, \citetalias{Essick:16}). For such a system, the parent mode's saturation is $E_{a,{\rm s}} = E_{\rm th}$ and for $\Eoff \gg E_{a,{\rm s}}$, the daughter mode's equilibrium is
\begin{align}
\frac{E_{b,{\rm s}}}{E_0} &\simeq \sqrt{\frac{\gamma_c\omega_b}{\gamma_b \omega_c}} \Big{|} \frac{U_a}{2 \kappa_{abc}}\Big{|} \nonumber \\
&\simeq 6.7 \times 10^{-16} \left(\frac{2\mratio}{1+\mratio}\right) \left( \frac{P_{\rm orb}}{50\,{\rm min}} \right)^{-7.7}.
\label{eq:E_dght_ss}
\end{align}
Comparing Equations (\ref{eq:E_th}) and (\ref{eq:E_dght_ss}), we see that $E_{b,{\rm s}}\simeq E_{c,{\rm s}} \gg E_{a, {\rm s}}$. As a result, the leading order drive to the granddaughters will be via daugher-granddaughter three-mode coupling rather than parent-granddaughter coupling at higher nonlinear orders. We therefore include multiple generations in our networks but only account for three-mode coupling between adjacent generations.\footnote{Four-mode coupling can be important for the $p$-$g$ instability~\citep{Venumadhav:14, Weinberg:16}.  However, that is a nonresonant instability whereas here we focus on the resonant parametric instability.}

Energy is stored not only in each individual mode ($\propto E_a$) but also in the three-mode couplings [$\propto {\rm Re}\left[\kappa_{abc}q_a q_b q_c\right]$; see Equation~(\ref{eq:E_a_nl_def})]. However, the latter makes a negligible contribution to the total energy at saturation.  We can easily see this for a three-mode system, since at saturation the total nonlinear energy is $\sim \kappa_{abc} \sqrt{E_{a, \rm s}} E_{b, \rm s} \ll E_{b, \rm s}$. To see roughly why this also holds for our large mode networks, note that at saturation the nonlinear forces approximately balance the linear forces.  By Equation (\ref{eq:ode_amp}), this implies $|\sum_{bc}\kappa_{abc} q_b^\ast q_c^\ast|$ approximately equals $|U_a|$ for parent modes and $(\gamma/\omega_a) |q_a|$ for the other modes.  Since both are $\ll|q_a|$,  it follows that $|\sum_{bc}\kappa_{abc} q_a q_b q_c| \ll E_a$ and therefore the nonlinear energy is a small contribution to the total energy.

\subsection{Standing waves vs. traveling waves}
\label{sec:sw_vs_tw}
The relations above and our mode network calculations assume that the modes are all standing waves.  In order to be a standing wave, a mode's linear damping time must be longer than its group travel time through the propagation cavity (which here spans much of the WD radius; see Figure~\ref{fig:prop_diag}), $T_a = 2 \int \diff r v_{\rm grp}^{-1}=2\int \diff r (\diff \omega_a / \diff k_r)^{-1}$, where $v_{\rm grp}$ is the mode's group velocity. Otherwise, it is a traveling wave.  Defining the inverse group travel time $\alpha_a = 2\pi/T_a$, we find 
\begin{equation}
\alpha_a \simeq 0.28 \frac{l_a}{n_a^2} \omega_0 = 0.097 \frac{l_a}{n_a^2}\ {\rm rad\,s^{-1}}. 
\label{eq:WKB_alpha_a}
\end{equation}
In Figure~\ref{fig:gam}, we compare $\alpha_a$ to the linear damping rate of modes.  We find that the standing wave condition $\gamma_a < \alpha_a$ is satisfied for
\begin{equation}
n_a \lesssim 1200  \left(\frac{l_a}{2}\right)^{1/4}, \text{ i.e., } P_{\rm orb}\lesssim 1320\left(\frac{l_a}{2}\right)^{-3/4}\,{\rm min},
\end{equation}
which is true of all the modes in our networks.

Another necessary condition for standing wave is that the shear $|\diff \xi_r /\diff r| \simeq |k_r \xi_r|$ be everywhere less than unity, where $\xi_r Y_{lm}$ is the radial component of the physical Lagrangian displacement $\vect{\xi}$. If a g-mode's shear exceeds unity, it is strongly nonlinear and overturns the local stratification and breaks~(see, e.g., \citealt{Goodman:98, Barker:11b}).

\citeauthor{Fuller:12a} (2012a; hereafter \citetalias{Fuller:12a}) and \citeauthor{Burkart:13} (2013; hereafter \citetalias{Burkart:13}) use this local wave-breaking condition to address the onset of nonlinear tidal effects in WD binaries.  They show that at sufficiently short orbital periods, the tide excites internal gravity waves that are initially linear deep within the WD but become nonlinear and break as they approach the stellar surface.\footnote{It is interesting to note that whereas the the local wave-breaking occurs at the surface, the global three-mode coupling happens mostly in the core region. See Appendix~\ref{sec:three_mode_cpl} and Figure~\ref{fig:kappa_abc}. This is different from the case of solar models \citepalias{Weinberg:12}.}

We first evaluate the wave-breaking condition \emph{assuming a standing wave}, i.e., a g-mode.  Using the approach described in Appendix~\ref{sec:shr_prfl}, we find that a g-mode's shear exceeds unity if its energy exceeds [see Equation~(\ref{eq:WKB_shr})]
\begin{equation}
\frac{E_{\rm brk}}{E_0} = 3.6\times10^{-12}\left(\frac{P_{\rm orb}}{50 \, {\rm min}}\right)^{-2}. 
\label{eq:Ebrk}
\end{equation}
In Figure~\ref{fig:E_th_vs_P}, we show $E_{\rm brk}$ as a function of $P_{\rm orb}$.  We find that $\Eoff $ first exceeds $E_{\rm brk}$ at $\Porb \simeq  10\textrm{ min}$.  Moreover, even highly resonant parent modes are unlikely to break before  $\Porb \simeq 10\textrm{ min}$.  That is because the parent is parametrically unstable ($E_{a, \rm lin} > E_{\rm th}$) out to $\Porb \approx 150\textrm{ min}$ (Figure~\ref{fig:E_th_vs_P}) and excites secondary modes which prevent it from reaching the wave-breaking limit (see Section~\ref{sec:saturation}).

\begin{figure}
   \centering
   \includegraphics[width=0.45\textwidth]{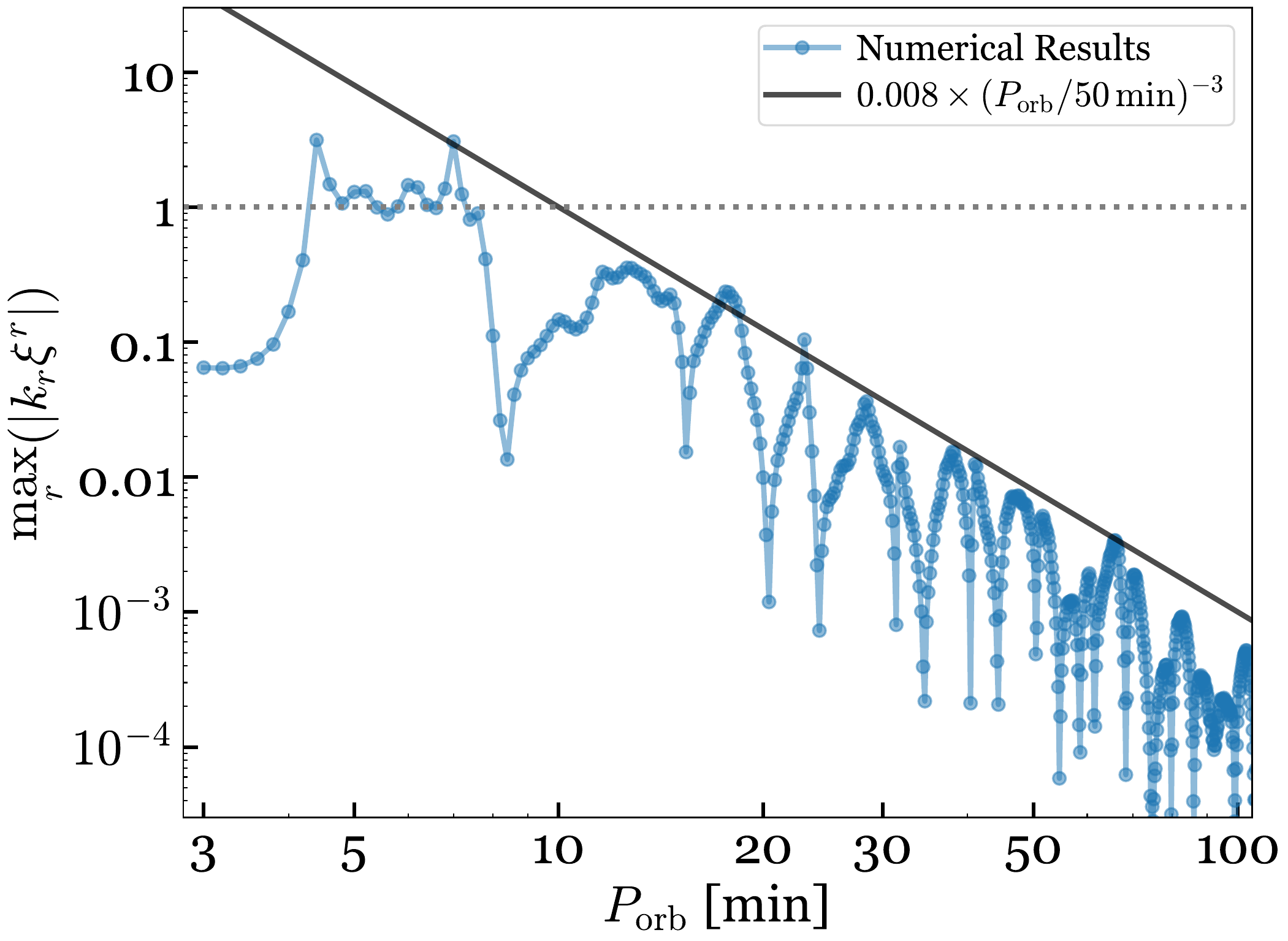} 
   \caption{Maximum local shear $|k_r \xi_r|$ from the traveling wave solution for a non-rotating WD with $T_{\rm eff}=9000\,{\rm K}$. }
   \label{fig:shear_TW}
\end{figure}

We now evaluate the wave-breaking condition \emph{assuming a traveling wave} rather than a standing wave.  Specifically, we use the approach described in \citetalias{Fuller:12a} (see Appendix~\ref{sec:TW} for a brief synopsis) to find the traveling-wave solution of the linear inhomogeneous tidal equations [Equations~(\ref{eq:xi_r_inh}) - (\ref{eq:xi_h_inh})].  Just as the standing wave assumption is valid only if $\textrm{max}|k_r \xi_r| < 1$, the traveling wave assumption is valid only if $\textrm{max}|k_r \xi_r| > 1$.  In Figure~\ref{fig:shear_TW} we show $\textrm{max}|k_r \xi_r| $ computed under the traveling-wave assumption for a non-rotating WD with $T_{\rm eff}=9000\,{\rm K}$. We find that the upper envelope of the shear $\propto P_{\rm orb}^{-3}$ (which we explain in Appendix~\ref{sec:TW}) and reaches unity at $P_{\rm orb}\simeq10\, {\rm min}$, consistent with the results assuming a standing wave.  

\emph{The weakly nonlinear regime of this study therefore spans a large range of orbital periods ($10 \lesssim P_{\rm orb}/\textrm{min} \lesssim 150$).
Evaluating the global, multi-mode dynamics  in this regime is essential for understanding the impact of tidal dissipation on WD binaries.}

Our analysis assumes that all modes, not just the parent, are standing waves and thus below the wave breaking threshold.  Since $E_{\rm th} \propto \omega^{3}$ while $E_{\rm brk} \propto \omega^2$, higher generation (i.e., lower frequency) modes have even smaller ratios of $E_{\rm th}$ to $E_{\rm brk}$ than the parent.  Thus, the excited daughters, granddaughters, etc. will likely become parametrically unstable and saturate before breaking (see also Appendix F in \citetalias{Essick:16}).  In practice, because we truncate our networks at the fifth generation and include only the most resonant pairs for each generation (since including more modes does not significantly increase the calculated $\Edot$; see Section~\ref{sec:dEdt_no_rot}), some high generation modes in our network can have shears that momentarily exceed unity.  However, at any given time these represent only a very small fraction of the excited modes and thus they are unlikely to modify the overall dynamics and dissipation.

It is also worth noting that the shear can be a sensitive function of the WD temperature. For example, in Appendix~\ref{sec:shr_prfl} we show that for $T_{\rm eff} =18000\,{\rm K}$ the maximum shear is about an order of magnitude larger than for $9000\,{\rm K}$. On the other hand, this is compensated by the tidal synchronization which decreases the driving frequency (see Section~\ref{sec:sync} and Appendix~\ref{sec:TW}). The orbital period where the dynamical tide transitions from weakly to strongly nonlinear is therefore still $\approx 10\textrm{ min}$ when the effects of both temperature and synchronization are taken into account. 

\section{Numerical implementation}
\label{sec:num}
The modes in our networks oscillate near their eigenfrequencies and have small linear detunings $\Delta_a$ (parents) or nonlinear detunings $\Delta_{bc}$ (daughters, granddaughters, etc.).  We can therefore factor out the fast-oscillations by transforming coordinates to $c_a = q_a \exp(\imag \omega_a t)$, similar to the approach of previous mode network studies (\citealt{Brink:05}, \citetalias{Essick:16}).   The parent mode amplitude Equation (\ref{eq:parent_amp_eqn}) is then 
\begin{equation}
\dot{c}_a  + \gamma_a c_a = \imag \omega_a |U_a| \e^{-\imag \Delta_a t} +\imag \omega_a \sum_{bc}\kappa_{abc} c_b^\ast c_c^\ast \e^{\imag \Delta'_{bc}t},
\end{equation}
and similarly for the other modes, where $\Delta'_{bc}=\omega_a + \omega_b +\omega_c$.   We initialize our networks by starting each mode at its linear tidal energy and a random phase.  
We implemented the calculations in \texttt{Python} and used the \texttt{NUMBA} package~\citep{Lam:15} to enhance the computational performance. 

Initially, the amplitudes of the unstable daughters will grow exponentially at a characteristic rate (see \citetalias{Weinberg:12})
\begin{equation}
\Gamma_{\rm nl} \simeq \omega_a \kappa_{abc}\sqrt{\frac{E_a}{E_0}}.
\label{eq:Gamma_nl}
\end{equation}
This allows us to define a characteristic nonlinear growth timescale 
\begin{equation}
T_{\rm nl} \equiv \frac{1}{\Gamma_{\rm nl}} \sim 0.12 \left(\frac{P_{\rm orb}}{50\,{\rm min}}\right)^{3.7} \,{\rm yr}, 
\label{eq:tau_nl}
\end{equation}
where the numerical value is for a parent mode at an initial energy $\Eoff$.

We find that the mode networks saturate and reach a nonlinear equilibrium over a few nonlinear growth times $T_{\rm nl}$. This is much shorter than the GW-induced orbital decay timescale
\begin{equation}
T_{\rm gw} = \frac{\Omega_{\rm orb}}{\dot{\Omega}_{\rm orb, gw}} = 4.8\times10^{7} \left(\frac{P_{\rm orb}}{50\,{\rm min}}\right)^{8/3}\,{\rm yr},
\label{eq:T_gw}
\end{equation}
where the numerical value assumes a typical WD binary with $M = M' =  0.6\,M_\odot$. The timescale $T_{\rm nl}$ is also shorter than the time it takes for the GW orbital decay to change the three-mode detuning $\Delta_{bc}$ by an amount $(\gamma_b + \gamma_c)\simeq2\gamma_b$ (see Section~\ref{sec:nonlinear_threshold}),
\begin{equation}
T_{\rm det} = \frac{\gamma_b}{\dot{\Omega}_{\rm orb, gw}} \simeq 150 \left(\frac{P_{\rm orb}}{50\,{\rm min}}\right)^{19/6} \,{\rm yr}.
\label{eq:T_delta}
\end{equation}
Therefore, the particular parametrically unstable pairs that are most resonant and thus have the lowest $E_{\rm th}$ do not change on a timescale of a few $T_{\rm nl}$.  We therefore only construct our mode networks once for each $P_{\rm orb}$ we consider.

In order to construct our mode networks, we search for the daughters, granddaughters, etc. with the lowest threshold energies.   Numerically, we find that a network's total energy dissipation rate $\dot{E}_{\rm diss}$ converges once we include five mode generations constructed as follows. The first generation (parents)  includes the two most linearly resonant modes.  The second generation (daughters) includes the three lowest threshold daughter pairs of each parent.  Since the two parent modes both oscillate at the tidal driving frequency, they usually have the same pair of most-resonant daughter modes and thus the second generation typically has 6 modes instead of 12.  The third through fifth generations include the single lowest threshold pair of each mode from the previous generation.  A typical network consists of 92 modes, with $(2, 6, 12, 24, 48)$ modes in each generation (since modes sometimes appear in more than one pair, some networks have slightly fewer than 92 modes).   We find that increasing the number of modes and generations does not significantly change the computed $\dot{E}_{\rm diss}$ (see Section~\ref{sec:dEdt_no_rot}).

 \begin{figure}
   \centering
   \includegraphics[width=0.45\textwidth]{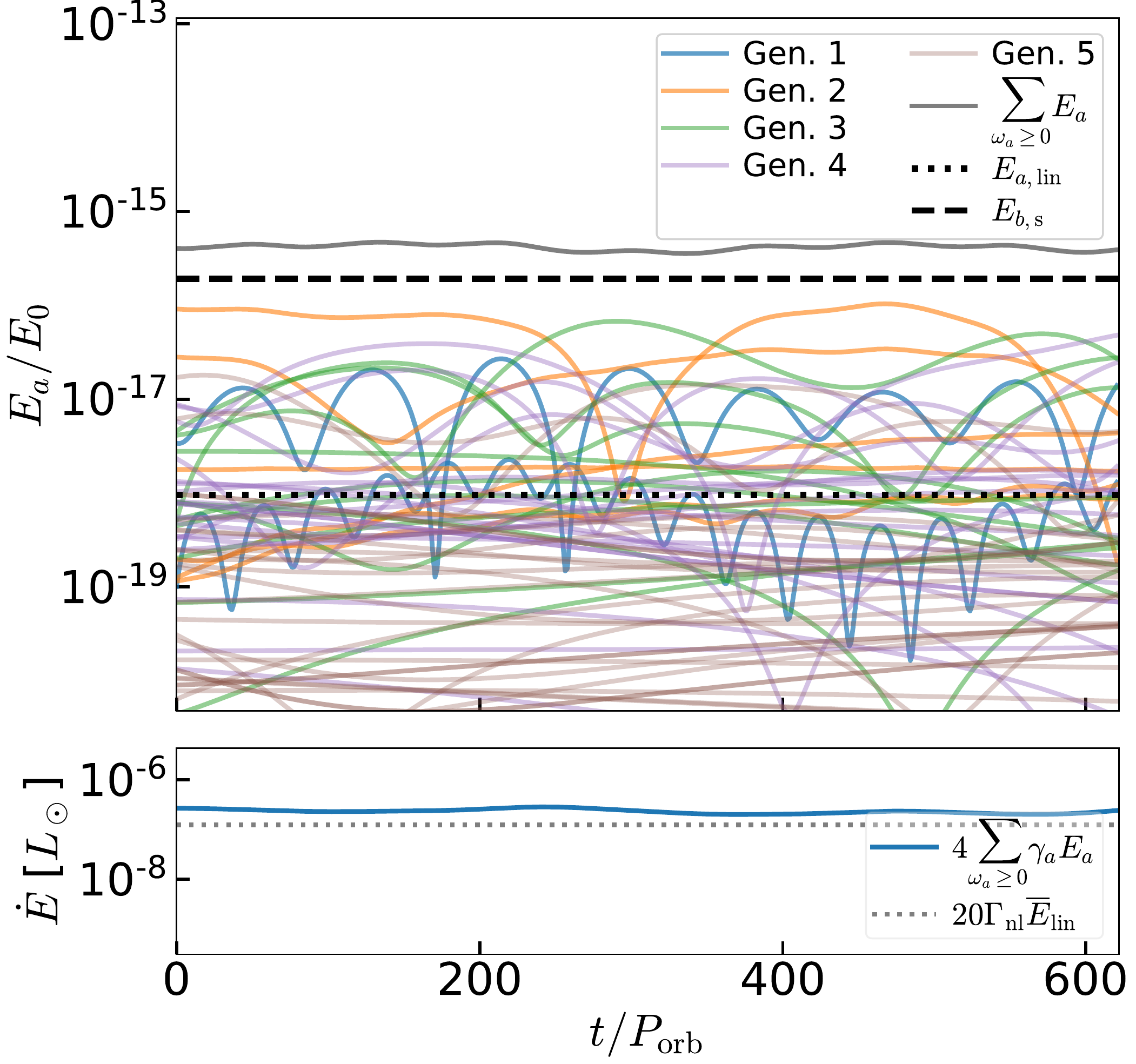} 
   \caption{Mode energy (upper panel) and  total energy dissipation rate (lower panel) as a function of time at an orbital period $P_{\rm orb}{=}59.46\,{\rm min}$. At this period, the most resonant parent has a detuning $|\Delta_a| = \omega_a / 3 n_a$.  We include 5 generations of modes and in the upper panel label the first (i.e., the parent) to the fifth generation of modes with blue, orange, green, purple, and brown lines, respectively. The grey-solid line is the total mode energy.  For comparison, we also show the mode energy according to linear theory [Equation~(\ref{eq:E_lin}); black-dotted line] and the three-mode equilibrium energy of the daughter modes (black-dashed line) estimated using Equation~(\ref{eq:E_dght_ss}). In the lower panel, the blue-solid line is the numerically computed total dissipation rate $\Edot$ and the grey-dotted line is 20 times the product of the maximally detuned parent mode energy $\Eoff$ and the corresponding three-mode growth rate [Equation~(\ref{eq:Gamma_nl})].  }
   \label{fig:E_Edot_60}
\end{figure}

A collective instability can occur if daughters form large sets of mutually coupled pairs \citepalias{Weinberg:12}.  Collectively unstable daughters initially grow much more rapidly than the isolated pairs described above.   However, in our problem the collective instability threshold $E_{\rm th, col}$ is higher than the isolated pair instability threshold $E_{\rm th}$.  \citetalias{Essick:16} found that the parents, whose linear energy might be well above $E_{\rm th, col}$, reach a nonlinear equilibrium at an energy below $E_{\rm th, col}$ due to their coupling to isolated pairs. As a result, they found that the collective pairs eventually decay away and thus do not enhance the net dissipation in the system.   We expect similar dynamics here and therefore do not include collectively unstable pairs in our networks.

\section{Results}
\label{sec:results}

Having described the formalism and numerical methods in Sections~\ref{sec:formalism} and~\ref{sec:num}, we now describe the results of our coupled mode network simulations.  In Section~\ref{sec:saturation}, we show representative examples of the mode dynamics on short timescales.  In Section~\ref{sec:dEdt_no_rot}, we show how the total energy dissipation rate depends on orbital period over a wide range of orbital separations.  In Section~\ref{sec:semianalytic} we describe semi-analytic models that accurately capture the scalings found in the network simulations.  In all three sections, we assume a non-rotating WD in an equal mass binary.  In Section~\ref{sec:sync}, we consider a rotating WD and study the impact of the tide on the spin evolution and synchronization of the binary.

\begin{figure}
   \centering
   \includegraphics[width=0.45\textwidth]{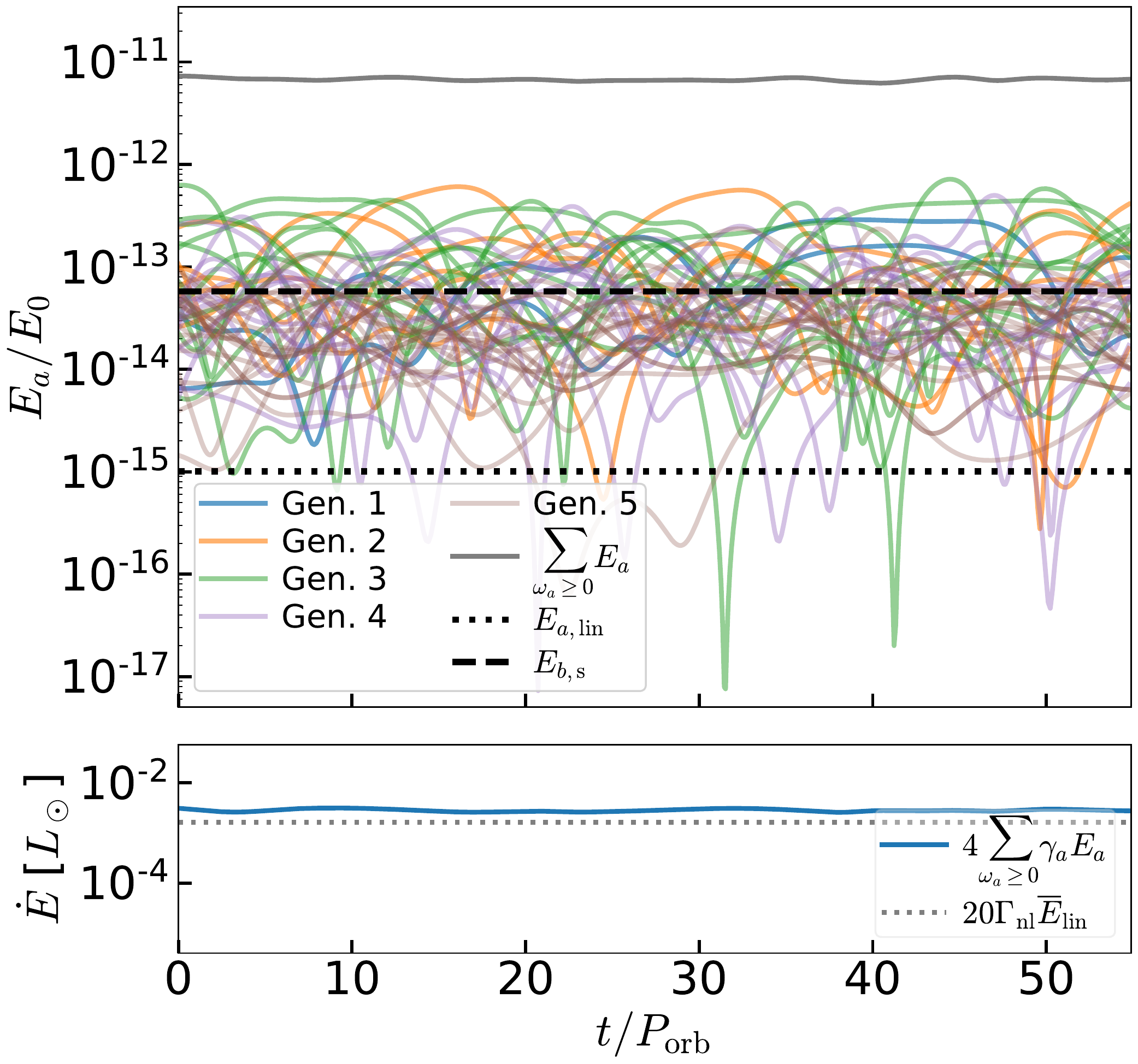} 
   \caption{Same as Figure~\ref{fig:E_Edot_60} but at $P_{\rm orb}{=}28.48\,{\rm min}$.}
   \label{fig:E_Edot_30}
\end{figure}

\subsection{Mode dynamics on short timescales}
\label{sec:saturation}
In the top panel of Figure~\ref{fig:E_Edot_60}  we show a zoomed-in view of the energy $E_a(t)$ of each mode  in our network over a  duration of approximately one nonlinear growth timescales $T_{\rm nl}$ [Equation (\ref{eq:tau_nl})] at an orbital period near $P_{\rm orb}\simeq 60\textrm{ min}$.  The top panel of Figure~\ref{fig:E_Edot_30} is similar except  at $P_{\rm orb}\simeq 30\textrm{ min}$.  In both figures, the precise periods are chosen in order that the most resonant parent mode has a detuning $|\Delta_a| = \omega_a / 3 n_a$, which is somewhat far from a resonance peak ($|\Delta_a| = \omega_a / 2 n_a$ half-way between adjacent resonance peaks).  The solid grey line in each figure shows the total mode energy  $E_{\rm tot} = \sum_{a} E_a$. Although an individual mode's energy can vary by orders of magnitude over time, we find that over  duration of a few $T_{\rm nl}$, the system settles into a quasi-equilibrium state with $E_{\rm tot}\approx\textrm{constant}$. Thus, there is a balance between the time-averaged tidal power driving the parents and the net thermal dissipation from mode damping. We also find  that $E_{\rm tot} \gg E_{\rm lin}$, where $E_{\rm lin}$ is the total energy according to linear theory [dotted black line; Equation~(\ref{eq:E_lin})].

We note that there is no energy hierarchy in the mode generation.   
In fact, modes from different generations alternatively dominate the system's energy in a limit-cycle-like (or even chaotic) manner with a variation timescale shorter than $T_{\rm nl}$. 

\begin{figure}
   \centering
   \includegraphics[width=0.45\textwidth]{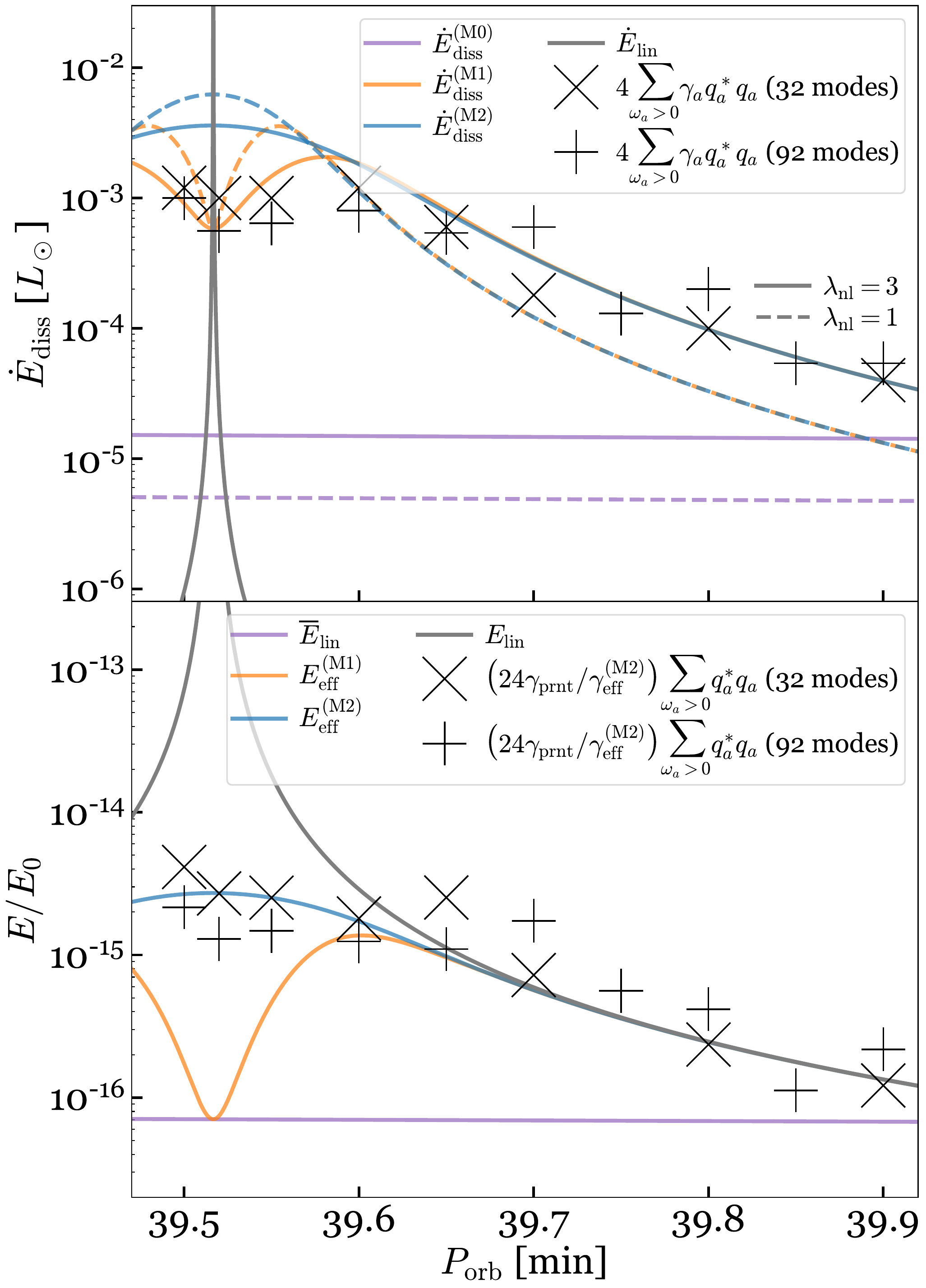} 
   \caption{Top panel: Tidal dissipation rate $\Edot$ as a function of orbital period near $\Porb=40\textrm{ min}$ assuming a non-rotating WD.  The period range shown spans $\simeq 0.5$ min, which corresponds to half of the separation between two adjacent linear resonances.  We show $\Edot$ computed with the standard 92-mode network ($+$ symbols) and a 32-mode network ($\times$ symbols).  The colored lines show $\Edot$ according to linear theory (grey line) and models M0, M1, and M2 (purple, orange, and blue) for $\lambda_{\rm nl}=1$ (dashed lines) and $\lambda_{\rm nl}=3$ (solid lines); we see that the latter value provides the best fit to the numerical results.  Bottom panel: Similar to the top panel but showing energy rather than $\Edot$.  We show the total mode energy $E_{\rm tot}$ in the two networks, the parent linear energy $E_{\rm lin}$ (grey line), the maximally detuned parent linear energy $\Eoff$ (purple line), and the effective energies $E_{\rm eff}$ of models M1 and M2 (orange and blue lines).  Note that $E_{\rm tot}$ has been multiplied by a factor of $24\gamma_a/\gamma_{\rm eff}^{\rm (M2)}\ll 1$, where  $\gamma_a$ is the linear damping rate of the parent mode.  We do this in order to be able to show it on the same scale as $E_{\rm lin}$ and $E_{\rm eff}$. 
   }
   \label{fig:dE_lorentzian}
\end{figure}

In the bottom panels of  Figures~\ref{fig:E_Edot_60} and \ref{fig:E_Edot_30} we show the numerically computed energy dissipation rate $\dot{E}_{\rm diss} = 4\sum_{\omega_a>0} \gamma_a E_a$ on short timescales.  Similar to $E_{\rm tot}$, we find that  $\dot{E}_{\rm diss} \approx \textrm{constant}$.  

\begin{figure}
   \centering
   \includegraphics[width=0.45\textwidth]{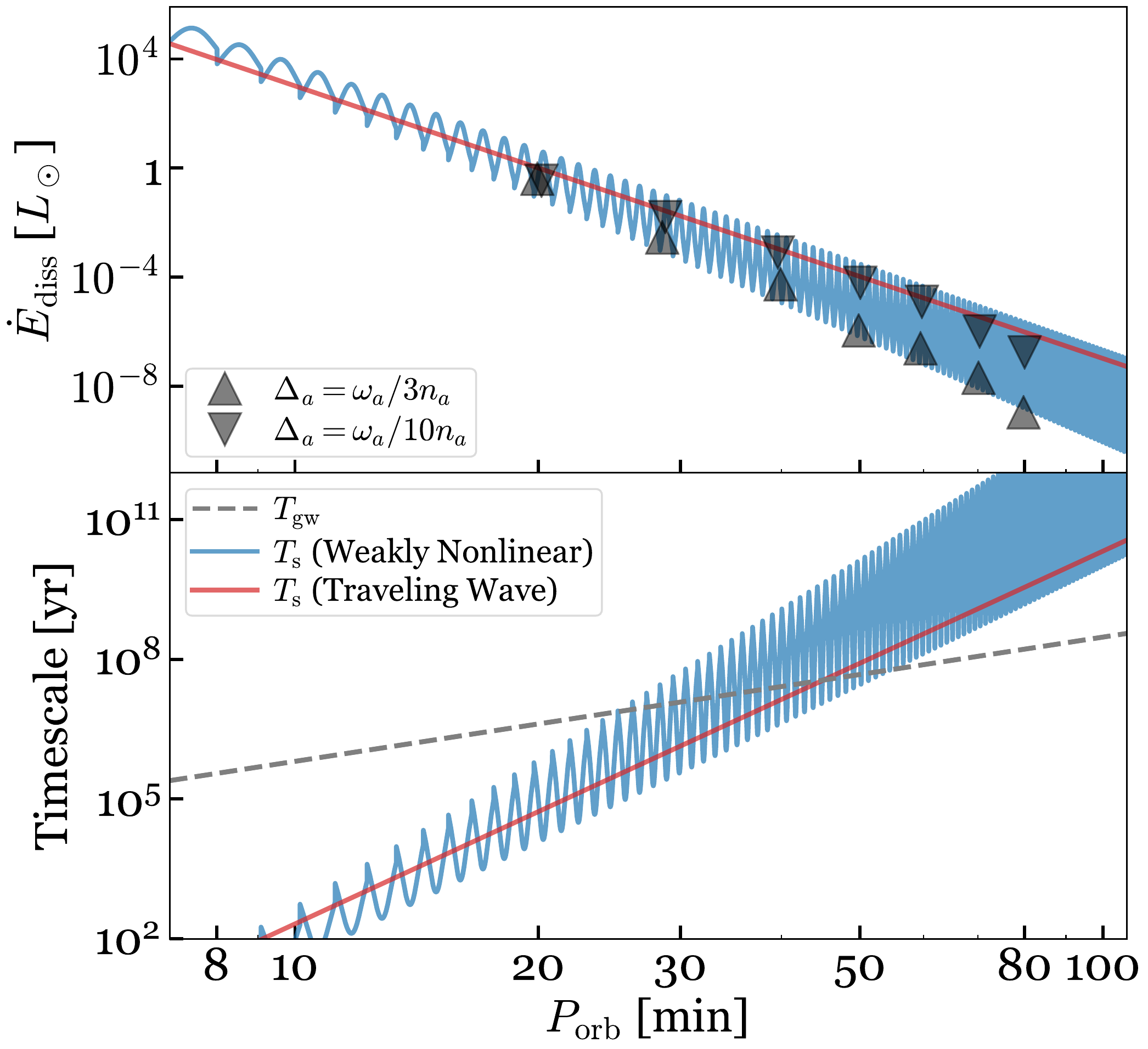} 
   \caption{Top panel: Energy dissipation rate $\dot{E}_{\rm diss}$ as a function of orbital period. The up (down) triangles show results of the 92-mode network at $\Porb$ where the parent has a relatively large (small) detuning $|\Delta_a|=\omega/3n_a$ ($\omega/10n_a$). The blue line shows results of the weakly nonlinear model M2 and the red line shows results of the traveling-wave model. Bottom panel: The spin-up timescale $T_{\rm s}$ for model M2 (blue line) and the traveling-wave model (red line).  The dashed line is the GW-induced orbital decay timescale $T_{\rm gw}$. In both panels we assume a non-rotating WD and use $\lambda_{\rm nl}{=}3$ when evaluating M2. 
}
   \label{fig:Edot_Tsync_no_rot}
\end{figure}

\subsection{Energy dissipation as a function of orbital period}
\label{sec:dEdt_no_rot}

We now use the numerical results at individual $\Porb$ to determine the time-averaged nonlinear dissipation as a function of orbital separation. 

In the upper panel of Figure~\ref{fig:dE_lorentzian}, we show $\dot{E}_{\rm diss}$ over a narrow range of orbital period near $P_{\rm orb}\simeq 40 \textrm{ min}$.  The range is chosen to  span half the distance between two adjacent linear resonances. We find that $\dot{E}_{\rm diss}$ is many orders of magnitude larger than the linear energy dissipation rate $\dot{E}_{\rm lin}$ (solid grey line) except when extremely near the resonance peak.  Although the nonlinear dissipation is much less sensitive to distance from resonance than the linear dissipation, it does still vary significantly with $\Delta_a$.  In going from on-resonance to half-way between resonance, $\Edot$ decreases by a factor of $\simeq 10$ at $P_{\rm orb}\simeq 40 \textrm{ min}$. As we show below, it is even more sensitive to $\Delta_a$ at larger $P_{\rm orb}$.

In Figure~\ref{fig:dE_lorentzian}, we show results for our standard 92-mode network with $(2,6,12,24,48)$ modes in each generation, and a 32-mode network with $(2,2,4,8,16)$ modes in each generation (see Section~\ref{sec:num}).  The $\dot{E}_{\rm diss}$ of each  agree to within a factor of about two.  From this experiment and others we performed, we conclude that our 92-mode network is sufficiently large to adequately capture the full nonlinear dissipation. 

In Figure~\ref{fig:Edot_Tsync_no_rot}, we show $\Edot$ over a wide range of $P_{\rm orb}$.  The triangles are the results from a series of mode network simulations, with the upward (downward) triangles corresponding to $P_{\rm orb}$ when the most resonant parent has a relatively large (small) detuning $|\Delta_a| =\omega_a / 3n_a$ ($\omega_a / 10n_a$).  We find that the difference in $\Edot$ between peaks and trough decreases considerably with decreasing $P_{\rm orb}$; the difference is a factor of $\sim 10^3$ at $P_{\rm orb}\simeq 80\textrm{ min}$ while it is only a factor of $\sim 30$  at $P_{\rm orb}\simeq 20\textrm{ min}$. 

From the numerical results, we see that the dissipation scales approximately as $\Edot \propto P_{\rm orb}^{-13}$ when the detuning is large.  Since the typical linear damping of the modes in the network scales approximately\footnote{This ignores the fact that the dissipation at different instants may be dominated by modes from different generations. Therefore, at a given $P_{\rm orb}$, the linear damping among different modes can vary by factors of $\mathcal{O}(10)$. } as $\gamma_a \propto \Porb^{-2}$, the total mode energy at large detuning  $E_{\rm tot} \propto \Porb^{-11}$.  As a result, although the simple three-mode daughter equilibrium energy  $E_{b,\rm s} \propto \Porb^{-7.7}$ [Equation~(\ref{eq:E_dght_ss})] roughly equals $E_{\rm tot}$ at $P_{\rm orb}\approx 60\textrm{ min}$ (Figure~\ref{fig:E_Edot_60}), it is significantly smaller than $E_{\rm tot}$ at $\Porb\approx 30\textrm{ min}$ (Figure~\ref{fig:E_Edot_30}).  

\subsection{Semi-analytic models  of the dissipation rate}
\label{sec:semianalytic}

Since the mode network integrations are computationally expensive, it is useful to have a semi-analytic model calibrated to the numerical results that can provide an estimate of $\Edot$ over the full range of $\Porb$.  Here we consider models in which the energy dissipation rate is approximated as
\begin{equation}
\Edot = 4  \gamma_{\rm eff} E_{\rm eff},
\end{equation}
where $E_{\rm eff}$ is an effective energy and $\gamma_{\rm eff}$ is an energy-dependent effective damping rate.  The factor of 4 accounts for the two frequency signs and the fact that $\gamma_{\rm eff}$ is the amplitude, rather than energy, damping rate.

\subsubsection{Model 0}
\label{sec:M0}

In Model 0 (M0), our simplest model, we assume that 
\begin{align}
\gamma_{\rm eff}^{({\rm M}0)} &=
\Gamma_{\rm nl} = \lambda_{\rm nl} \omega_a \kappa_{abc} \sqrt{\frac{\Eoff}{E_0}},
\nonumber \\
E_{\rm eff} ^{({\rm M}0)} &= \Eoff ,
\end{align}
which implies
\begin{equation}
\Edot ^{({\rm M}0)} = 4 \gamma_{\rm eff}^{({\rm M}0)} E_{\rm eff} ^{({\rm M}0)}  \propto \left(\frac{2\mratio}{1+\mratio}\right)^3\Omega_{\rm orb}^6 \omega_a^{7.0},
\label{eq:dE_diss_M0_scaling}
\end{equation}
where $\Gamma_{\rm nl}$ is given by Equation~(\ref{eq:tau_nl}) evaluated at a parent energy $E_a=\Eoff$, modes $b,c$ are the fastest growing daughter pair, and $\lambda_{\rm nl}$ is a dimensionless constant whose value is determined by fitting to the numerical results. We separated our $\Edot^{({\rm M}0)}$ expression into the part that depends on $\Omega_{\rm orb}$ and the part that depends on the eigenfrequency $\omega_a$ [which is further related to the driving frequency $\omega=m(\Omega_{\rm orb}-\Omega_{\rm s})\simeq \omega_a$].  The $\Omega_{\rm orb}$ dependence arises from terms that scale with the overall tidal amplitude $\epsilon$ [Equation~(\ref{eq:epsilon})], while the $\omega_a$ dependence arises from terms that depend on the internal structure of the resonant parent modes (e.g., $Q_a$, $\gamma_a$, $\kappa_{abc}$, etc.).  Separating the expression for energy dissipation rates in this way will be useful when we consider a rotating WD and tidal synchronization in Section~\ref{sec:sync}.

 Model 0 is similar to one proposed in \citet{Kumar:96} who studied nonlinear mode damping in tidal capture binaries.  In their analysis, the binary is on a highly eccentric orbit and the parent is excited from essentially zero energy to a linear energy $E_{a, \rm lin}$ during pericenter passage (using the method of \citealt{Press:77}).  
They argue that $\Edot \approx 4\Gamma_{\rm nl} E_{a, \rm lin}$ because that is the maximum rate at which the fastest growing daughter pair can drain energy from an undriven parent that has an initial energy $E_{a, \rm lin}$.

In the bottom panels of  Figures~\ref{fig:E_Edot_60} and \ref{fig:E_Edot_30} we compare M0 to the network simulations.  Although M0 can match the simulations at both $\Porb\simeq 30\textrm{ min}$ and $60\textrm{ min}$, the agreement is only good at the large detuning $\Delta_a = \omega_a/3n_a$ assumed in both figures.  Since  $\Edot ^{({\rm M}0)}$ is independent of $\Delta_a$, it cannot account for the significant variation of $\Edot$ with $\Delta_a$ seen in the numerical simulations (see Figures~\ref{fig:dE_lorentzian} and~\ref{fig:Edot_Tsync_no_rot}).   This failure is perhaps not surprising since here, unlike the highly eccentric orbit of the tidal capture problem, there is a continuous, $\Delta_a$-dependent interaction between the parent's tidal driving and nonlinear damping.

\subsubsection{Model 1}
\label{sec:model_1}
 
In order to construct models that depend on $\Delta_a$, we next consider effective energies with Lorentzian profiles of the form
\begin{equation} 
\frac{E_{\rm eff}}{E_0} = \frac{\omega_a^2}{\Delta_a^2 + \gamma_{\rm eff}^2} U_a^2.
\label{eq:E_eff_model}
\end{equation}
This is similar to the expression for linear energy [Equation~(\ref{eq:E_lin})] except that the linear damping rate  $\gamma_a$ is replaced by the effective damping rate $\gamma_{\rm eff}$.  

Since M0 gives a reasonable approximation to the dissipation rate when $\Delta_a$ is large, we construct models by starting from M0 and using an iterative approach to improve upon it.  Specifically, starting with the maximally-detuned linear energy of the parent $\Eoff$, 
we first define the 0th order expressions \begin{align}
\gamma_{\rm eff}^{(0)}&= \lambda_{\rm nl} \omega_a \kappa_{abc} \sqrt{\frac{\Eoff}{E_0}},
\nonumber \\
\frac{E_{\rm eff}^{(0)}}{E_0} &= \frac{\omega_a^2}{\Delta_a^2 + \left[\gamma_{\rm eff}^{(0)}\right]^2} U_a^2.
\end{align}
We then use these to evaluate the next order expressions, which define our Model 1 (M1)
\begin{align}
\gamma_{\rm eff}^{({\rm M}1)}&= \lambda_{\rm nl} \omega_a \kappa_{abc} \sqrt{\frac{E_{\rm eff}^{(0)}}{E_0}},
\nonumber \\
\frac{E_{\rm eff}^{({\rm M}1)}}{E_0} &= \frac{\omega_a^2}{\Delta_a^2 + \left[\gamma_{\rm eff}^{({\rm M}1)}\right]^2} U_a^2.
\end{align}
Note that $E_{\rm eff}^{({\rm M}1)}$ is \emph{not} the total energy stored in the nonlinear network (see the upper panels of Figures~\ref{fig:E_Edot_60} and \ref{fig:E_Edot_30}). Instead, the total energy is greater than $E_{\rm eff}^{({\rm M}1)}$ by a factor of $\mathcal{O}\left(\gamma_{\rm eff} / \gamma_a \right) \gg1$, as shown in the lower panel of Figure~\ref{fig:dE_lorentzian}.

\subsubsection{Model 2}
\label{sec:model_2}

An alternative and perhaps more natural choice of energy at which to evaluate $\gamma_{\rm eff}$ is $E_{\rm eff}$ itself.  This choice defines our Model 2 (M2), namely 
\begin{align}
\gamma_{\rm eff}^{\rm (M2)} &= \lambda_{\rm nl} \omega_a \kappa_{abc} \sqrt{\frac{E_{\rm eff}^{\rm (M2)}}{E_0}},
\nonumber \\
\frac{E_{\rm eff}^{\rm (M2)}}{E_0} &= \frac{\omega_a^2 }{\Delta_a^2 + \left[\gamma_{\rm eff}^{\rm (M2)}\right]^2}U_a^2
\nonumber \\&
= \frac{-\Delta_a^2 + \sqrt{\Delta_a^4 + 4\lambda_{\rm nl}^2 \omega_a^4 \kappa_{abc}^2 U_a^2}}{2\lambda_{\rm nl}^2 \omega_a^2 \kappa_{abc}^2}.
\label{eq:E_eff_model_M2}
\end{align}
The second equality in the effective energy expression follows by solving the quadratic equation for $E_{\rm eff}$. Note that if we keep performing the iteration process we used in M1, it will eventually converge to M2. 

It will be useful to have the M2 scaling relations for the effective energy and the energy dissipation rate when the parent mode is exactly on resonance.  We find
\begin{align}
\frac{E_{\rm eff}^{\rm (M2)}}{E_0}  \left({\Delta_a {=} 0}\right)  &= \frac{U_a}{\lambda_{\rm nl} \kappa_{\rm abc}} \propto \left(\frac{2\mratio}{1+\mratio}\right)\Omega_{\rm orb}^2 \omega_a^{5.7},
\\
\dot{E}_{\rm diss}^{\rm (M2)} \left({\Delta_a {=} 0}\right)  &\propto \left(\frac{2\mratio}{1+\mratio}\right)^{3/2}\Omega_{\rm orb}^3 \omega_a^{7.5}. 
\label{eq:dE_diss_M2_scaling}
\end{align}
Note that the resonant effective energy scales with the orbital frequency as $\Omega_{\rm orb}^2$, whereas the linear tidal energy scales as $\Omega_{\rm orb}^4$. The difference is due to the fact that the nonlinear damping term $\gamma_{\rm eff}$ is itself a function of tidal energy, whereas the linear damping $\gamma_a$ is independent of $\Omega_{\rm orb}$. We will use Equation~(\ref{eq:dE_diss_M2_scaling})  in Section~\ref{sec:sync} to address the possibility of resonant locking (as studied in \citetalias{Burkart:13} for linear tides) in the weakly nonlinear tide regime.

\subsubsection{Traveling-wave limit}

In Section~\ref{sec:sw_vs_tw} we showed that in the traveling wave regime ($P_{\rm orb} \lesssim 10\textrm{ min}$), the internal gravity waves excited in the core reach such large amplitudes that they become strongly nonlinear and break near the stellar surface.      Although the focus of our study is instead  weakly nonlinear mode coupling in the standing wave regime ($10 \lesssim \Porb/\textrm{min} \lesssim 150$), it is nonetheless instructive to compare the predictions of the two regimes as if one or the other applied at all $\Porb$.

The tidal evolution in the traveling wave regime was studied in detail by \citetalias{Fuller:12a} (see also \citetalias{Burkart:13}). In Appendix~\ref{sec:TW} we review key aspects of the traveling-wave solution and show that it gives an energy dissipation
\begin{equation}
\dot{E}_{\rm diss}^{\rm (tw)} \simeq \hat{f} \omega E_0  \left(\frac{\mratio}{1+\mratio}\right)^2 \left(\frac{\Omega_{\rm orb}}{\omega_0}\right)^4\left(\frac{\omega}{\omega_0}\right)^5 \propto \Omega_{\rm orb}^{4} \omega^{6},
\label{eq:dE_diss_TW_scaling}
\end{equation}
where $\omega = 2(\Omega_{\rm orb} - \Omega_{\rm s})$ is the frequency at which the wave is forced (there are no resonances) and $\hat{f}$ is a dimensionless quantity that characterizes the overall strength of the  dissipation. Based on our WD model, we find $\hat{f}\simeq 20$ (see Figure~\ref{fig:F_omega}), which agrees well with the value obtained by \citetalias{Fuller:12a} for a similar model.  The above equation is related to the tidal energy transfer rate   (see Equation~(42) in \citetalias{Fuller:12a}) by $\dot{E}^{\rm (tw)}_{\rm tide}= \left(\Omega_{\rm orb}/\omega\right)\dot{E}_{\rm diss}^{\rm (tw)}$.

The models adopted by \citetalias{Fuller:12a} and  \citetalias{Burkart:13} are effectively linear models since $\dot{E}_{\rm diss}\propto\epsilon^2\propto\Omega_{\rm orb}^4$. In our weakly nonlinear models, by contrast, $\gamma_{\rm eff}$ is itself a function of $\epsilon$ and thus the dissipation does not scale as $\Omega_{\rm orb}^4$ [see, e.g., Equation~(\ref{eq:dE_diss_M2_scaling})]. We will show in Section~\ref{sec:sync} that this can result in a substantially different spin evolution. 

\subsubsection{Comparison of tidal dissipation models}
\label{sec:compare_Edot}

In the upper panel of Figure~\ref{fig:dE_lorentzian} we show that   M1 and M2 provide a good fit to the mode network simulations for $\lambda_{\rm nl}\simeq 3$.  Moreover, they provide a much better fit than  M0, especially at small $|\Delta_a|$.  Although near exact resonance M1 provides a better fit than M2,  we show in Appendix~\ref{sec:NL_model_compare} that the tidal synchronization and heating are similar in the two models.  We therefore adopt M2 as our fiducial model, as its analytic form is simpler than M1's. 

In the top panel of Figure~\ref{fig:Edot_Tsync_no_rot} we show $\Edot$ of M2 over a wide range of $\Porb$.  We see that it agrees well with the network simulations both in terms of its overall $\Porb$ scaling and the high frequency oscillations with $\Delta_a$ (modulo the slight overestimate near the resonance peaks, as noted above).    It also helps explain why the oscillations decrease in amplitude at smaller $\Porb$; namely, $\gamma_{\rm eff}$ increases and becomes comparable to the maximum detuning, which smears out the resonance peaks.

In Figure~\ref{fig:Edot_Tsync_no_rot} we also show the dissipation $\dot{E}_{\rm diss}^{\rm (tw)}$ if we treat the dynamical tide as a traveling wave at large $\Porb$ (even though it is a standing wave).  We find that the the traveling-wave solution appears to trace the upper envelope of the weakly nonlinear solution.  However, this is merely a coincidence. Indeed, comparing Equations~(\ref{eq:dE_diss_M2_scaling}) and (\ref{eq:dE_diss_TW_scaling}) we see that they only have similar scaling when the spin rate is fixed at zero so that $\omega_a \simeq 2\Omega_{\rm orb}$. When we consider the tidal synchronization problem, they in fact have qualitatively different behaviors.

\subsection{Tidal synchronization and heating}
\label{sec:sync}
 
We now study the tidal synchronization and heating of the WD by using  the calculation of $\Edot$ to solve for the tidal torque $\tau_{\rm tide}$ [Equation~(\ref{eq:tau_tide})] and thereby determine $\dot{\Omega}_{\rm s}$ and $\dot{\Omega}_{\rm orb}$ [Equations~(\ref{eq:dOmega_s}) and (\ref{eq:dOmega_orb})].  We use $\Edot$ as given by model M2 with $\lambda_{\rm nl}=3$ since it provides a useful analytic form that agrees well with the numerical results (Section~\ref{sec:semianalytic}).  

\begin{figure}
   \centering
   \includegraphics[width=0.45\textwidth]{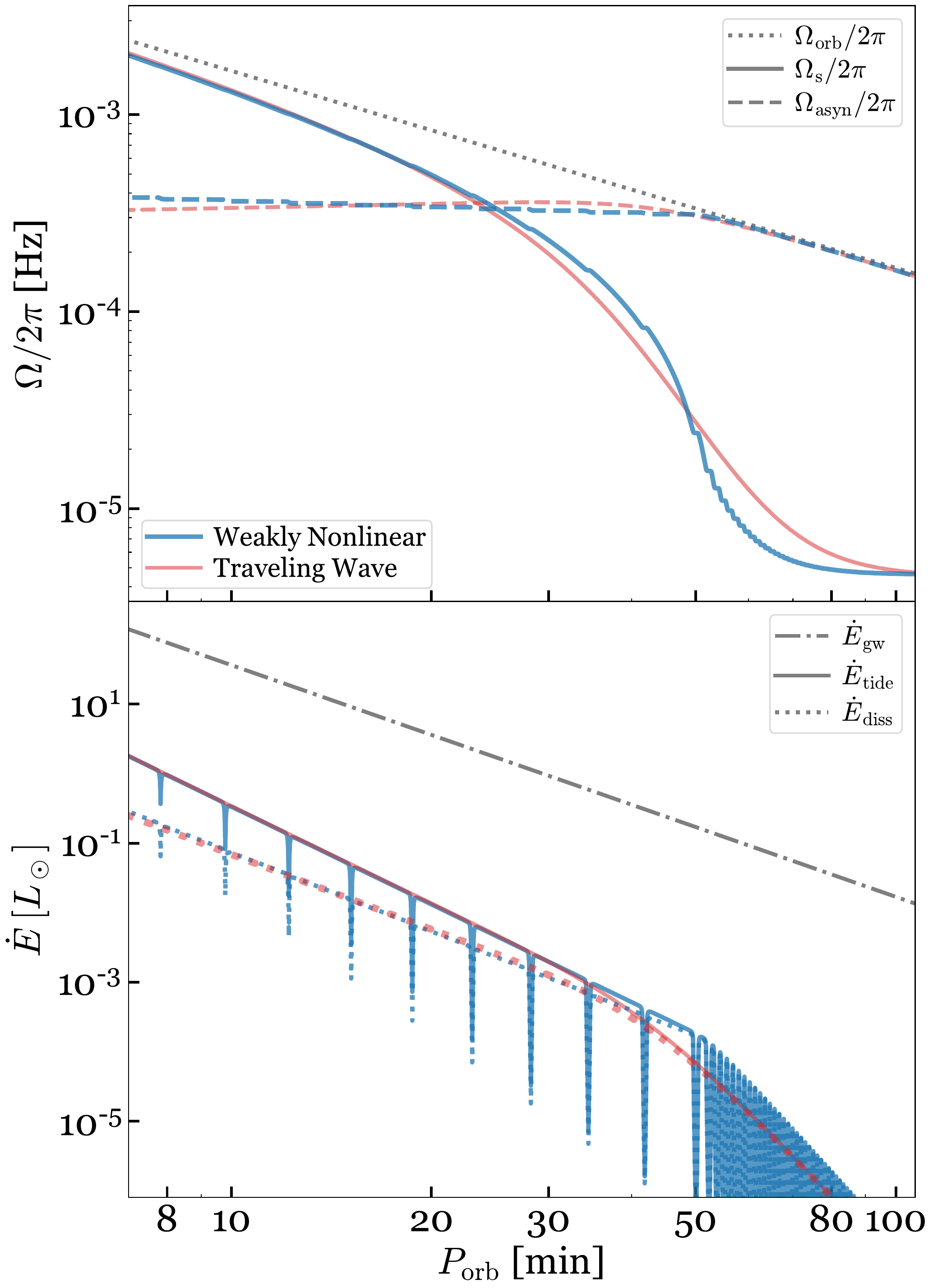} 
   \caption{Top panel: Spin frequency $\Omega_{\rm s}$ (solid lines), asynchronicity $\Omega_{\rm asyn}=\Omega_{\rm orb}-\Omega_{\rm s}$ (dashed lines), and orbital frequency $\Omega_{\rm orb}$ (dotted lines) as a function of orbital period $\Porb$. Bottom panel: Tidal power $\dot{E}_{\rm tide}$ (solid line), tidal dissipation rate inside the star $\Edot$ (dotted line), and GW power $\dot{E}_{\rm gw}$ (dash-dotted line) as a function of $\Porb$.  In both panels, the blue lines correspond to the weakly nonlinear model M2 (assuming $\lambda_{\rm nl}=3$) and the red lines correspond to the traveling wave model. 
}
   \label{fig:spin_power_NLeff_3}
\end{figure}

In the top panel of Figure~\ref{fig:spin_power_NLeff_3} we show the evolution of $\Omega_{\rm s}$ and the asynchronicity $\Omega_{\rm asyn}\equiv \Omega_{\rm orb} - \Omega_{\rm s}$ as a function of $P_{\rm orb}$.  We initialize the frequencies at $\Omega_{\rm orb} =2\pi/\left(240\,{\rm min}\right)$ and $\Omega_{\rm s}=\Omega_{\rm orb}/30$, although we 
find that the synchronization calculation is insensitive to the initial conditions as long as both frequencies are initially small.  
Initially both $\Omega_{\rm s}$ and $\Omega_{\rm asyn}$ increase as the orbit decays but at a critical orbital period $P_{\rm c} \simeq 50\textrm{ min}$ the spin-up has an inflection point and $\Omega_{\rm asyn}$ becomes nearly constant.  This is because at $P_{\rm c}$, the spin-up timescale 
\begin{equation}
T_{\rm s} =  \frac{\Omega_{\rm orb}}{\dot{\Omega}_{\rm s}}
\end{equation}
 first becomes smaller than the orbital decay timescale $T_{\rm gw} =  \Omega_{\rm orb}/ \dot{\Omega}_{\rm orb, gw}$ [Equation~(\ref{eq:T_gw})], as shown in the bottom panel of Figure~\ref{fig:Edot_Tsync_no_rot}.  By evaluating $\Edot$ using model M2 at a resonance $\Delta_a=0$ [see Equation~(\ref{eq:dE_diss_M2_scaling})], where the dissipation has a local maximum and thus $T_{\rm s}$ has a local minimum, we find that the condition $T_{\rm s}=T_{\rm gw}$ is first satisfied at
\begin{equation}
P_{\rm c} = 55 \left(\frac{\lambda_{\rm nl}}{3}\right)^{-0.09}\textrm{ min}.
\label{eq:Pc_nl}
\end{equation} 
Although model M2 slightly overestimates $\Edot$ at resonances (see Section~\ref{sec:semianalytic}), this estimate of $P_{\rm c}$ is robust owing to the weak dependence on $\lambda_{\rm nl}$.  

For $P_{\rm orb} < P_{\rm c}$, the spin frequency $\Omega_{\rm s}$ continues to increase as the orbit decays.  Meanwhile, the asynchronicity $\Omega_{\rm asyn}$ is nearly constant, although importantly it continues to increase, albeit slowly.  This continual increase implies that the system never acquires a resonance lock.  In a resonance lock, the
tidal torque causes the tidal forcing frequency $|\omega|= 2\Omega_{\rm asyn}$ to remain constant even as the orbit shrinks \citep{Witte:99}.  \citetalias{Burkart:13} found that resonance locks should occur universally in WD binaries, whether the parent is a standing wave or a traveling wave.  However, their study did not account for nonlinear mode coupling, which we find prevents resonance locks from forming in the standing-wave regime ($\Porb \ga 10\textrm{ min}$; Section~\ref{sec:sw_vs_tw}).  This is because $\dot{\Omega}_{\rm s} \propto \Edot$ and based on model M2, $\Edot \propto \Omega_{\rm orb}^3 \omega_a^{7.5}$ at perfect resonance $\Delta_a = 0$  [Equation~(\ref{eq:dE_diss_M2_scaling})].  Thus, if $\omega_a\simeq \omega$ remains at a constant value near $4\pi / P_c$ for $\Porb < P_{\rm c}$, we have $T_{\rm s}\propto \Omega_{\rm orb}^{-2}$ and since $T_{\rm gw} \propto \Omega_{\rm orb}^{-8/3}$ even the maximal tidal torque at $\Delta_a = 0$ is insufficient to maintain a resonance lock as $\Omega_{\rm orb}$ increases.

\begin{figure}
   \centering
   \includegraphics[width=0.45\textwidth]{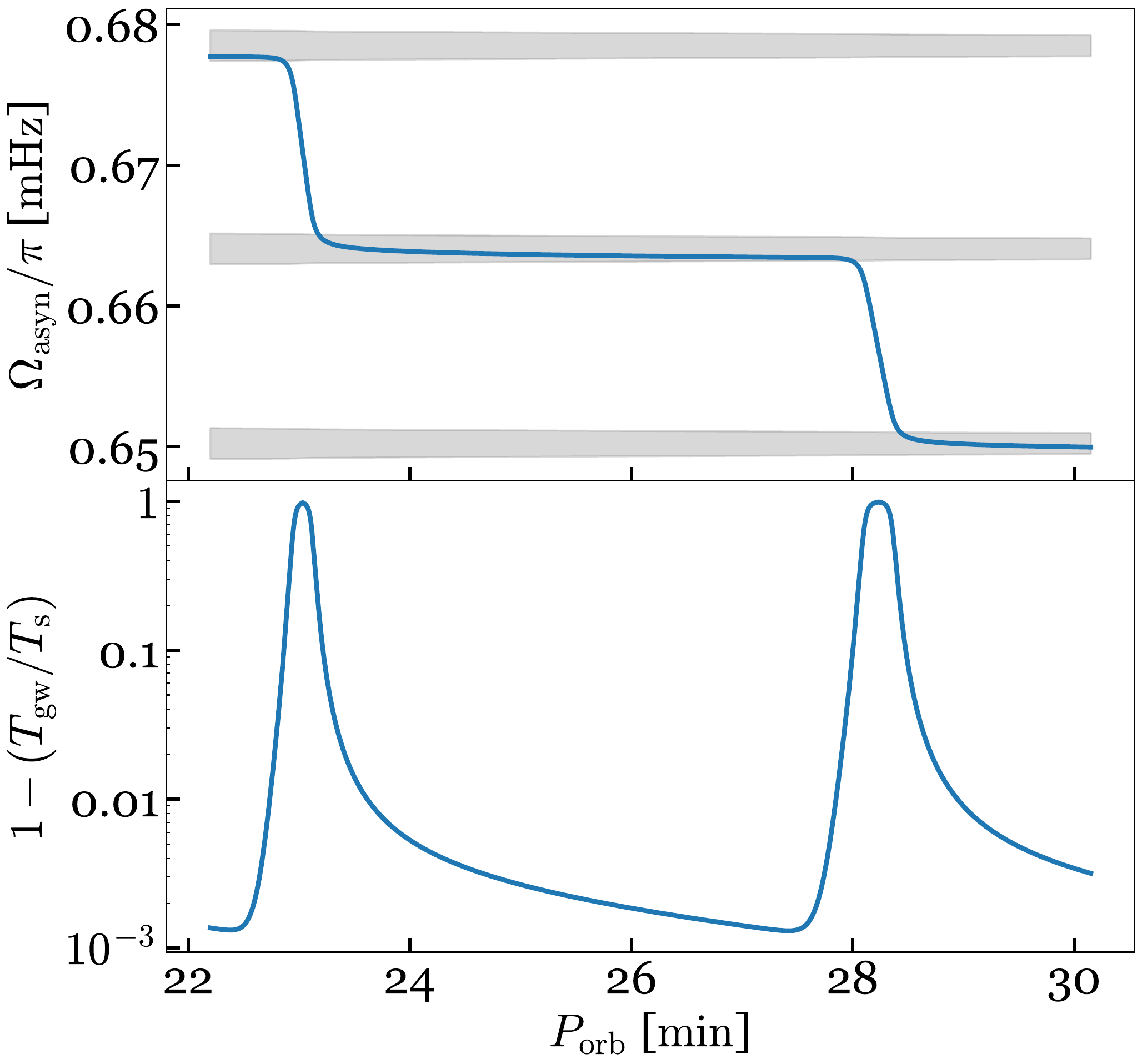} 
   \caption{A zoomed-in view of the spin evolution in the weakly nonlinear model when $P_{\rm orb}<P_{\rm c}^{\rm (nl)}$. The top panel shows the evolution of the asynchronicity, $\Omega_{\rm asyn}/\pi$. The shaded regions (from bottom to top) are centered on the eigenfrequencies of the $n_a=48,\ 47,\ 46$ g-modes, with a width determined by the effective nonlinear damping rate $\gamma_{\rm eff}$ (which increases slightly with decreasing $\Porb$; see Section~\ref{sec:semianalytic}). The bottom panel shows $1-(T_{\rm gw}/T_{\rm s})$, where $T_{\rm s}$ is the spin-up timescale and $T_{\rm gw}$ is the GW-induced orbital decay timescale .}
   \label{fig:spin_zoom_in}
\end{figure}

To better illustrate why resonance locks do not form, in Figure~\ref{fig:spin_zoom_in} we show a zoomed-in view of the spin evolution over three consecutive resonances.  The top panel shows $\Omega_{\rm asyn}$ and the bottom panel shows $1-T_{\rm gw} / T_{\rm s}$.  We see that at resonances (shaded regions), the tidal torque is nearly strong enough to keep $\Omega_{\rm asyn}$  constant and a lock almost forms.  However,  the torque is not quite sufficient to maintain synchronization as the orbit decays (as evidenced by the weaker $\Omega_{\rm orb}$ scaling of $T_{\rm s}\propto \Omega_{\rm orb}^{-2}$ than $T_{\rm gw} \propto \Omega_{\rm orb}^{-8/3}$).  As a result, $\Omega_{\rm asyn}$ slowly increases and the driving frequency gradually moves away from the mode resonance.  This in turn reduces the torque and  increases $T_s$ until at some point (top edge of shaded regions in Figure~\ref{fig:spin_zoom_in})  the detuning becomes greater than the effective damping, $\Delta_a \ga \gamma_{\rm eff}$.  The torque then drops dramatically, $\Omega_{\rm s}$ stops increasing, and $\Omega_{\rm asyn}$ increases rapidly (entirely due to the GW-induced orbital decay). Eventually, $\Omega_{\rm asyn}$ gets so large that it hits the next mode resonance and the cycle begins again.

In the bottom panel of Figure~\ref{fig:spin_power_NLeff_3} we show $\Edot$ and $\dot{E}_{\rm tide}$ as a function of $\Porb$. While the weakly nonlinear model has a heating rate that is overall quite similar to the traveling-wave model (see also Section~\ref{sec:tw_sync} below), it has brief but significant dips. Each dip corresponds to a transition from one resonant mode to the next  (Figure~\ref{fig:spin_zoom_in}), during which the tidal heating is much less than the traveling wave prediction given by Equation~(\ref{eq:dE_diss_post_sync}) below. 

In order to estimate the full width of each dip, first note that the driving frequency changes by $\Delta \omega = |\partial \omega_a/\partial n_a|\simeq \omega_a/n_a$ when evolving through the dip, where $n_a$ ($\omega_a$) is the radial order (eigenfrequency) of the mode prior to the transition. During the dip the orbit evolves much faster than the spin, and therefore $\Delta \omega \simeq m\Delta \Omega_{\rm orb}=2\pi m |\Delta P_{\rm orb}|/P_{\rm orb}^2$. Since, as noted above, $\Omega_{\rm asyn}$ (and hence the driving frequency) evolve slowly for $P_{\rm orb}<P_{\rm c}$, we have $\omega_a \simeq \omega (P_{\rm orb}=P_{\rm c}) \simeq 2\pi m /P_{\rm c}$. Therefore, the width of the dip, i.e., the amount by which the orbital period changes during the dip, is 
\begin{equation}
    |\Delta P_{\rm orb}| \simeq \frac{P_{\rm orb}^2}{n_a P_{\rm c}}.
    \label{eq:dip_width}
\end{equation} 

As we discuss in Section~\ref{sec:EMobs}, the dips may have direct observational consequences, and may provide an explanation for the observed luminosity of the CO WD in J0651~(\citealt{Hermes:12}).

\subsubsection{Comparison with traveling wave limit}
\label{sec:tw_sync}

As with $\Edot$ in Section~\ref{sec:compare_Edot}, it is useful to compare these weakly nonlinear results to the traveling wave results (even though the dynamical tide is a standing wave at $\Porb \ga 10\textrm{ min}$).  According to the latter, $\Edot^{({\rm tw})} \propto \Omega_{\rm orb}^{4} \omega^{6}$ [Equation~(\ref{eq:dE_diss_TW_scaling})] .  Thus, unlike our weakly nonlinear results, $T_{\rm s}^{({\rm tw})}\propto \Omega_{\rm orb}^{-3}$ is steeper than $T_{\rm gw} \propto \Omega_{\rm orb}^{-8/3}$ and for $\Porb < P_{\rm c}^{({\rm tw})}$ the asynchronicity is almost perfectly constant at a value $\Omega_{\rm asyn} \simeq 2\pi/ P_{\rm c}^{({\rm tw})}$.   Using our traveling wave solution (Appendix~\ref{sec:TW}), we find
\begin{equation}
P_{\rm c}^{\rm (tw)} = 45\left(\frac{\hat{f}}{20}\right)^{3/16}\,{\rm min}. 
\label{eq:P_c_tw}
\end{equation}
More specifically, by plugging Equations~(\ref{eq:tau_tide}), (\ref{eq:dOmega_s})-(\ref{eq:dOmega_gw}), and (\ref{eq:dE_diss_TW_scaling}) into the condition $\dot{\Omega}_{\rm s}\simeq\dot{\Omega}_{\rm orb}$, we find
\begin{equation}
\omega = 2\Omega_{\rm asyn} \propto \Omega_{\rm orb}^{-1/15}. 
\label{eq:Omasyn_vs_Omorb}
\end{equation}
We thus see that even when the tidal torque is a smooth power-law of the frequency, the asynchronicity can stay very nearly constant (it in fact decreases very slightly with increasing $\Omega_{\rm orb}$ to compensate for the excess tidal torque and maintain synchronization).  \citetalias{Burkart:13} argued that the torque needs to be a ``jagged" function of the driving frequency $\omega$ in order to maintain a resonance lock.  While we agree that that is necessary in order to maintain an exact lock, i.e., $\dot{\Omega}_{\rm asyn}=0$, Equation~(\ref{eq:Omasyn_vs_Omorb}) implies that even a torque that is a smooth, power-law  function of $\omega$ has $\dot{\Omega}_{\rm asyn}\simeq 0$ and thus will, in effect, result in a lock.

\subsubsection{Tidal heating when synchronous}

When tidal synchronization does occur, the condition $\dot{\Omega}_{\rm s} \simeq \dot{\Omega}_{\rm orb}$ implies that the tidal energy transfer rate is dictated by the GW-induced decay rate and is essentially independent of the microphysics governing the dissipation process. In particular, 
\begin{equation}
\dot{E}_{\rm tide} \simeq I_{\rm WD}\Omega_{\rm orb} \dot{\Omega}_{\rm orb, gw} \simeq \frac{3}{2} I_{\rm WD} \Omega_{\rm orb}^2 \frac{\dot{E}_{\rm pp}}{E_{\rm orb}}, 
\label{eq:dE_tide_post_sync}
\end{equation}
where in the second equality we use the relation $\dot{\Omega}_{\rm orb, gw}/\Omega_{\rm orb}\simeq(3/2) \dot{E}_{\rm pp}/E_{\rm orb}$, with $\dot{E}_{\rm pp}$ the point-particle GW power and $E_{\rm orb}={-}GMM'/2D$ the orbital energy. The tidal heating rate is then given by
\begin{equation}
\dot{E}_{\rm diss} \simeq \frac{2\pi}{\Omega_{\rm orb}P_{\rm c}}\dot{E}_{\rm tide} \simeq \frac{2\pi}{P_{\rm c}} I_{\rm WD}\dot{\Omega}_{\rm orb,gw},
\label{eq:dE_diss_post_sync}
\end{equation}
with $I_{\rm WD}$ and $P_{\rm c}$ the only free parameters. 

Note that even if we do not use the simple power-law fitting formula for the traveling-wave dissipation but take into account the scattering in the internal structure (see Figure~\ref{fig:F_omega}), the post-synchronization heating rate should still be a smooth function of frequency as demonstrated in Figure 14 of \citetalias{Fuller:12a}. Varying the microphysics of the dissipation process (i.e., changing $\hat{f}$ or $\lambda_{\rm nl}$) only affects the post-synchronization heating rate through a change in the location of $P_{\rm c}$, which by Equation~(\ref{eq:dE_diss_post_sync}) only changes the overall magnitude of the dissipation rate. And since $P_{\rm c}$ only depends weakly on $\hat{f}$ and $\lambda_{\rm nl}$ [see Equations~(\ref{eq:Pc_nl}) and~(\ref{eq:P_c_tw})], the observed  tidal heating should have a relatively small scattering for different CO WDs at a given orbital period $P_{\rm orb}< P_{\rm c}$. We discuss the observational implications of this in more detail in the next section. 

\section{Observational Signatures}
\label{sec:obs_signatures}

\subsection{In electromagnetic radiation}
\label{sec:EMobs}

Tidal dissipation converts a fraction of the orbital energy into heat. In Appendix~\ref{sec:lin_diss} we argue that the majority of the heat should be deposited at locations sufficiently close to the WD's surface where the thermal diffusion timescale is much shorter than the orbital decay timescale. As a result, we would expect the tidal heating to be instantly manifested at the surface and to play a significant role in determining the luminosity of WDs in compact binaries. This is especially true for systems with orbital periods $\lesssim 20\,{\rm min}$, as we may expect the tidal heating to exceed the WD's intrinsic cooling (for a typical CO WD with an age of 1 Gyr, the luminosity due to its cooling is about $10^{-3}\,L_{\odot}$; \citealt{Salaris:97}). Thus it would be particularly interesting to compare our prediction of the tidal heating rate to the observed luminosity of the CO WD\footnote{We focus here only on the CO WD which is consistent with our background stellar model. We leave for future study the case of weakly nonlinear dynamical tides of a He WD.} in the 13-min system J0651 ($L=1.0\times10^{-3}L_\odot$ and $T_{\rm eff}=8700\,{\rm K}$;~\citealt{Hermes:12}).

We first consider the heating rates calculated under the traveling-wave model, which is appropriate for $\Porb \la 10\textrm{ min}$ (see Section~\ref{sec:sw_vs_tw}). As shown in \citetalias{Fuller:12a} and \citetalias{Burkart:13}, the traveling-wave calculation would predict a heating rate higher than the observed luminosity of the CO WD in J0651 by about a factor of $10$. However, one of the key features of the traveling-wave model is that the heating rate should be a relatively smooth function of $\Porb$ with little scatter. Because of the synchronization condition $\dot{\Omega}_s \simeq  \dot{\Omega}_{\rm orb}$, the heating rate is dictated by the GW radiation and should thus follow a smooth power-law with respect to period. The only free parameters are the moment of inertia of the WD, $I_{\rm WD}$, and the asynchronicity period, $P_{\rm c}^{\rm (tw)}$ [see Equation~(\ref{eq:dE_diss_post_sync})]. The uncertainty in $I_{\rm WD}$ should be relatively small. Meanwhile, to increase $P_{c}^{\rm (tw)}$ by a factor of 10 (in order to explain the luminosity of J0651), it would require an increase of $\hat{f}$, the characteristic traveling-wave dissipation rate, by a factor of $2.2\times10^5$ [see Equation~(\ref{eq:P_c_tw})]! 

On the other hand, our nonlinear model offers a potential explanation of the observed luminosity of J0651 (though it may not be the only explanation). Recall from Figure~\ref{fig:spin_power_NLeff_3} that the nonlinear model (blue traces) has a heating rate that is overall similar to the traveling-wave prediction when $10\,{\rm min}{<}P_{\rm orb}{<}20\,{\rm min}$, except that there are dips in the nonlinear heating model when the asynchronicity transitions from one mode's resonance to the next. 

\begin{figure}
   \centering
   \includegraphics[width=0.45\textwidth]{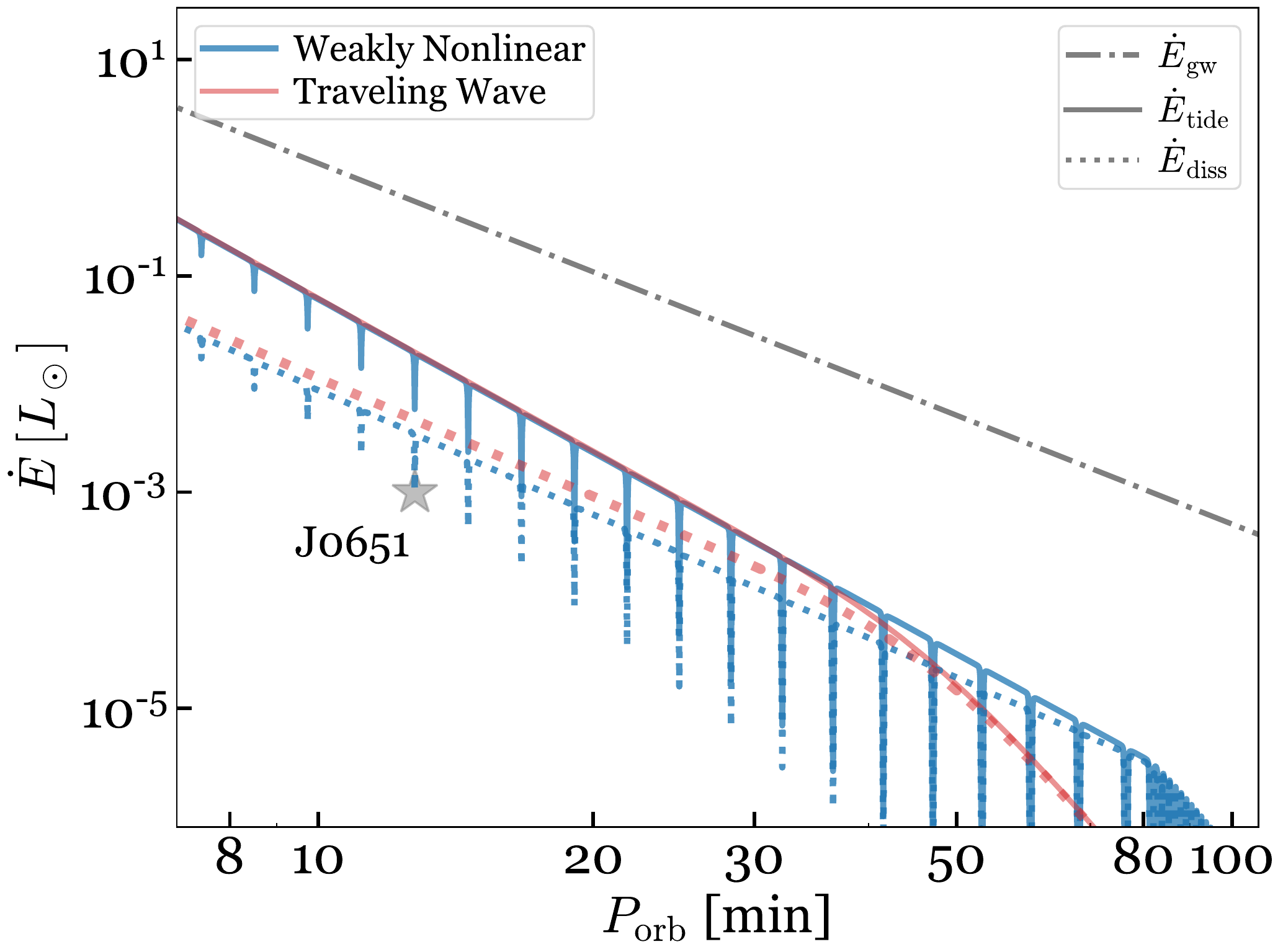} 
   \caption{Similar to the bottom panel of Figure~\ref{fig:spin_power_NLeff_3} but for parameters corresponding to a J0651-like binary.  We assume $M=0.5\,M_\odot$, $M'=0.25\,M_\odot$ and let the radius of the primary be $R=9.9\times10^{-8}\,{\rm cm}$ but keep the other parameters controlling the primary WD's internal structure the same as our main model. The star symbol is at the observed period  and luminosity  of J0651~\citep{Hermes:12}.}
   \label{fig:power_J0651}
\end{figure}

In Figure~\ref{fig:power_J0651},  we  repeat the tidal heating calculation as we have done in Section~\ref{sec:sync}. To generate the plot, we have adjusted the overall tidal amplitude $\epsilon$ according to J0651~\citep{Hermes:12} so that $M=0.5\,M_\odot$ and $M'=0.25\,M_\odot$ for the primary and the secondary, respectively, and $R=1.4\times10^{-2}R_\odot = 9.9\times10^{-8}\,{\rm cm}$ (for the primary; the secondary is treated as a point mass). The other parameters determining the internal structure of the primary WD are left the same as our main WD model (see Section~\ref{sec:eom}; this should be a good approximation as our model has a similar mass and effective temperature as the CO WD in J0651). We find a surprisingly good agreement between our nonlinear model and the observation.\footnote{It is also interesting to note that when the companion becomes less massive, the weakly nonlinear model has a greater critical period than the traveling-wave model, $P_{\rm c}^{\rm (nl)} > P_{\rm c}^{\rm (tw)}$ when $\mratio<1$. This is because the weakly nonlinear mode has a tidal dissipation rate that scales with the mass-ratio $\mratio$ as $\left[2\mratio/(1+\mratio)\right]^{3/2}$ whereas in the traveling-wave model the scaling is $\left[2\mratio/(1+\mratio)\right]^{2}$. See Equations~(\ref{eq:dE_diss_M2_scaling}) and (\ref{eq:dE_diss_TW_scaling}). }

While the exact match between our model and observation in Figure~\ref{fig:power_J0651} is a coincidence of our background model, we can nonetheless estimate the probability of observing such a system. In order to produce the low luminosity of J0651, it requires a system to be undergoing transition from one resonant mode to the other (see Figure~\ref{fig:spin_zoom_in}). In Figure~\ref{fig:power_J0651} it corresponds to the transition from parent mode $n_a=67$ to $n_a=66$. Thus the frequency difference between the two modes can be estimated as $|\partial \omega_a/\partial n_a |\simeq \omega_a/n_a$, corresponding to a width of the dip in terms of orbital period of $P^2_{\rm orb}/\left(n_aP_{\rm c}\right)\simeq 0.031\,{\rm min}$ [Equation~(\ref{eq:dip_width})]. This gives the analytical approximation of the full width of the dip, and numerically we find a width of $0.021\,{\rm min}$ inside which the luminosity is within a factor of 2 of the local minimum. 
Meanwhile, the typical separation between two dips is about $1.7\,{\rm min}$ (the three dips closest to J0651 are respectively at orbital periods of $11.2\,{\rm min}$, $12.8\,{\rm min}$, and $14.5\,{\rm min}$). Therefore, the probability of finding a system at a dip in the tidal heating is thus estimated to be $0.021/1.7\simeq1.2\%$. 

We note that the parameter space can be further expanded if one takes into account the scattering in, e.g., the tidal overlap of the parent mode $Q_a$, and/or the three-mode coupling coefficient $\kappa_{abc}$ as they can make the dips deeper and hence a larger range of $P_{\rm orb}$ would be consistent with the observation. Note that the scattering in the internal structure affects the tidal heating only when $T_{\rm s}\gg T_{\rm gw}$, and therefore has little effect in the traveling-wave limit as argued above.

However, we cannot readily explain the luminosity of the recent observed 7-min system J1539~\citep{Burdge:19}. The model only allows for extra scattering towards the lower luminosity side of the traveling-wave model, which cannot be used to explain the higher than expected temperature of the CO WD in J1539.  Moreover, the very low luminosity and temperature of the secondary WD in that system likely fall below our estimates (though we have not yet computed nonlinear effects in He WD models). In general, it is difficult for tidal heating models to simultaneously explain the high luminosity of the primary and the low luminosity of the secondary in J1539, so it is likely that other effects such as ongoing mass transfer are occurring in that system.

Looking towards future, the nature of tidal dissipation can be better constrained when more compact WD binaries are observed by campaigns like the ELM~\citep{Brown:16} and ZTF~\citep{Graham:19} surveys. Whereas the traveling-wave model predicts the luminosities should follow a smooth power-law with respect to the orbital period with small scatter, in the nonlinear model we might expect occasional dips in the luminosity that are $\mathcal{O}(10)$ times fainter than the prediction of a smooth power-law. The probability of seeing an under-luminous system is estimated to be a few percent, with the CO WD in J0651 potentially being one such example.  A complication is that some WDs may be born at short orbital periods and still radiating their natal thermal energy, adding upward scatter to the observed temperatures. More discoveries at very short orbital periods ($P < 15 \, {\rm min}$) where tidal heating dominates the luminosity will help test these ideas.

\subsection{In gravitational waves}
\label{sec:GWobs}

The tidal interaction may lead to signatures in GWs that are potentially observable for proposed GW observatories like LISA~\citep{Kupfer:18,Korol:20} and TianGO~\citep{Kuns:19}, whose detectability we estimate here. Our focus will be on systems that are sufficiently compact  that their frequency chirping can be resolved by GW observatories over $\sim$5 years. Moreover, we want the source to be individually resolvable instead of being part of the confusion foreground. This typically  requires the system to start at a GW frequency $f_{\rm gw} > 3\,{\rm mHz}$, which corresponds to an orbital period $P_{\rm orb} < 11\,{\rm min}$. For those systems, the traveling-wave limit studied by \citetalias{Fuller:12a} begins to apply, as shown in Section~\ref{sec:sw_vs_tw}. In fact, the best constraints on the tide will be derived from systems that are so compact that they are close to the onset of mass-transfer.\footnote{For a typical 0.6-0.6 WD binary, the onset of the Roche-lobe overflow corresponds to a GW frequency of 30\,{\rm mHz} and $P_{\rm orb}=1.1\,{\rm min}$.} In part, these systems are intrinsically louder in GW radiation compared to the less compact ones. Furthermore, as argued in Equation~(\ref{eq:dE_tide_post_sync}), we have $\dot{E}_{\rm tide}/|\dot{E}_{\rm pp}|\propto \Omega_{\rm orb}^{4/3}$, and thus tidal effects play an increasingly important role relative to the point-particle GW radiation as the orbital frequency increases. More importantly, these systems will experience a significant amount of frequency evolution, which allows us to disentangle the point-particle effects and the tidal effects even if we do not know the binary's chirp mass a priori. 
To address this quantitatively, we will focus on binaries with $P_{\rm orb}\lesssim 5\,{\rm min}$ and adopt the Fisher information matrix to estimate the detectability of parameters, especially the WD moment of inertia, $I_{\rm WD}$. Our study compliments that by \cite{Piro:19} who focused on systems at longer orbital periods near  $P_{\rm orb}\simeq 10\,{\rm min}$. Such systems experience much less frequency evolution and thus only the leading-order frequency derivatives can be resolved.\footnote{Assuming a 5-year observation,  the frequency resolution is $6.3\,{\rm nHz}$. Over this period, a system initially at $P_{\rm orb}\simeq 10\,{\rm min}$ ($f_{\rm gw}\simeq 3\,{\rm mHz}$) evolves only $\sim 25\,{\rm nHz}$. In comparison, the  systems we consider in this Section will evolve by an amount ranging from $\sim1\,{\rm \mu Hz}$ if the initial GW frequency is $f_{\rm gw}=10\,{\rm mHz}$, to $\sim 100\,{\rm \mu Hz}$ if the initial frequency is $f_{\rm gw}=30\,{\rm mHz}$.} 

While in the case of inspiraling neutron star binaries, the leading order effect on the gravitational waveform is due to the equilibrium tide~\citep{Flanagan:08}, it plays a comparatively minor role in the case of double-WDs. To see why, first note that the energy of the equilibrium tide can be written as \citepalias{Burkart:13}
\begin{equation}
\frac{E_{\rm eq}}{E_0} = k_{\rm eq} \epsilon^2, 
\end{equation}
where the constant $k_{\rm eq} \equiv 2\sum_a W_a^2 Q_a^2\simeq 0.07$, which is largely dominated by the f-mode contribution.
The internal dissipation of the equilibrium tide induces a negligible tidal lag~\citep{Willems:10}. Instead, the dominant contribution to the tidal lag is the GW-induced orbital decay (see, e.g., \citealt{Lai:94}). The associated energy transfer rate into the equilibrium tide (to raise the tidal bulge) is thus given by
\begin{align}
\dot{E}_{\rm eq} &= \frac{2}{3} \frac{\dot{\Omega}_{\rm orb}}{\Omega_{\rm orb}} E_{\rm eq} \nonumber \\
&= 2.0\times10^{-5} L_\odot \left(\frac{k_{\rm eq}}{0.1}\right) \left(\frac{P_{\rm orb}}{10 \,{\rm min}}\right)^{-20/3}.
\end{align}
This is negligible compared to the energy transfer rate due to the dynamic tide (see Figure~\ref{fig:spin_power_NLeff_3}), and we therefore ignore the effect of the equilibrium tide in the following discussion (see however the last paragraph of this Section).

Since the systems we consider in this Section are \emph{in the traveling-wave regime}, we expect the WD's spin to be well-synchronized with the orbit and thus $\dot{\Omega}_{\rm orb} \simeq \dot{\Omega}_{\rm s}$. We can solve for the excess frequency evolution due to the dynamical tide $\dot{\Omega}_{\rm tide} \equiv \dot{\Omega}_{\rm orb} - \dot{\Omega}_{\rm orb, gw}$ by using the relations given in Section~\ref{sec:energy_and_AM_transfer} and the fact that  the post-synchronization dynamical tide is essentially controlled by a single parameter, $I_{\rm WD}$ (Section~\ref{sec:tw_sync}), and is therefore insensitive to the details of the tidal interaction (namely, the value of $\hat{f}$).  We find 
\begin{equation}
\dot{\Omega}_{\rm tide} = \dot{\Omega}_{\rm orb} - \dot{\Omega}_{\rm orb, gw} = \left(\frac{3I_{\rm WD}/I_{\rm orb}}{1- 3I_{\rm WD}/I_{\rm orb}}\right)\dot{\Omega}_{\rm orb, gw}. 
\end{equation}
where $I_{\rm orb}$ is the orbital moment of inertia, and for future convenience we express it in terms of $\mathcal{M}_c$ and $\Omega_{\rm orb}$ as
\begin{equation}
    I_{\rm orb} = \mu D^2 = \frac{G^{2/3}\mathcal{M}_{\rm c}^{5/3}}{\Omega_{\rm orb}^{4/3}}. 
    \label{eq:I_orb}
\end{equation}

The Fisher matrix analysis is most conveniently done in the frequency domain. This requires finding the phase $\Psi(f_{\rm gw})$ of the GW waveform in the frequency domain, which is related to the time-domain GW phase $\phi(t)$ as \citep{Cutler:94}
\begin{equation}
\Psi(f_{\rm gw}) = 2\pi f_{\rm gw} t(f_{\rm gw})  - \phi\left[t(f_{\rm gw})\right] -\frac{\pi}{4}. 
\end{equation}
Separating the GW frequency evolution into the point-particle contribution $\dot{f}_{\rm pp} = \dot{\Omega}_{\rm orb, gw}/\pi$ and the tide-induced contribution $\dot{f}_{\rm tide} = \dot{\Omega}_{\rm tide}/\pi$, we have 
\begin{align}
t(f_{\rm gw}) &= \int^{f_{\rm gw}} \frac{\diff f}{\dot{f}}  = \int^{f_{\rm gw}} \frac{\diff f}{\dot{f}_{\rm pp} + \dot{f}_{\rm tide}}, \nonumber \\
&= t_{\rm pp}(f_{\rm gw}) - \int^{f_{\rm gw}} \frac{3I_{\rm WD}}{I_{\rm orb}} \frac{\diff f}{\dot{f}_{\rm pp}},
\end{align}
where $t_{\rm pp} (f_{\rm gw}) = \int \diff f_{\rm gw}/\dot{f}_{\rm pp}$ is the time as a function of GW frequency without tidal effects and we use Equation~(\ref{eq:I_orb}) to derive the last equality. The lower limit of the integration (not shown) is set to be the initial frequency of the waveform. 
Similarly, the time-domain phase can be evaluated as 
\begin{align}
\phi\left[t(f_{\rm gw})\right] &= 2\pi\int^{\rm f_{\rm gw}}\frac{f}{\dot{f}} \diff f  \nonumber \\
&= \phi_{\rm pp} \left[ t(f_{\rm gw})\right] - 2\pi\int^{f_{\rm gw}}\frac{3I_{\rm WD}}{I_{\rm orb}}\frac{f}{\dot{f}_{\rm pp}} \diff f,
\end{align}
where $\phi_{\rm pp}$ is the point-particle phase. Consequently, we have
\begin{align}
\Psi(f_{\rm gw}) = &\Psi_{\rm pp}(f_{\rm gw}) \nonumber \\
&- 2\pi \left(f_{\rm gw} \int^{f_{\rm gw}} \frac{3I_{\rm WD}}{I_{\rm orb}} \frac{\diff f}{\dot{f}_{\rm pp}} - \int^{f_{\rm gw}} \frac{3I_{\rm WD}}{I_{\rm orb}} \frac{f}{\dot{f}_{\rm pp}} \diff f\right). 
\label{eq:Psi_gw}
\end{align}

Even at the onset of mass-transfer ($f_{\rm gw}\simeq30\,{\rm mHz}$ for a typical 0.6-0.6 WD binary), the orbital velocity  $(v_{\rm orb}/c)^2 \lesssim 10^{-4}$. Thus the leading-order quadruple formula is sufficient to describe the point-particle phase $\Psi_{\rm pp}$, which is given by
\begin{equation}
\Psi_{\rm pp}(f_{\rm gw}) = 2\pi f_{\rm gw} t_c - \phi_c -\frac{\pi}{4} - \frac{3}{4}\left(\frac{8\pi G}{c^3} \mathcal{M}_c f_{\rm gw}\right)^{5/3},
\end{equation}
where $t_c$ and $\phi_c$ are constants of integration. Since the orbital moment of inertia $I_{\rm orb}$ can be viewed as function of $f_{\rm gw}$, with the chirp mass $\mathcal{M}_{c}$ a parameter [see Equation~(\ref{eq:I_orb})], we can  construct the frequency-domain strain waveform $h(f_{\rm gw})$ with 5 parameters,\footnote{Here we focus on the detectability of intrinsic parameters, so we have dropped the inclination, polarization, and sky location of the source, and use the sky-averaged sensitivity curves of LISA and TianGO~(see figure 1 of \citealt{Kuns:19}).} $\left\{r, \mathcal{M}_c, t_c, \phi_c, I_{\rm WD}\right\}$, where $r$ is the distance to the source.  Note that so far we only included the tidal effect from one of the WDs; in reality,  both WDs contribute to the phase shift and the quantity we measure will be the sum of their moments of inertia.\footnote{Note that the WD's moment of inertia enters the phase linearly. Therefore the parameter estimation uncertainty on the moment of inertia, $\Delta I_{\rm WD}$, is independent of the magnitude of $I_{\rm WD}$, whereas the fractional error $\Delta I_{\rm WD}/I_{\rm WD}$ decreases as $I_{\rm WD}$ increases.}

\begin{figure}
   \centering
   \includegraphics[width=0.45\textwidth]{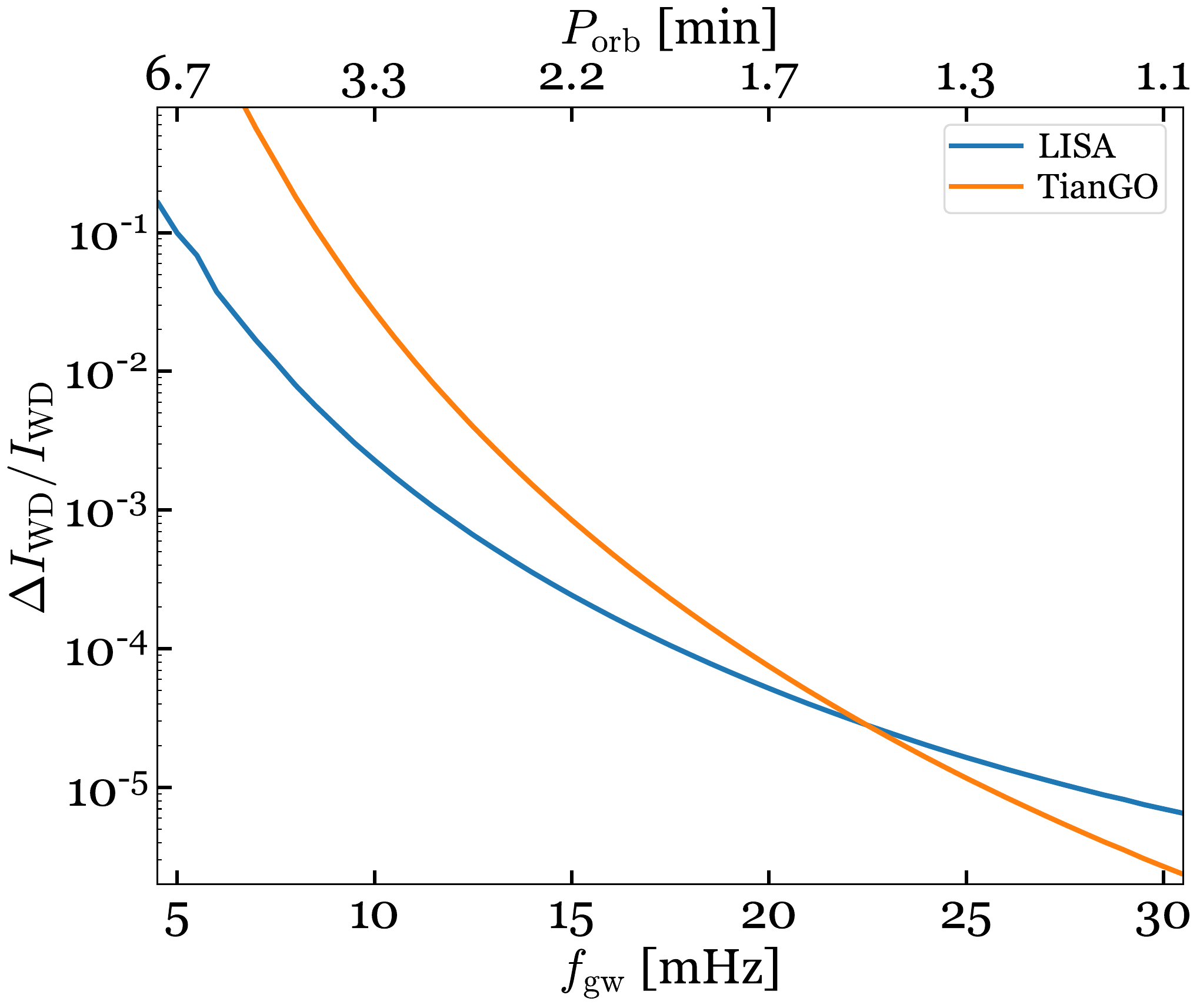} 
   \caption{Fractional uncertainty on the WD's moment of inertia as a funtion of the binary's GW frequency over an observational period of 5 years. We assume the binary is at a distance of 10\,kpc and we averaged over its sky location. Note we focus here on binaries with very short orbital periods, and hence assume that the system is in the traveling-wave regime and the spin is synchronized with the orbit ($\dot{\Omega}_{\rm s}=\dot{\Omega}_{\rm orb}$). }
   \label{fig:PE}
\end{figure}

In Figure~\ref{fig:PE} we show the fractional measurement uncertainty of the WD's moment of inertia as a function of the binary's GW frequency. The x-axis gives the binary's \emph{initial} GW frequency, and we show results assuming a 5-year observation.  We fix the source distance at 10\,kpc and adopt the Fisher matrix technique to calculate the parameter estimation uncertainty using both the sky-averaged LISA (blue) and TianGO (orange) sensitivities.  We see that for a WD at $f_{\rm gw}=10\,{\rm mHz}$, LISA can already constrain the moment of inertia to better than $0.1\%$. For a system close to the onset of Roche-lobe overflow, the statistical uncertainty in $I_{\rm WD}$ with TianGO's sensitivity can be as small as $10^{-6}$. In reality, this precision may not be reached because the modeling assumptions of this Section (such as the assumption of spin-orbit synchronization) introduce systematic errors.  
Nonetheless, it is clear that future space-based GW observatories will be able to detect the tide's contribution to the orbital decay, which will constrain the WD moments of inertia and theories of tidal dissipation.

While here we have treated the moment of inertia $I_{\rm WD}$ as a free parameter, in the near future we may have a sufficiently reliable model of WDs (especially after verifying the tidal effect after the first few detections by LISA and/or TianGO with high statistical accuracy). If, for example, we can treat the moment of inertia as a function of mass, $I_{\rm WD}=I_{\rm WD}(M)$, instead of as a free parameter, then the tidal effect will allow us to measure the component masses of the binary instead of just chirp mass [see Equation~(\ref{eq:Psi_gw})]. This will help improve our understanding of Type-Ia supernovae and their progenitors as it allows us to measure a  binary's total mass and determine whether it is super- or sub-Chandrasekhar.  This possibility was  also demonstrated by~\citet{Kuns:19}. 

Lastly, while here we focus on the effects of dynamical tides on the secular evolution of the binary, \citet{McNeill:19} recently proposed an alternative method of constraining the WD structure using the equilibrium tide. Specifically, the equilibrium tide introduces a non-dissipative radial force $-\partial H /\partial D{\sim} {\rm Re}\left[q_a^\ast U\right]$ [see Equation~(\ref{eq:Hamiltonian})]. This force causes a \emph{non-secular} oscillation of the orbital eccentricity, which generates GWs at both $\Omega_{\rm orb}$ and $3\Omega_{\rm orb}$ in addition to the main $2\Omega_{\rm orb}$ harmonic. Since the effect proposed by \citet{McNeill:19} operates on a timescale of $P_{\rm orb}$ whereas the dynamical tide is manifested over a much longer timescale $\sim T_{\rm gw}$, we expect the two effects to be complementary to each other. The eccentricity excited by the equilibrium tide might further enhance the dynamical tide's dissipation rate, as the spin is not synchronized with the first and third orbital harmonics, and it could thus further excite waves inside the WDs. We plan to study this interaction in the future.  

\section{Summary and discussion}
\label{sec:conclusion}

In this Paper, we studied the effects of nonlinear dynamical tides in compact WD binaries that inspiral due to GW radiation. Our focus was on the weakly nonlinear regime which we showed covers the orbital period range $10 \lesssim \Porb / \textrm{min} \lesssim 150$ (see Figure~\ref{fig:E_th_vs_P}).  In this range, parent modes resonantly driven by the linear tide are so energetic that they excite secondary waves through the three-mode parametric instability.  At longer periods linear theory applies, and at shorter periods the parents are driven to such large amplitudes that they become  strongly nonlinear and break near the WD's surface.  Such parents are therefore  traveling waves rather than standing waves.

To study the weakly nonlinear regime, we carried out a suite of numerical integrations of coupled mode networks over a wide  range of orbital periods.  The duration of each integration was a few nonlinear growth times $T_{\rm nl}\ll T_{\rm gw}$, where $T_{\rm gw}$ is the GW decay timescale. On this timescale, the system settled into a quasi-steady state in which the total mode energy and energy dissipation rate approached constant values. We considered mode networks with 32 and 92 modes, both consisting of five mode generations, and found that they converged on similar  values for the total energy dissipation rate.  The computed dissipation rates are orders of magnitude larger than that predicted by linear theory. 

Based on the mode network integrations, we  constructed phenomenological models that provided a robust estimate of the nonlinear dissipation rate as a function of the system parameters (Section~\ref{sec:dEdt_no_rot}). In the models, the total energy dissipation rate is given by the product of an effective damping rate $\gamma_{\rm eff}$ and an effective energy $E_{\rm eff}$. The effective damping is characterized by the three-mode parametric  growth rate $\Gamma_{\rm nl}$ [Equation~(\ref{eq:Gamma_nl})] which is itself a function of mode energy. The effective energy has a Lorentzian profile like the linear tide energy but with $\gamma_{\rm eff}$ replacing the linear damping rate $\gamma_a$ [see  Equations~(\ref{eq:E_lin}) and~(\ref{eq:E_eff_model_M2})]. They are approximately equal when the frequency detuning is large ($\Delta_a > \gamma_{\rm eff}\gg \gamma_a$), while $E_{\rm eff}$ is always much smaller than the total mode energy in the nonlinear network [their ratio is $\mathcal{O}(\gamma_a/\Gamma_{\rm nl})\ll 1$]. 

We used the dissipation models to analyze the tidal synchronization and heating of a CO WD as a function of orbital separation. Although the trajectories in the weakly nonlinear, standing wave regime are similar to what previous studies found by (incorrectly) assuming a traveling wave at $\Porb > 10\textrm{ min}$,  there are some important differences. The most significant difference is that in the weakly nonlinear analysis, there are brief dips in the tidal heating rate that are $10-100$ times below the traveling-wave  estimates (see Figures~\ref{fig:spin_power_NLeff_3} and Figures~\ref{fig:power_J0651}). This is because in our weakly nonlinear model, tidal synchronization can only be approximately achieved for a finite duration near a resonance  peak (Figure~\ref{fig:spin_zoom_in}). Once the tidal torque at resonance becomes insufficient to synchronize the spin with the orbit, the asynchronicity $\Omega_{\rm asyn}=\left(\Omega_{\rm orb} - \Omega_{\rm s}\right)$ increases, and the mode moves out of resonance. As a result, the total tidal torque and heating rate drop significantly until the next mode becomes resonant. 

These dips offer a potential explanation for the observed luminosity of the CO WD in J0651 (see Figure~\ref{fig:power_J0651}), which is about 10 times fainter than predicted by the traveling-wave model. On the other hand,   the probability of finding a WD in such a state is only a few percent based on the width and spacing of the dips.   The recently observed 7-min system J1539 has an especially high luminosity that cannot be explained by our model, although it is likely in the traveling wave regime and other non-tidal effects, such as ongoing or previous mass transfer, are likely at play in this system.

More generally, we predict that most WD binaries with orbital periods between about $10\textrm{ min}$ and $20\textrm{ min}$ will have a luminosity $L$ consistent with the traveling-wave model and follow a power-law scaling with respect to the orbital period, $L\simeq \dot{E}_{\rm diss}\propto P_{\rm orb}^{-11/3}$ [Equation~(\ref{eq:dE_diss_post_sync})].  However, we expect $\mathcal{O}\left(1\%\right)$ will be outliers that are 10 times dimmer. Future surveys should be able to test this idea.

Lastly, we considered the impact of dynamical tides on the GW signal. Since the loudest sources will have $\Porb < 10\textrm{ min}$, in this part of the analysis we adopted the traveling-wave model and assumed that the WD spin would be synchronized with the orbit. We showed that under these assumptions the only free parameter impacting the GW signal is the moment of inertia of the WD (or the sum of the moments of inertia if the tides in both WDs are taken into account). We found that the moment of inertia should be constrained to better than $1\%$ with future space-based GW observatories like LISA or TianGO. 

Our mode coupling formalism and network integrations assumed that all the excited modes are standing waves.  Although we showed that the parent mode  does not break for $\Porb \ga 10\textrm{ min}$ and is therefore a standing wave, it is less clear whether the same is true of the secondary waves that the parent excites.  Since the shear increases with increasing wavenumber, the secondary waves break at a smaller energy than the parent.  On the other hand, they are parametrically unstable to three mode coupling at a smaller energy than the parent.  As our network integrations show, three mode coupling can suppress mode amplitudes and prevent them from reaching wave breaking energies (e.g., we find that the parent's energy at resonance peaks is suppressed by orders of magnitude compared to the linear value as a result of three mode coupling; see Figure~\ref{fig:dE_lorentzian}).  Addressing this issue in detail requires a formalism that allows for a mix of coupled standing waves and traveling waves.  Such an analysis might be especially important for very hot WDs since the shear and linear damping rates increase with increasing temperature and thus the tide is more likely to excite traveling waves.

Throughout our analysis, we only accounted for the spin's effect on the Doppler shift of the tidal driving frequency but ignored Coriolis and centrifugal effects of rotation on the WD's oscillation modes. We also assumed that the WD can maintain a solid-body rotation throughout its evolution.  In the future, it would be interesting to carry out a more rigorous and comprehensive treatment of rotation in the weakly nonlinear regime. Nonetheless, the study by~\citet{Fuller:14} suggests that such a treatment is unlikely to change our general conclusions.

\section*{Acknowledgements}
The authors thank Yanbei Chen and Dong Lai for the valuable discussions. This work made use of the High Performance Computing resources at MIT Kavli Institute. HY is supported by the Sherman Fairchild foundation. NNW acknowledges support from the NSF through grant AST-1909718.

\bibliographystyle{mnras}
\bibliography{ref}{}

\begin{thebibliography}{}
\makeatletter
\relax
\def\mn@urlcharsother{\let\do\@makeother \do\$\do\&\do\#\do\^\do\_\do\%\do\~}
\def\mn@doi{\begingroup\mn@urlcharsother \@ifnextchar [ {\mn@doi@}
  {\mn@doi@[]}}
\def\mn@doi@[#1]#2{\def\@tempa{#1}\ifx\@tempa\@empty \href
  {http://dx.doi.org/#2} {doi:#2}\else \href {http://dx.doi.org/#2} {#1}\fi
  \endgroup}
\def\mn@eprint#1#2{\mn@eprint@#1:#2::\@nil}
\def\mn@eprint@arXiv#1{\href {http://arxiv.org/abs/#1} {{\tt arXiv:#1}}}
\def\mn@eprint@dblp#1{\href {http://dblp.uni-trier.de/rec/bibtex/#1.xml}
  {dblp:#1}}
\def\mn@eprint@#1:#2:#3:#4\@nil{\def\@tempa {#1}\def\@tempb {#2}\def\@tempc
  {#3}\ifx \@tempc \@empty \let \@tempc \@tempb \let \@tempb \@tempa \fi \ifx
  \@tempb \@empty \def\@tempb {arXiv}\fi \@ifundefined
  {mn@eprint@\@tempb}{\@tempb:\@tempc}{\expandafter \expandafter \csname
  mn@eprint@\@tempb\endcsname \expandafter{\@tempc}}}

\bibitem[\protect\citeauthoryear{{Amaro-Seoane} et~al.,}{{Amaro-Seoane}
  et~al.}{2017}]{Amaro-Seoane:17}
{Amaro-Seoane} P.,  et~al., 2017, arXiv e-prints, \href
  {https://ui.adsabs.harvard.edu/abs/2017arXiv170200786A} {p. arXiv:1702.00786}

\bibitem[\protect\citeauthoryear{Barker}{Barker}{2011}]{Barker:11b}
Barker A.~J.,  2011, Monthly Notices of the Royal Astronomical Society, 414,
  1365

\bibitem[\protect\citeauthoryear{{Brink}, {Teukolsky}  \& {Wasserman}}{{Brink}
  et~al.}{2005}]{Brink:05}
{Brink} J.,  {Teukolsky} S.~A.,   {Wasserman} I.,  2005, \mn@doi [\prd]
  {10.1103/PhysRevD.71.064029}, \href
  {https://ui.adsabs.harvard.edu/abs/2005PhRvD..71f4029B} {71, 064029}

\bibitem[\protect\citeauthoryear{{Brown}, {Gianninas}, {Kilic}, {Kenyon}  \&
  {Allende Prieto}}{{Brown} et~al.}{2016}]{Brown:16}
{Brown} W.~R.,  {Gianninas} A.,  {Kilic} M.,  {Kenyon} S.~J.,   {Allende
  Prieto} C.,  2016, \mn@doi [\apj] {10.3847/0004-637X/818/2/155}, \href
  {https://ui.adsabs.harvard.edu/abs/2016ApJ...818..155B} {818, 155}

\bibitem[\protect\citeauthoryear{{Burdge} et~al.,}{{Burdge}
  et~al.}{2019}]{Burdge:19}
{Burdge} K.~B.,  et~al., 2019, \mn@doi [\nat] {10.1038/s41586-019-1403-0},
  \href {https://ui.adsabs.harvard.edu/abs/2019Natur.571..528B} {571, 528}

\bibitem[\protect\citeauthoryear{{Burkart}, {Quataert}, {Arras}  \&
  {Weinberg}}{{Burkart} et~al.}{2013}]{Burkart:13}
{Burkart} J.,  {Quataert} E.,  {Arras} P.,   {Weinberg} N.~N.,  2013, \mn@doi
  [\mnras] {10.1093/mnras/stt726}, \href
  {https://ui.adsabs.harvard.edu/abs/2013MNRAS.433..332B} {433, 332}

\bibitem[\protect\citeauthoryear{{Clayton}}{{Clayton}}{2012}]{Clayton:12}
{Clayton} G.~C.,  2012, Journal of the American Association of Variable Star
  Observers (JAAVSO), \href
  {https://ui.adsabs.harvard.edu/abs/2012JAVSO..40..539C} {40, 539}

\bibitem[\protect\citeauthoryear{Cutler \& Flanagan}{Cutler \&
  Flanagan}{1994}]{Cutler:94}
Cutler C.,  Flanagan E.~E.,  1994, \mn@doi [Phys. Rev. D]
  {10.1103/PhysRevD.49.2658}, 49, 2658

\bibitem[\protect\citeauthoryear{{Dan}, {Rosswog}, {Br{\"u}ggen}  \&
  {Podsiadlowski}}{{Dan} et~al.}{2014}]{Dan:14}
{Dan} M.,  {Rosswog} S.,  {Br{\"u}ggen} M.,   {Podsiadlowski} P.,  2014,
  \mn@doi [\mnras] {10.1093/mnras/stt1766}, \href
  {https://ui.adsabs.harvard.edu/abs/2014MNRAS.438...14D} {438, 14}

\bibitem[\protect\citeauthoryear{{Essick} \& {Weinberg}}{{Essick} \&
  {Weinberg}}{2016}]{Essick:16}
{Essick} R.,  {Weinberg} N.~N.,  2016, \mn@doi [\apj]
  {10.3847/0004-637X/816/1/18}, \href
  {https://ui.adsabs.harvard.edu/abs/2016ApJ...816...18E} {816, 18}

\bibitem[\protect\citeauthoryear{{Fenn}, {Plewa}  \& {Gawryszczak}}{{Fenn}
  et~al.}{2016}]{Fenn:16}
{Fenn} D.,  {Plewa} T.,   {Gawryszczak} A.,  2016, \mn@doi [\mnras]
  {10.1093/mnras/stw1831}, \href
  {https://ui.adsabs.harvard.edu/abs/2016MNRAS.462.2486F} {462, 2486}

\bibitem[\protect\citeauthoryear{{Ferrario}, {de Martino}  \&
  {G{\"a}nsicke}}{{Ferrario} et~al.}{2015}]{Ferrario:15}
{Ferrario} L.,  {de Martino} D.,   {G{\"a}nsicke} B.~T.,  2015, \mn@doi [\ssr]
  {10.1007/s11214-015-0152-0}, \href
  {https://ui.adsabs.harvard.edu/abs/2015SSRv..191..111F} {191, 111}

\bibitem[\protect\citeauthoryear{{Flanagan} \& {Hinderer}}{{Flanagan} \&
  {Hinderer}}{2008}]{Flanagan:08}
{Flanagan} {\'E}.~{\'E}.,  {Hinderer} T.,  2008, \mn@doi [\prd]
  {10.1103/PhysRevD.77.021502}, \href
  {https://ui.adsabs.harvard.edu/abs/2008PhRvD..77b1502F} {77, 021502}

\bibitem[\protect\citeauthoryear{{Fuller} \& {Lai}}{{Fuller} \&
  {Lai}}{2011}]{Fuller:11}
{Fuller} J.,  {Lai} D.,  2011, \mn@doi [\mnras]
  {10.1111/j.1365-2966.2010.18017.x}, \href
  {http://adsabs.harvard.edu/abs/2011MNRAS.412.1331F} {412, 1331}

\bibitem[\protect\citeauthoryear{{Fuller} \& {Lai}}{{Fuller} \&
  {Lai}}{2012a}]{Fuller:12a}
{Fuller} J.,  {Lai} D.,  2012a, \mn@doi [\mnras]
  {10.1111/j.1365-2966.2011.20320.x}, \href
  {https://ui.adsabs.harvard.edu/abs/2012MNRAS.421..426F} {421, 426}

\bibitem[\protect\citeauthoryear{{Fuller} \& {Lai}}{{Fuller} \&
  {Lai}}{2012b}]{Fuller:12b}
{Fuller} J.,  {Lai} D.,  2012b, \mn@doi [\apjl] {10.1088/2041-8205/756/1/L17},
  \href {https://ui.adsabs.harvard.edu/abs/2012ApJ...756L..17F} {756, L17}

\bibitem[\protect\citeauthoryear{{Fuller} \& {Lai}}{{Fuller} \&
  {Lai}}{2013}]{Fuller:13}
{Fuller} J.,  {Lai} D.,  2013, \mn@doi [\mnras] {10.1093/mnras/sts606}, \href
  {http://adsabs.harvard.edu/abs/2013MNRAS.430..274F} {430, 274}

\bibitem[\protect\citeauthoryear{{Fuller} \& {Lai}}{{Fuller} \&
  {Lai}}{2014}]{Fuller:14}
{Fuller} J.,  {Lai} D.,  2014, \mn@doi [\mnras] {10.1093/mnras/stu1698}, \href
  {https://ui.adsabs.harvard.edu/abs/2014MNRAS.444.3488F} {444, 3488}

\bibitem[\protect\citeauthoryear{{Goodman} \& {Dickson}}{{Goodman} \&
  {Dickson}}{1998}]{Goodman:98}
{Goodman} J.,  {Dickson} E.~S.,  1998, \mn@doi [\apj] {10.1086/306348}, \href
  {https://ui.adsabs.harvard.edu/abs/1998ApJ...507..938G} {507, 938}

\bibitem[\protect\citeauthoryear{{Graham} et~al.,}{{Graham}
  et~al.}{2019}]{Graham:19}
{Graham} M.~J.,  et~al., 2019, \mn@doi [\pasp] {10.1088/1538-3873/ab006c},
  \href {https://ui.adsabs.harvard.edu/abs/2019PASP..131g8001G} {131, 078001}

\bibitem[\protect\citeauthoryear{{Hermes} et~al.,}{{Hermes}
  et~al.}{2012}]{Hermes:12}
{Hermes} J.~J.,  et~al., 2012, \mn@doi [\apjl] {10.1088/2041-8205/757/2/L21},
  \href {https://ui.adsabs.harvard.edu/abs/2012ApJ...757L..21H} {757, L21}

\bibitem[\protect\citeauthoryear{{Iben} \& {Tutukov}}{{Iben} \&
  {Tutukov}}{1984}]{Iben:84}
{Iben} I. J.,  {Tutukov} A.~V.,  1984, \mn@doi [\apjs] {10.1086/190932}, \href
  {https://ui.adsabs.harvard.edu/abs/1984ApJS...54..335I} {54, 335}

\bibitem[\protect\citeauthoryear{{Iben}, {Tutukov}  \& {Fedorova}}{{Iben}
  et~al.}{1998}]{Iben:98}
{Iben} Icko J.,  {Tutukov} A.~V.,   {Fedorova} A. r.~V.,  1998, \mn@doi [\apj]
  {10.1086/305972}, \href
  {https://ui.adsabs.harvard.edu/abs/1998ApJ...503..344I} {503, 344}

\bibitem[\protect\citeauthoryear{{Korol} et~al.,}{{Korol}
  et~al.}{2020}]{Korol:20}
{Korol} V.,  et~al., 2020, arXiv e-prints, \href
  {https://ui.adsabs.harvard.edu/abs/2020arXiv200210462K} {p. arXiv:2002.10462}

\bibitem[\protect\citeauthoryear{{Kumar} \& {Goodman}}{{Kumar} \&
  {Goodman}}{1996}]{Kumar:96}
{Kumar} P.,  {Goodman} J.,  1996, \mn@doi [\apj] {10.1086/177565}, \href
  {https://ui.adsabs.harvard.edu/abs/1996ApJ...466..946K} {466, 946}

\bibitem[\protect\citeauthoryear{{Kuns}, {Yu}, {Chen}  \& {Adhikari}}{{Kuns}
  et~al.}{2019}]{Kuns:19}
{Kuns} K.~A.,  {Yu} H.,  {Chen} Y.,   {Adhikari} R.~X.,  2019, arXiv e-prints,
  \href {https://ui.adsabs.harvard.edu/abs/2019arXiv190806004K} {p.
  arXiv:1908.06004}

\bibitem[\protect\citeauthoryear{{Kupfer} et~al.,}{{Kupfer}
  et~al.}{2018}]{Kupfer:18}
{Kupfer} T.,  et~al., 2018, \mn@doi [\mnras] {10.1093/mnras/sty1545}, \href
  {https://ui.adsabs.harvard.edu/abs/2018MNRAS.480..302K} {480, 302}

\bibitem[\protect\citeauthoryear{{Lai}}{{Lai}}{1994}]{Lai:94}
{Lai} D.,  1994, \mn@doi [\mnras] {10.1093/mnras/270.3.611}, \href
  {https://ui.adsabs.harvard.edu/abs/1994MNRAS.270..611L} {270, 611}

\bibitem[\protect\citeauthoryear{Lam, Pitrou  \& Seibert}{Lam
  et~al.}{2015}]{Lam:15}
Lam S.~K.,  Pitrou A.,   Seibert S.,  2015, in Proceedings of the Second
  Workshop on the LLVM Compiler Infrastructure in HPC. LLVM ’15.
Association for Computing Machinery, New York, NY, USA,
  \mn@doi{10.1145/2833157.2833162}, \url
  {https://doi.org/10.1145/2833157.2833162}

\bibitem[\protect\citeauthoryear{{Luo} et~al.,}{{Luo} et~al.}{2016}]{Luo:16}
{Luo} J.,  et~al., 2016, \mn@doi [Classical and Quantum Gravity]
  {10.1088/0264-9381/33/3/035010}, \href
  {https://ui.adsabs.harvard.edu/abs/2016CQGra..33c5010L} {33, 035010}

\bibitem[\protect\citeauthoryear{{McNeill}, {Mardling}  \&
  {M{\"u}ller}}{{McNeill} et~al.}{2019}]{McNeill:19}
{McNeill} L.~O.,  {Mardling} R.~A.,   {M{\"u}ller} B.,  2019, arXiv e-prints,
  \href {https://ui.adsabs.harvard.edu/abs/2019arXiv190109045M} {p.
  arXiv:1901.09045}

\bibitem[\protect\citeauthoryear{{Nelemans}, {Portegies Zwart}, {Verbunt}  \&
  {Yungelson}}{{Nelemans} et~al.}{2001}]{Nelemans:01}
{Nelemans} G.,  {Portegies Zwart} S.~F.,  {Verbunt} F.,   {Yungelson} L.~R.,
  2001, \mn@doi [\aap] {10.1051/0004-6361:20010049}, \href
  {https://ui.adsabs.harvard.edu/abs/2001A&A...368..939N} {368, 939}

\bibitem[\protect\citeauthoryear{{Paxton}, {Bildsten}, {Dotter}, {Herwig},
  {Lesaffre}  \& {Timmes}}{{Paxton} et~al.}{2011}]{Paxton:11}
{Paxton} B.,  {Bildsten} L.,  {Dotter} A.,  {Herwig} F.,  {Lesaffre} P.,
  {Timmes} F.,  2011, \mn@doi [\apjs] {10.1088/0067-0049/192/1/3}, \href
  {https://ui.adsabs.harvard.edu/abs/2011ApJS..192....3P} {192, 3}

\bibitem[\protect\citeauthoryear{{Paxton} et~al.,}{{Paxton}
  et~al.}{2013}]{Paxton:13}
{Paxton} B.,  et~al., 2013, \mn@doi [\apjs] {10.1088/0067-0049/208/1/4}, \href
  {https://ui.adsabs.harvard.edu/abs/2013ApJS..208....4P} {208, 4}

\bibitem[\protect\citeauthoryear{{Paxton} et~al.,}{{Paxton}
  et~al.}{2015}]{Paxton:15}
{Paxton} B.,  et~al., 2015, \mn@doi [\apjs] {10.1088/0067-0049/220/1/15}, \href
  {https://ui.adsabs.harvard.edu/abs/2015ApJS..220...15P} {220, 15}

\bibitem[\protect\citeauthoryear{{Paxton} et~al.,}{{Paxton}
  et~al.}{2018}]{Paxton:18}
{Paxton} B.,  et~al., 2018, \mn@doi [\apjs] {10.3847/1538-4365/aaa5a8}, \href
  {https://ui.adsabs.harvard.edu/abs/2018ApJS..234...34P} {234, 34}

\bibitem[\protect\citeauthoryear{{Piro}}{{Piro}}{2019}]{Piro:19}
{Piro} A.~L.,  2019, \mn@doi [\apjl] {10.3847/2041-8213/ab44c4}, \href
  {https://ui.adsabs.harvard.edu/abs/2019ApJ...885L...2P} {885, L2}

\bibitem[\protect\citeauthoryear{{Polin}, {Nugent}  \& {Kasen}}{{Polin}
  et~al.}{2019a}]{Polin:20}
{Polin} A.,  {Nugent} P.,   {Kasen} D.,  2019a, arXiv e-prints, \href
  {https://ui.adsabs.harvard.edu/abs/2019arXiv191012434P} {p. arXiv:1910.12434}

\bibitem[\protect\citeauthoryear{{Polin}, {Nugent}  \& {Kasen}}{{Polin}
  et~al.}{2019b}]{Polin:19}
{Polin} A.,  {Nugent} P.,   {Kasen} D.,  2019b, \mn@doi [\apj]
  {10.3847/1538-4357/aafb6a}, \href
  {https://ui.adsabs.harvard.edu/abs/2019ApJ...873...84P} {873, 84}

\bibitem[\protect\citeauthoryear{{Press} \& {Teukolsky}}{{Press} \&
  {Teukolsky}}{1977}]{Press:77}
{Press} W.~H.,  {Teukolsky} S.~A.,  1977, \mn@doi [\apj] {10.1086/155143},
  \href {https://ui.adsabs.harvard.edu/abs/1977ApJ...213..183P} {213, 183}

\bibitem[\protect\citeauthoryear{{Raskin}, {Scannapieco}, {Fryer},
  {Rockefeller}  \& {Timmes}}{{Raskin} et~al.}{2012}]{Raskin:12}
{Raskin} C.,  {Scannapieco} E.,  {Fryer} C.,  {Rockefeller} G.,   {Timmes}
  F.~X.,  2012, \mn@doi [\apj] {10.1088/0004-637X/746/1/62}, \href
  {https://ui.adsabs.harvard.edu/abs/2012ApJ...746...62R} {746, 62}

\bibitem[\protect\citeauthoryear{{Salaris}, {Dom{\'\i}nguez},
  {Garc{\'\i}a-Berro}, {Hernanz}, {Isern}  \& {Mochkovitch}}{{Salaris}
  et~al.}{1997}]{Salaris:97}
{Salaris} M.,  {Dom{\'\i}nguez} I.,  {Garc{\'\i}a-Berro} E.,  {Hernanz} M.,
  {Isern} J.,   {Mochkovitch} R.,  1997, \mn@doi [\apj] {10.1086/304483}, \href
  {https://ui.adsabs.harvard.edu/abs/1997ApJ...486..413S} {486, 413}

\bibitem[\protect\citeauthoryear{{Shen}, {Kasen}, {Miles}  \&
  {Townsley}}{{Shen} et~al.}{2018}]{Shen:18}
{Shen} K.~J.,  {Kasen} D.,  {Miles} B.~J.,   {Townsley} D.~M.,  2018, \mn@doi
  [\apj] {10.3847/1538-4357/aaa8de}, \href
  {https://ui.adsabs.harvard.edu/abs/2018ApJ...854...52S} {854, 52}

\bibitem[\protect\citeauthoryear{{Shiode}, {Quataert}  \& {Arras}}{{Shiode}
  et~al.}{2012}]{Shiode:12}
{Shiode} J.~H.,  {Quataert} E.,   {Arras} P.,  2012, \mn@doi [\mnras]
  {10.1111/j.1365-2966.2012.21130.x}, \href
  {https://ui.adsabs.harvard.edu/abs/2012MNRAS.423.3397S} {423, 3397}

\bibitem[\protect\citeauthoryear{{Timmes}, {Townsend}, {Bauer}, {Thoul},
  {Fields}  \& {Wolf}}{{Timmes} et~al.}{2018}]{Timmes:18}
{Timmes} F.~X.,  {Townsend} R. H.~D.,  {Bauer} E.~B.,  {Thoul} A.,  {Fields}
  C.~E.,   {Wolf} W.~M.,  2018, \mn@doi [\apjl] {10.3847/2041-8213/aae70f},
  \href {https://ui.adsabs.harvard.edu/abs/2018ApJ...867L..30T} {867, L30}

\bibitem[\protect\citeauthoryear{{Toonen}, {Nelemans}  \& {Portegies
  Zwart}}{{Toonen} et~al.}{2012}]{Toonen:12}
{Toonen} S.,  {Nelemans} G.,   {Portegies Zwart} S.,  2012, \mn@doi [\aap]
  {10.1051/0004-6361/201218966}, \href
  {https://ui.adsabs.harvard.edu/abs/2012A&A...546A..70T} {546, A70}

\bibitem[\protect\citeauthoryear{{Townsend} \& {Teitler}}{{Townsend} \&
  {Teitler}}{2013}]{Townsend:13}
{Townsend} R.~H.~D.,  {Teitler} S.~A.,  2013, \mn@doi [\mnras]
  {10.1093/mnras/stt1533}, \href
  {https://ui.adsabs.harvard.edu/abs/2013MNRAS.435.3406T} {435, 3406}

\bibitem[\protect\citeauthoryear{{Townsend}, {Goldstein}  \&
  {Zweibel}}{{Townsend} et~al.}{2018}]{Townsend:18}
{Townsend} R.~H.~D.,  {Goldstein} J.,   {Zweibel} E.~G.,  2018, \mn@doi
  [\mnras] {10.1093/mnras/stx3142}, \href
  {https://ui.adsabs.harvard.edu/abs/2018MNRAS.475..879T} {475, 879}

\bibitem[\protect\citeauthoryear{{Unno}, {Osaki}, {Ando}, {Saio}  \&
  {Shibahashi}}{{Unno} et~al.}{1989}]{Unno:89}
{Unno} W.,  {Osaki} Y.,  {Ando} H.,  {Saio} H.,   {Shibahashi} H.,  1989,
  {Nonradial oscillations of stars}.
Univ. Tokyo Press, Tokyo

\bibitem[\protect\citeauthoryear{{Venumadhav}, {Zimmerman}  \&
  {Hirata}}{{Venumadhav} et~al.}{2014}]{Venumadhav:14}
{Venumadhav} T.,  {Zimmerman} A.,   {Hirata} C.~M.,  2014, \mn@doi [ApJ]
  {10.1088/0004-637X/781/1/23}, \href
  {http://adsabs.harvard.edu/abs/2014ApJ...781...23V} {781, 23}

\bibitem[\protect\citeauthoryear{{Webbink}}{{Webbink}}{1984}]{Webbink:84}
{Webbink} R.~F.,  1984, \mn@doi [\apj] {10.1086/161701}, \href
  {https://ui.adsabs.harvard.edu/abs/1984ApJ...277..355W} {277, 355}

\bibitem[\protect\citeauthoryear{{Weinberg}}{{Weinberg}}{2016}]{Weinberg:16}
{Weinberg} N.~N.,  2016, \mn@doi [ApJ] {10.3847/0004-637X/819/2/109}, \href
  {http://adsabs.harvard.edu/abs/2016ApJ...819..109W} {819, 109}

\bibitem[\protect\citeauthoryear{{Weinberg}, {Arras}, {Quataert}  \&
  {Burkart}}{{Weinberg} et~al.}{2012}]{Weinberg:12}
{Weinberg} N.~N.,  {Arras} P.,  {Quataert} E.,   {Burkart} J.,  2012, \mn@doi
  [\apj] {10.1088/0004-637X/751/2/136}, \href
  {https://ui.adsabs.harvard.edu/abs/2012ApJ...751..136W} {751, 136}

\bibitem[\protect\citeauthoryear{{Willems}, {Deloye}  \& {Kalogera}}{{Willems}
  et~al.}{2010}]{Willems:10}
{Willems} B.,  {Deloye} C.~J.,   {Kalogera} V.,  2010, \mn@doi [\apj]
  {10.1088/0004-637X/713/1/239}, \href
  {https://ui.adsabs.harvard.edu/abs/2010ApJ...713..239W} {713, 239}

\bibitem[\protect\citeauthoryear{{Witte} \& {Savonije}}{{Witte} \&
  {Savonije}}{1999}]{Witte:99}
{Witte} M.~G.,  {Savonije} G.~J.,  1999, \aap, \href
  {https://ui.adsabs.harvard.edu/abs/1999A&A...350..129W} {350, 129}

\bibitem[\protect\citeauthoryear{{Wu} \& {Goldreich}}{{Wu} \&
  {Goldreich}}{2001}]{Wu:01}
{Wu} Y.,  {Goldreich} P.,  2001, \mn@doi [\apj] {10.1086/318234}, \href
  {https://ui.adsabs.harvard.edu/abs/2001ApJ...546..469W} {546, 469}

\makeatother
\end{thebibliography}

\appendix
\section{Asymptotic Relations}
\label{sec:WKB_relations}
In this Appendix we present various asymptotic relations used in our calculations. For future convenience, we expand the Lagrangian displacement vector field $\vect{\xi}_a$ as 
\begin{equation}
\vect{\xi}_a (\vect{r}) = \left[\xi_a^r(r) \vect{e}_{r} + \xi_a^h(r) r \nabla \right] Y_{l_a m_a }(\theta, \phi),
\end{equation}
where $\vect{e}_r$ is the unit vector along the radial direction. The radial and horizontal motions can thus be characterized by $\xi_a^r(r)$ and $\xi_a^h(r)$, respectively. 

\subsection{Shear profile}
\label{sec:shr_prfl}
\begin{figure}
   \centering
   \includegraphics[width=0.45\textwidth]{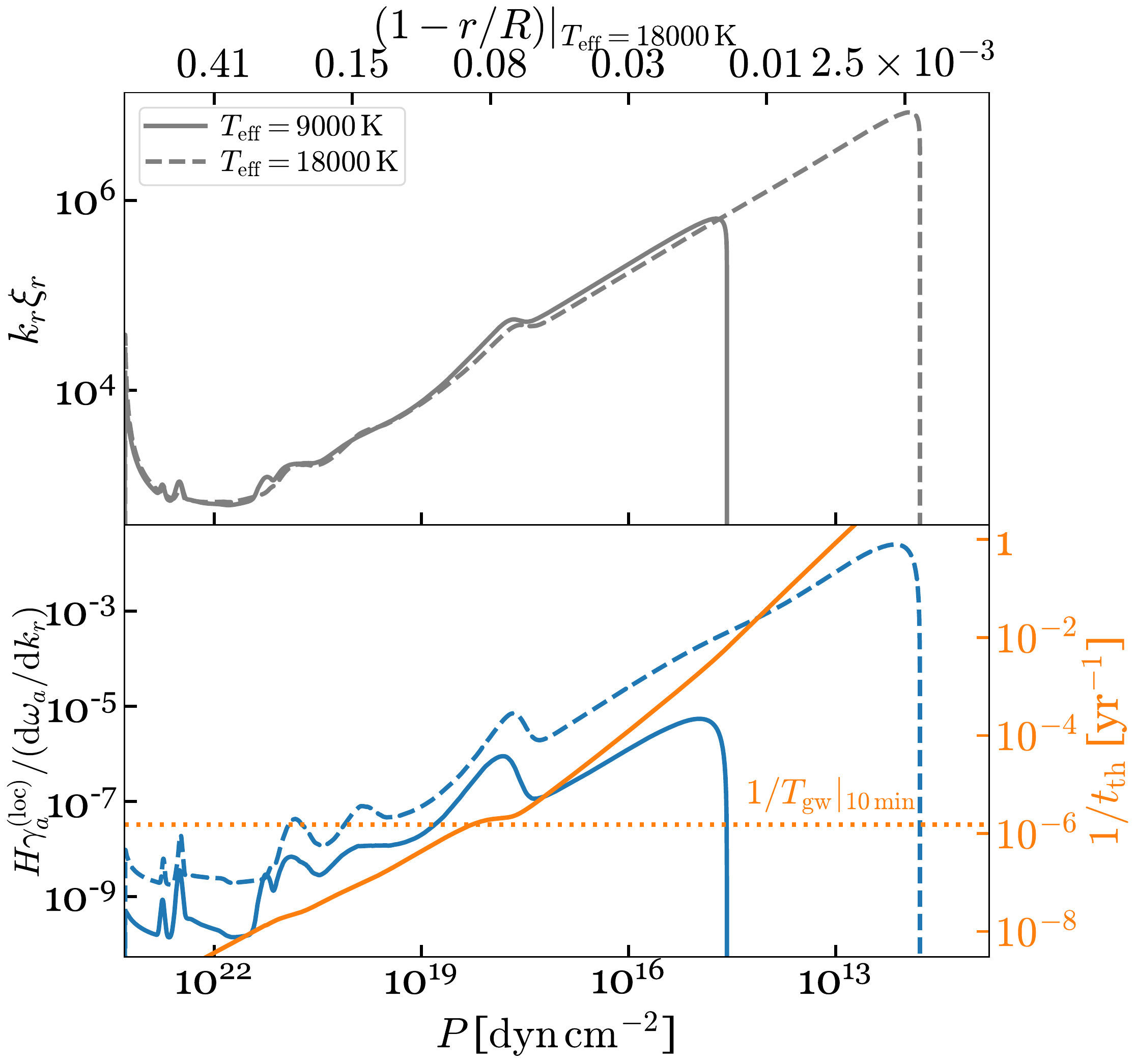} 
   \caption{Top panel: Profile of the shear envelope $k_r A$ for a normalized mode with $q_a=1$. Bottom panel: Local damping rate weighted by the time the wave spends per scale height (blue line) and timescales for thermal diffusion (solid orange line) and GW-driven orbital decay (dotted orange line).  We show results for both our standard CO WD model with $T_{\rm eff}=9000\textrm{ K}$ (solid lines) and a hotter CO WD model with $T_{\rm eff}=18000\textrm{ K}$ (dashed lines).  We set $\omega_a = 0.01 \omega_0$ here; for the range of interest, varying $\omega_a$ modifies the overall scale but has almost no effect on the shape of the profile.}
   \label{fig:shear_profile}
\end{figure}

For a high-order g-mode normalized according to Equation~(\ref{eq:xi_norm}), the radial displacement $\xi_a^{r}$ can be approximated by the WKB relation~\citep{Unno:89}
\begin{equation}
\xi^r_a \simeq A  \sin \phi,
\end{equation}
where the phase $\phi=\int k_r \diff r + \pi/4$ and the amplitude 
\begin{equation}
A^2 = \left(\frac{E_0}{\int \mathcal{N} \diff \ln r}\right)\frac{1}{\rho r^3 \mathcal{N}}.
\label{eq:WKB_wave_amp}
\end{equation}
Consequently, we can approximate the envelope of the shear as a function of radius as $|k_r \xi_a^r| \sim k_r A$ with the wavenumber $k_r$ given by Equation~(\ref{eq:kr_sq}).

In Figure~\ref{fig:shear_profile} we show the shear profile. When evaluating $k_r$, we assume a reference frequency $\omega_a = 0.01 \omega_0$. Notice that the amplitude $A$ is determined purely by background quantities while the wavenumber $k_r \propto 1/\omega_a$ when $\omega_a^2 \ll \mathcal{N}^2, S_l^2$. Therefore, different g-modes (as well as the traveling wave solution) will have essentially the same shape as the shear envelope, which peaks at a radius  $r=0.994 R$ and a pressure $P=5.4\times10^{14}\textrm{ dyn cm}^{-2}$ for the $T_{\rm eff} = 9000\,K$ model we consider in the main text (corresponding to the solid traces in Figure~\ref{fig:shear_profile}). The maximum shear over the star (which we used to derive the threshold energy of local wave-breaking in Figure~\ref{fig:E_th_vs_P}) can then be expressed as 
\begin{equation}
\max_r |q_a k_r \xi_a^r| = \max_r |q_a k_r A| \simeq 6.4\times10^5 q_a \left(\frac{0.01 \omega_0}{\omega_a}\right),
\label{eq:WKB_shr}
\end{equation}
where the numerical value is for the $T_{\rm eff} = 9000\,{\rm K}$ model.  

While in the main text we focus on a single WD model with $T_{\rm eff}{=}9000\,{\rm K}$, in Figure~\ref{fig:shear_profile} we also consider a model with $T_{\rm eff}{=}18000\,{\rm K}$. As the WD becomes hotter, the radiative zone extends closer to the surface. Thus, a gravity wave of a given frequency propagates out to smaller pressures and its peak shear can be greater. Consequently, tidal heating might further accelerate the transition between the weakly-nonlinear tidal interaction to the traveling-wave limit as the binary's separation decreases. Assessing this possibility requires a study that couples tidal effects to the adjustment of the WD's internal structure.

\subsection{Linear dissipation}
\label{sec:lin_diss}
We now describe our calculation of the dissipation rates, which closely follows the work of \citetalias{Burkart:13}.  

There are two types of mode damping that are potentially relevant in the WD. The first is due to electron conduction and radiative diffusion, which can be estimated as 
\begin{equation}
\gamma_{a}^{(\rm diff)}= \frac{\omega_a^2}{E_0} \int \chi k_r^2 \left[(\xi_a^r)^2 + l (l+1) (\xi_a^h)^2\right] \rho r^2 \diff r,
\label{eq:gam_a_global}
\end{equation}
where the thermal diffusivity  
\begin{equation}
\chi = \frac{16 \sigma T^3 }{3 \kappa \rho^2 C_P}.
\end{equation}
Here $\kappa$ and $C_P$ are the opacity and the specific heat at constant total pressure. 
In addition to thermal diffusion, a g-mode can also be damped by convective turbulence, whose dissipation rate we estimate as 
\begin{align}
\gamma_a^{\rm (turb)} &=  \frac{\omega_a^2}{E_0} \int \nu_{\rm turb} \times \nonumber \\
& \left[0.23 \left(\frac{\diff \xi_a^r}{\diff r}\right)^2 + 0.084 l(l+1) \left(\frac{\diff \xi_a^h}{\diff r}\right)^2\right] \rho r^2 \diff r,
\end{align}
where $\nu_{\rm turb}$ is the effective turbulent viscosity~\citep{Shiode:12}, 
\begin{equation}
\nu_{\rm turb} = L_{\rm c} v_{\rm c} \min \left[10 \left(\frac{2\pi L_{\rm c}}{\omega_a v_{\rm c}} \right)^2, \left(\frac{2\pi L_{\rm c}}{\omega_a v_{\rm c}} \right), 2.4\right]. 
\end{equation}
Here $L_{\rm c}$ and $v_{\rm c}$ are the convective luminosity and velocity according to mixing length theory (which are given by our \texttt{MESA} model). 

In Figure~\ref{fig:gam} we present the dissipation rates as a function of the mode radial order (bottom axis) and frequency (top axis). The dots are exact values calculated under the prescription described in this Section, and the solid lines are the power-law fits given in Section~\ref{sec:eom}. The blue and green lines are the dissipation due to thermal diffusion and turbulent damping, and the orange line is the inverse of a mode's group-travel time [see Equation~(\ref{eq:WKB_alpha_a})]. We see that for modes with $n_a\gtrsim 20$, the dissipation is dominated by thermal diffusion. Those are the modes most relevant for the tidal synchronization calculation. For modes with lower radial orders, the turbulent damping becomes significant. We do not include the contribution of $\gamma_a^{({\rm turb})}$ in our mode network calculations since modes with $n_a \lesssim 20$ are not excited once the tidal synchronization is taken into account (see Section~\ref{sec:sync}). 

The quantity presented in Equation~(\ref{eq:gam_a_global}) is the \emph{global} damping rate. To better understand the tidal heating process, it is also instructive to study the \emph{local} heating rate, $\gamma_a^{\rm (loc)} {=} \chi k_r^2/2$. Specifically, the global damping rate  can be viewed as the integral of the local rate weighted by the time the wave packet spends at each radius~\citep{Goodman:98},
\begin{align}
\gamma_a^{\rm (diff)} &\simeq \frac{2}{T_a} \int  \gamma_a^{\rm (loc)} \frac{\diff r}{\diff \omega_a/\diff k_r} \nonumber \\
&=\frac{2}{T_a}\int \gamma_a^{\rm (loc)} \frac{H}{\diff \omega_a/\diff k_r} \diff \ln P,
\label{eq:gam_a_local}
\end{align}
where $T_a = 2\int \diff r (\diff \omega_a / \diff k_r)^{-1}$ is the wave's group travel time and $H \equiv P/g\rho$ is the pressure scale height.\footnote{Using the WKB amplitude of a mode [see Equation~(\ref{eq:WKB_wave_amp})] together with the property that $l(l+1)\xi^h_a/r \sim k_r \xi^r_a \gg \xi^r_a$, it can be shown that $\gamma_a^{\rm (loc)} \left(\partial  k_r/ \partial \omega_a\right)$ is proportional to the integrand of Equation~(\ref{eq:gam_a_global}).} In the second line, we reversed the limits of integration so that it corresponds to increasing $\ln P$. 
In the lower panel of Figure~\ref{fig:shear_profile}, we show the integrand of Equation~(\ref{eq:gam_a_local}). Note that by presenting it in this form (local damping rate weighted by the time the wave-packet spends per pressure scale height), we expect an equal contribution to the total damping per $\diff \ln P$. The figure assumes a reference frequency $\omega_a  = 0.01 \omega_0$ and the solid- and dashed-blue lines represent the $T_{\rm eff}=9000\,{\rm K}$ and $T_{\rm eff}=18000\,{\rm K}$ models, respectively. Note that similar to the shear profile, the reference frequency $\omega_a$ only affects the overall magnitude of the damping but does not change its shape. The solid-orange line shows the inverse of the local thermal diffusion timescale
\begin{equation}
t_{\rm th} = \frac{PC_P T}{gF}, 
\end{equation}
for the $T_{\rm eff}=9000\,{\rm K}$ model, where $F$ is the radiation flux (the $T_{\rm eff}=18000\,{\rm K}$ model has a similar $t_{\rm th}^{-1}$ profile). As a reference, the dotted-orange line is the inverse of the GW decay timescale for a binary at $P_{\rm orb}=10\,{\rm min}$ [see Equation~(\ref{eq:T_gw})]. 

As Figure~\ref{fig:shear_profile} shows, the peak of the weighted local damping rate happens near the WD surface at a location close to the peak of the shear. The typical thermal timescale at the peak is less than 1000 years, and all the heat deposited at $P\lesssim10^{18}\,{\rm dyne/cm^{2}}$ has  $t_{\rm th} < T_{\rm gw}$. Therefore, a significant portion of the tidal heating can diffuse out through the surface layers and hence affect the observed luminosity of WDs in compact binaries. 

\subsection{Tidal overlap}
\label{sec:tidal_overlap}

For the high-order g-modes that we consider, a brute-force calculation of the tidal overlap $Q_{a}$ according to Equation~(\ref{eq:WKB_Q_a}) is subject to considerable numerical error as the modes are highly oscillatory. A more numerically accurate approach is to evaluate it in terms of the equilibrium tide solution [see also Equations~(\ref{eq:xi_r_eq}) and (\ref{eq:xi_h_eq}); \citetalias{Burkart:13}]
\begin{equation}
\vect{f}_1 \left[\vect{\xi}_{\rm eq}(\vect{r})\right] \equiv \rho W_{lm} \frac{GM'}{D^{l+1}}  \nabla \left(r^l Y_{lm}\right).
\end{equation}
We further note that the set of linear eigenmodes $\left\{\vect{\xi}_a\right\}$ forms a complete, orthonormal basis. This allows us to expand the equilibrium tide as 
\begin{equation}
\vect{\xi}_{\rm eq}(\vect{r}) = \sum x_a \vect{\xi}_a (\vect{r}). 
\label{eq:xi_eq_as_sum_xi_a}
\end{equation}
Applying first the $\vect{f}_1$ operation on both side of Equation~(\ref{eq:xi_eq_as_sum_xi_a}) and then contracting with $\vect{\xi}_a$ using the orthogonality Equation~(\ref{eq:xi_norm}), we can express the expansion coefficient $x_a$ in terms of the tidal overlap integral
\begin{equation}
x_a = \epsilon W_{lm} Q_{alm}.
\end{equation}
Plugging this into  Equation~(\ref{eq:xi_eq_as_sum_xi_a}) and contracting both sides with $\vect{\xi}_a$ then gives an alternate expression for the tidal overlap 
\begin{align}
Q_{a} = \frac{\omega_a^2}{W_{lm}E_0} \int \left[\xi_{\rm eq}^r \xi_{\rm a}^{r} + l(l+1) \xi_{\rm eq}^h \xi_a^h\right] \rho r^2 \diff r,
\label{eq:Q_vs_xi_eq}
\end{align}
where $\vect{\xi}_{\rm eq}$ is evaluated with $\epsilon=1$. Also here we focus on the spatial part of $\vect{\xi}$ only (dropping the temporal phase, which is complex); for adiabatic oscillations both $\vect{\xi}_{\rm eq}(\vect{r})$ and $\vect{\xi}_{a} (\vect{r})$ are real. 

In Figure~\ref{fig:Qn} we plot $Q_{a}$ given by Equation~(\ref{eq:Q_vs_xi_eq}). The blue dots are the exact values and the solid-orange line is the asymptotic fit we use in the nonlinear network calculations.

\subsection{Three-mode-coupling coefficient}
\label{sec:three_mode_cpl}

In this Section we calculate the three-mode-coupling coefficient $\kappa_{abc}$ for our WD model. For conciseness, we use $a_r (a_h)$ to represent the radial (horizontal) displacement $\xi_a^r$ ($\xi_a^h$) of mode $a$. We further define $\Lambda_a^2=l_a(l_a +1)$ and the angular integral
\begin{align}
T&=\left[\frac{2(l_a+1)(l_b+1)(l_c+1)}{4\pi}\right]^{1/2} \nonumber \\
&\times
   \begin{pmatrix} 
      l_a & l_b & l_c \\
      m_a & m_b & m_c \\
   \end{pmatrix}
  \begin{pmatrix} 
      l_a & l_b & l_c \\
      0 & 0 & 0 \\
   \end{pmatrix},
\end{align}
where the matrices are Wigner 3-$j$ symbols. 

While the exact expression for the coupling coefficient is complicated [see equations (A55)-(A62) in \citetalias{Weinberg:12}], numerically we find that terms (A56) and (A58) in \citetalias{Weinberg:12} dominate  (the two have about the same magnitude but  opposite sign).  This allows us to write
\begin{align}
2 E_0 &\frac{\diff\kappa_{abc}}{\diff r}\simeq Tr\Gamma_1 P\left[\nabla\cdot\vect{b}\nabla \cdot \vect{c} \left( \Lambda_a^2 a_h-4a_r\right) \right.  \nonumber \\
&\quad \left.+ \nabla\cdot\vect{c}\nabla \cdot \vect{a} \left( \Lambda_b^2 b_h-4b_r\right) + \nabla\cdot\vect{a}\nabla \cdot \vect{b} \left( \Lambda_c^2 c_h-4c_r\right)\right]  \nonumber \\
& + 4Trg\rho\left(\nabla\cdot \vect{a} b_r c_r+\nabla\cdot \vect{b} c_r a_r+\nabla\cdot \vect{c} a_r b_r\right),
\end{align}
where we use the fact that $4g\gg r \diff g/\diff r$.  We further simplify this equation by substituting  
\begin{align}
\Gamma_1 P \nabla \cdot \vect{\xi}&\simeq g \rho \xi^r_a - \omega_a^2 r \rho \xi^h_a 
\simeq g \rho \xi^r_a,
\end{align}
where in the first equality we make the Cowling approximation and in the second we use the fact that $g\sim \omega_0^2 r \gg \omega_a^2 r$.  We then have
\begin{align}
2 E_0 \frac{\diff \kappa_{abc}}{\diff r} 
&\simeq Tr\frac{P}{\Gamma_1 H^2}(\Lambda_a^2 a_h b_r c_r +\Lambda_b^2 b_h a_r c_r+\Lambda_c^2 c_h a_r b_r ) \nonumber \\
&\simeq Tr\frac{P}{\Gamma_1 H^2}\Lambda_a^2 a_h b_r c_r,\label{eq:kap_abc_approx}
\end{align}
where $H$ is the pressure scale height. We keep only the $b_r c_r$ term for the reason given in \citetalias{Weinberg:12}. Namely, for high-order modes $b_r c_r\propto \sin(\phi_b)\sin(\phi_c)=\cos(\phi_b-\phi_c)/2\simeq1/2$, which is roughly a constant given that $\phi_b\simeq \phi_c$. By comparison, the $b_r c_h $ term is much smaller since $b_r c_h\propto \sin(\phi_b-\phi_c)/2\ll 1$ for $\phi_b\simeq \phi_c$. Although for the WD model we start from a different point than \citetalias{Weinberg:12} for solar-type stars [the sum of equations (A56) and (A58) instead of equation (A56) alone], the final result we obtain reduces to equation (A63) in \citetalias{Weinberg:12}.

\begin{figure}
   \centering
   \includegraphics[width=0.45\textwidth]{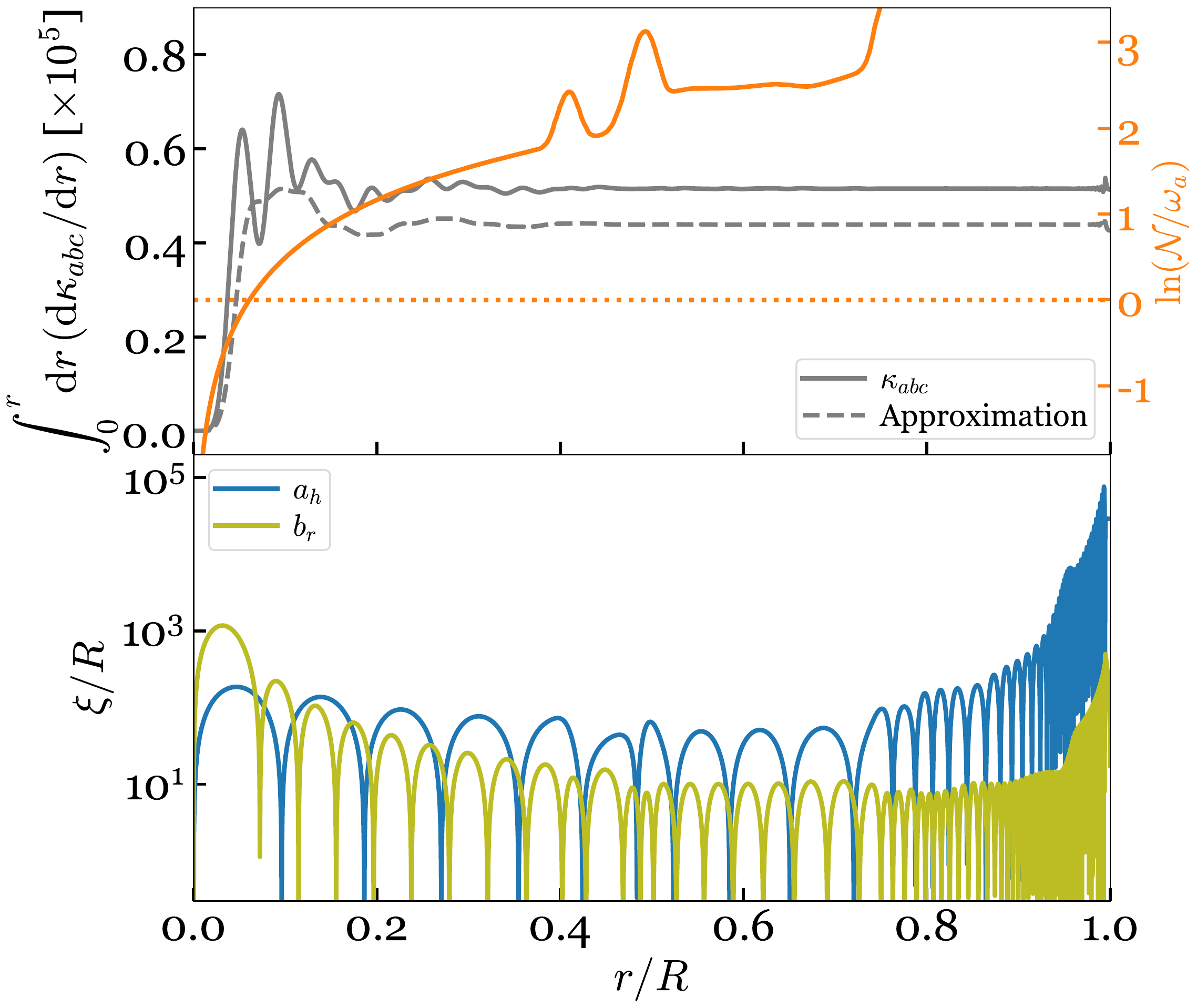} 
   \caption{Top panel: Cumulative three-mode coupling $\int^r  \diff r \left(\diff \kappa_{abc}/\diff r\right)$ (grey traces) and $\ln\left(\mathcal{N}/\omega_a\right)$ (orange traces). Note that most of the (global) mode coupling happens in the region $r \lesssim 0.1R$ despite the fact that the local shear peaks near the surface. Bottom panel: Lagrangian displacement of the parent mode's horizontal component $a_h$ (blue) and one of the daughter modes' radial component $b_r$ (olive).}
   \label{fig:cum_kappa_abc}
\end{figure}

In the top panel of Figure~\ref{fig:cum_kappa_abc} we show the cumulative three-mode-coupling coefficient $\int^r  \diff r \left(\diff \kappa_{abc}/\diff r\right)$. The solid-grey line is the result found by integrating all the terms in the exact expression for $\kappa_{abc}$ given in \citetalias{Weinberg:12} and the dashed-grey line is the result found by integrating the approximate expression given by Equation~(\ref{eq:kap_abc_approx}). Here the parent mode quantum numbers are $(l_a, m_a, n_a){=}(2, 2, 47)$ and the daughters are $(l_b, m_b, n_b){=}(2, 0, 94)$ and $(l_c, m_c, n_c){=}(2, -2, 96)$. 

Note that the three-mode coupling accumulates primarily in the core region (at $r \lesssim 0.1R$), near the inner boundary of the wave's propagation where $\omega_a \simeq  \mathcal{N}$ (see orange line in Figure~\ref{fig:cum_kappa_abc}). By contrast,  the shear and linear damping peak near the surface of the star where the modes are highly oscillatory (see Figure~\ref{fig:shear_profile}). 

We also show the horizontal displacement of the parent mode $a_h$ (blue) and the radial displacement of one of the daughters $b_r$ (olive) in the lower panel of Figure~\ref{fig:cum_kappa_abc}. Comparing with the top panel, we see that most of the contribution to three-mode coupling happens near the parent's inner turning point. In this region the parent transitions from being oscillatory to evanescent and the coupling adds coherently over a length scale of order $\simeq 0.1 R$ (although the parent is also evanescent at its outer turning point, that region contributes little to the global coupling since it is very near the surface where the density is small).  As we note above, since the daughters are spatially coherent, $b_r c_r \approx\textrm{constant}$, and it is the parent's spatial variations that matters most.  

In Figure~\ref{fig:kappa_abc} we show the coupling coefficient as a function of the parent mode's radial order (bottom axis) and frequency (top axis). The blue dots ($\kappa_{abc}$) are coefficients evaluated with the exact expression for daughter pairs with the smallest frequency detuning to the parent mode and satisfying $|n_b-n_c| < n_a$, and the orange dots ($\kappa_{abb}$) are evaluated for the most-resonant, self-coupled daughters (i.e., $b =c$). Here we restricted the daughters to have $l=2$, although the coupling to daughters with different $l$'s is similar. Indeed, as long as $|n_b-n_c| \lesssim n_a$, the coupling coefficients are approximately the same (after factoring out the angular dependence $T$) and can be treated as a function of the parent mode's $(l_a, n_a)$ alone (see also \citealt{Wu:01}, \citetalias{Weinberg:12}).  We also show the coupling coefficient evaluated with the approximate expression Equation~(\ref{eq:kap_abc_approx}) for self-coupled daughters (green dots). Lastly, the purple line is the asymptotic fit given by Equation~(\ref{eq:WKB_kap_abc}).

\section{Spin evolution with different nonlinear dissipation models}
\label{sec:NL_model_compare}

In the main text we use Model 2 (M2) as our fiducial model for the tidal dissipation rate (Section~\ref{sec:model_2}).  While M2 provides a good fit to the numerical results, near the resonance peaks Model 1 (M1; Section~\ref{sec:model_1}) provides a somewhat better fit (see Figure~\ref{fig:dE_lorentzian}), albeit at the expense of a less simple analytic form.  Here we compare the two models and show that they give similar results.

\begin{figure}
   \centering
   \includegraphics[width=0.45\textwidth]{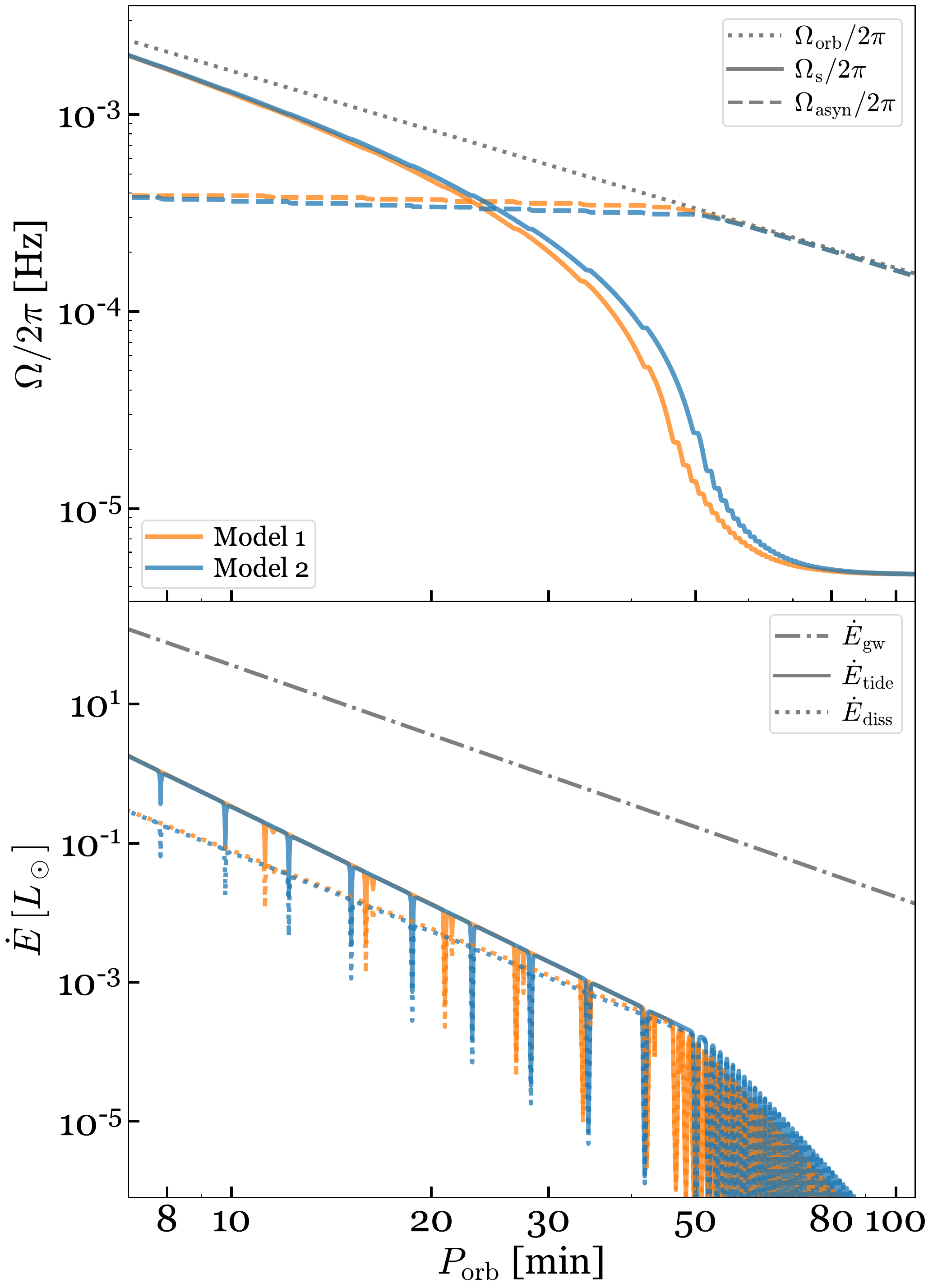} 
   \caption{Similar to Figure~\ref{fig:spin_power_NLeff_3} but now also showing the results of Model 1 (orange lines) in order to compare with Model 2 (blue lines). }
   \label{fig:spin_power_NL_comp}
\end{figure}

In Figure~\ref{fig:spin_power_NL_comp}, we show the M1 (orange lines) and M2 (blue line) trajectories for the evolution of the WD's spin and heating rates (similar to Figure~\ref{fig:spin_power_NLeff_3}). We see that they are very similar overall.  There are, however, two noticeable differences. First, the critical orbital period $P_{\rm c}$ when $T_{\rm s}=T_{\rm gw}$ is first satisfied is smaller for M1 than M2. This is because the maximum torque provided by a given parent mode is smaller in M1 than in M2 (see Figure~\ref{fig:dE_lorentzian}). Therefore, $P_{\rm c}$ corresponds to a smaller radial order $n_a$ of the parent mode in M1 than in M2. This also slightly reduces the temporal density and depth of the dips as both the mode density and the peak-to-trough spread of the dissipation rate (see Figure~\ref{fig:Edot_Tsync_no_rot}) decrease with decreasing $n_a$. 
Second, whereas in M2 the dips are single narrow lines that occur each time tidal synchronization is lost and the resonance transitions to a new mode, in M1 the dips are line doublets. This is because in M1, $\Edot$ has a more complicated dependence on parent detuning  $\Delta_a$; unlike M2 the peak $\Edot$ is not at $\Delta_a=0$ but instead at the shoulders with $\Delta_a \simeq \gamma_{\rm eff}$ (see Figure~\ref{fig:dE_lorentzian}). The major dip is still due to the transition from one resonant mode to the other (the same as in Model 2). The minor dip is due to the decrease of dissipation rate at exact resonance (with $\Delta_a=0$) compared to at the shoulder (with $\Delta_a\simeq \gamma_{\rm eff}^{\rm (M1)}$). The minor dips are not in the middle of two major ones because while $\Edot$ is symmetric about the resonance, the GW decay rate increases monotonically as $P_{\rm orb}$ decreases.

\section{Traveling-wave solution}
\label{sec:TW}
In the main text we compare the linear traveling wave results studied previously to our weakly nonlinear standing wave results.  In this Appendix, we describe the traveling-wave solution of the dynamical tide in more detail. Our approach closely follows that presented in \citetalias{Fuller:12a}. 

To linear order, the inhomogeneous equations describing the perturbed fluid flow are (see, e.g., \citealt{Lai:94})
\begin{align}
\left(r^2 \xi^r\right)'& = \frac{g}{c_s^2} r^2 \xi^r + \left[\frac{l(l+1)}{\omega^2} - \frac{r^2}{c_s^2}\right]\frac{\delta P}{\rho} + \frac{l(l+1)}{\omega^2} U,  \label{eq:xi_r_inh}\\
\left(\frac{\delta P}{\rho}\right)' &= (\omega^2 - \mathcal{N}^2) \xi^r+ \frac{\mathcal{N}^2}{g}\left(\frac{\delta P}{\rho}\right) - U', \label{eq:delP_rho_inh}\\
\xi^h(r) &= \frac{1}{\omega^2 r}\left[ \frac{\delta P}{\rho} + U \right],
\label{eq:xi_h_inh}
\end{align}
where primes denote partial derivatives with respect to radius, $\omega$ is the tidal forcing frequency, $\delta P$ is the Eulerian pressure perturbation, and in this section $U=U(r)= -W_{lm} \left( G M' /D\right) \left(r/D\right)^l$ is the radial dependence of the tidal potential. We neglect perturbations to the gravitational potential of the primary, i.e., we make the Cowling approximation, which is reasonable given the short wavelength of the dynamical tide for the orbital periods of interest.  

The displacement field can be expressed as sum of equilibrium tide and dynamical tide components, i.e., $\xi^{r} =\xi^{r}_{\rm eq} + \xi^{r}_{\rm dyn}$ and similarly for the horizontal displacement, where in the Cowling approximation
\begin{align}
\xi^r_{\rm eq} &= -\frac{U}{g},  \label{eq:xi_r_eq}\\
\xi^h_{\rm eq} &= -\frac{1}{l(l+1) r}\left(\frac{r^2 U}{g}\right)'. \label{eq:xi_h_eq}
\end{align}
In the traveling-wave limit, the shear at the outer boundary is assumed to be sufficiently large that the dynamical tide component  breaks locally at the location where $k_r \xi^r$ peaks (see Figure~\ref{fig:shear_profile}). To ensure that only an out-going wave exists, we impose a radiative outer boundary condition at the peak of $k_r \xi^r$ (corresponding to $r\simeq 0.994 R$ for the $T_{\rm eff}=9000\,{\rm K}$ WD model) given by 
\begin{equation}
\left(\xi^h - \xi^h_{\rm eq}\right)' = \left[\frac{-(\rho r^2/k_r)'}{2(\rho r^2/k_r)} - \imag k_r \right] \left(\xi^h - \xi^h_{\rm eq}\right). 
\end{equation}
At the inner boundary we apply the regularity condition $\omega^2 r \xi^r = \delta P/\rho + U$.  The set of inhomogeneous Equations (\ref{eq:xi_r_inh})-(\ref{eq:xi_h_inh}) can then be solved to obtain the perturbed displacement field. 

The wave carries a net angular momentum flux 
\begin{equation}
\dot{J}_z(r) = 2 m \omega^2 \rho r^3 {\rm Re}[\imag \left(\xi_{\rm dyn}^{r} \right)^\ast \xi_{\rm dyn}^h], 
\label{eq:ang_flux}
\end{equation}
which becomes a positive constant near the outer boundary since the wave is purely out-going. This constant corresponds to the tidal torque exerted on the star, which can be expressed as 
\begin{equation} 
\tau_{\rm tide} = E_0 \left(\frac{\mratio}{1+\mratio}\right)^2 \left(\frac{\Omega_{\rm orb}}{\omega_0}\right)^4 F(\omega),
\label{eq:tau_tide_tw}
\end{equation}
where the function $F(\omega)$ can be  approximated as [equation (78) in \citetalias{Fuller:12a}]
\begin{equation}
F(\omega) \simeq \hat{f} \left(\frac{\omega}{\omega_0}\right)^5,
\label{eq:F_omega_approx}
\end{equation}
with $\hat{f}$  a dimensionless constant that depends on the internal structure of the WD. 

\begin{figure}
   \centering
   \includegraphics[width=0.45\textwidth]{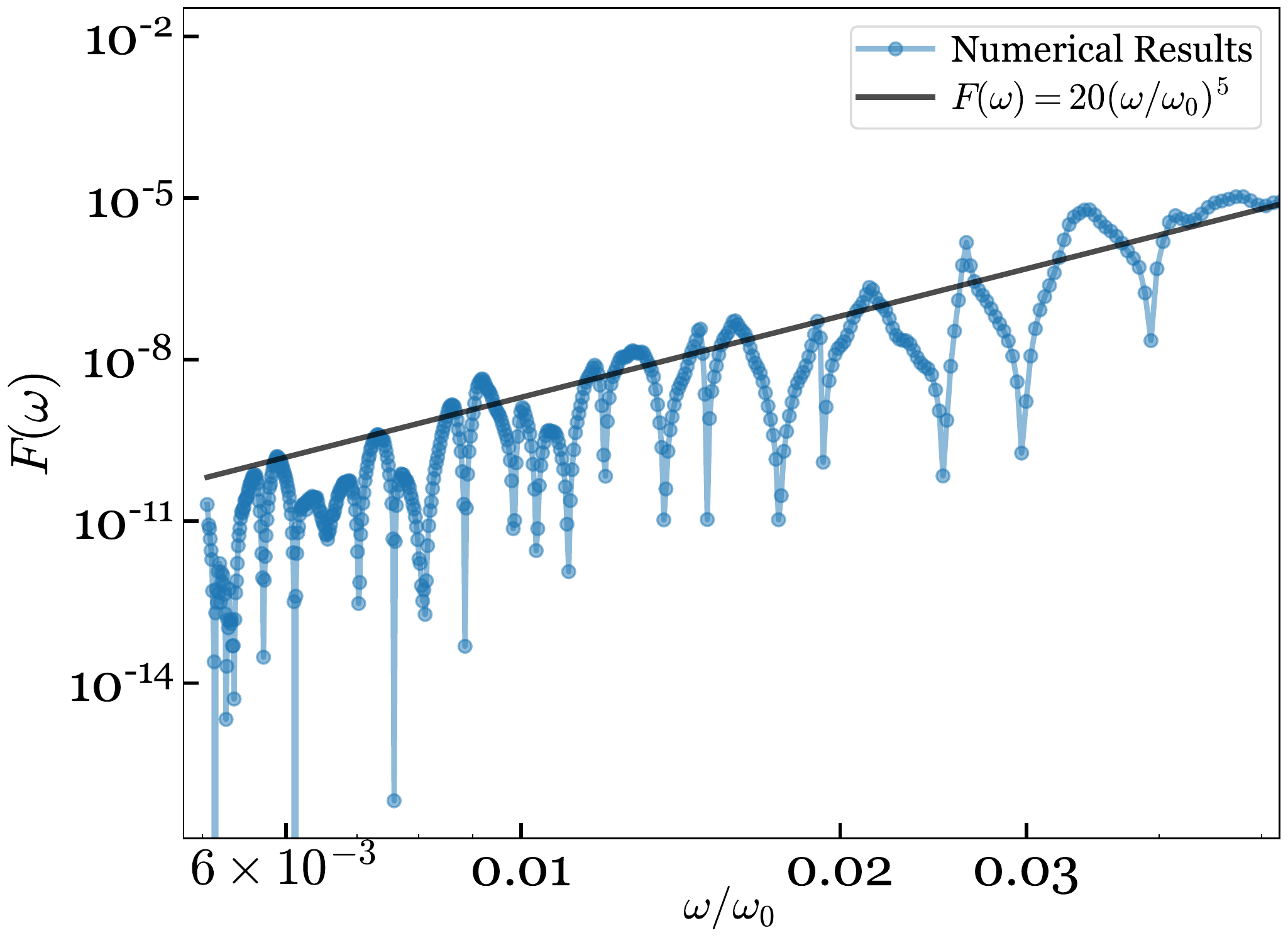} 
   \caption{The traveling wave tidal torque function $F(\omega)$ as a function of $\omega = m(\Omega_{\rm orb} - \Omega_{\rm s})$ [blue circles; see Equation~(\ref{eq:F_omega_approx})]. 
   The black line shows the approximation $\hat{f} (\omega/\omega_0)^5$ with $\hat{f} = 20$. The result is in good agreement with the calculation in \citetalias{Fuller:12a} for a similar WD model. }
   \label{fig:F_omega}
\end{figure}

In Figure~\ref{fig:F_omega}, we show the numerical calculation of $F(\omega)$ (blue dots) and the approximation given by  Equation~(\ref{eq:F_omega_approx}) for
our $T_{\rm eff}=9000\,{\rm K}$ WD model (solid-grey line). We find that $\hat{f}=20$ provides a reasonable fit to the numerical result.
This agrees well with the results of \citetalias{Fuller:12a} for their $0.6\,M_\odot$ WD model with $T_{\rm eff}{=}8720\,{\rm K}$. 

To understand the scaling of the traveling-wave shear shown in Figure~\ref{fig:shear_TW}, we first note that near the surface, the horizontal and radial components of the wave satisfies \citepalias{Fuller:12a}
\begin{equation}
\xi_{\rm dyn}^h = -\imag \frac{k_r r}{l (l+1)}\xi_{\rm dyn}^{r} \propto \frac{\xi_{\rm dyn}^r}{\omega}. 
\end{equation}
Combining this with Equations~(\ref{eq:ang_flux})-(\ref{eq:F_omega_approx}) gives
\begin{equation}
\xi_{\rm dyn}^r \propto \Omega_{\rm orb}^2 \omega^2,\ \text{and\ \ } k_r\xi_{\rm dyn}^r \propto \Omega_{\rm orb}^2 \omega. 
\end{equation}
For a non-rotating WD, $\omega = 2\Omega_{\rm orb}$, and we obtain the $\Porb^{-3}$ scaling shown in Figure~\ref{fig:shear_TW}.

\bsp	
\label{lastpage}
\end{document}